# Determination of Dielectric Functions and Exciton Oscillator Strength of Two-Dimensional Hybrid Perovskites

Baokun Song,[†,‡,#] Jin Hou,[$,&,#] Haonan Wang,[§] Siraj Sidhik,[$] Jinshui Miao,[†] Honggang Gu,[‡] Huiqin Zhang,[†] Shiyuan Liu,[‡,*] Zahra Fakhraai,[§] Jacky Even,[∥] Jean-Christophe Blancon,[$] Aditya D. Mohite,[$,&,*] Deep Jariwala[†,*]

[†]Department of Electrical and Systems Engineering, University of Pennsylvania, Philadelphia, PA 19104, USA

[‡]School of Mechanical Science and Engineering, Huazhong University of Science and Technology, Wuhan 430074, P. R. China

[$]Department of Chemical and Biomolecular Engineering Rice University, Houston, TX 77005, USA

[&]Department of Material Science and Nanoengineering Rice University, Houston, TX 77005, USA

[§]Department of Chemistry, University of Pennsylvania, Philadelphia, PA 19104, USA

[∥]Univ Rennes, INSA Rennes, CNRS, Institute FOTON, UMR 6082, Rennes F-35000, France

**ABSTRACT:**

Two-dimensional (2D) hybrid organic inorganic perovskite (HOIP) semiconductors have attracted widespread attention as a platform of next generation optoelectronic devices benefiting from their naturally occurring and tunable multiple quantum-well like (QW) structures, which enable a wide range of physical properties. Determining the intrinsic optical/electronic properties of 2D HOIPs is extremely important for further utility in photonic and optoelectronic devices. Here, we obtain the optical dielectric functions, complex refractive indices, and complex optical conductivities of both Ruddlesden-Popper (RP) and Dion-Jacobsen (DJ) phases of 2D HOIPs as a function of the perovskite QW thickness via spectroscopic ellipsometry over a broad energy range of 0.73–3.34 eV. We identify a series of feature peaks in the dielectric functions, and explain the evolution of ground state exciton peak with unit cell thickness and changing excitonic confinement. We observe extraordinary values of optical extinction and electric loss tangents at the primary excitonic resonances and provide their detailed comparison with other known excitonic materials. Our study is expected to lay foundation for understanding optical properties of pure phase 2D HOIPs, which will be helpful for the accurate modelling of their photonic and optoelectronic devices.

INTRODUCTION:

Two-dimensional hybrid organic-inorganic perovskites (2D HOIPs) are a class of layered materials fast emerging as promising semiconductors for optoelectronics and energy harvesting applications.[1–7] This surge in interest is due to several factors, namely (i) the physical properties of 2D HOIPs can be modulated by changing the structure (layer thickness and phase) and composition of the inorganic and organic components;[8] (ii) strong quantum and dielectric confinement of charges at room temperature resulting in important excitonic effects;[9,10] (iii) self-passivated layers with minimum electronic interaction between them that help maintain a direct band gap and high luminescence quantum yield in the bulk unlike several other layered 2D semiconductors;[11,12] (iv) enhanced stability in devices as compared to their 3D perovskite counterparts.[5,13,14]

The 2D HOIPs exist in a few structural phases, including the Ruddlesden-Popper (RP) phases, Dion-Jacobsen (DJ) phases, and the Aurivillius phases.[8] The general formulas of the prototypical members in the 2D RP and DJ phase compounds are $(BA)_2(MA)_{n-1}Pb_nI_{3n+1}$ and $A'(MA)_{m-1}Pb_mI_{3m+1}$, where BA = $CH_3(CH_2)_3NH_3$, MA = $CH_3NH_3$, A' = 4-(aminomethyl)piperidinium (4AMP), and the integers $n$, $m$ determine the thickness of the lead-iodide perovskite layers.[15,16] The 2D HOIPs can be described as quantum-well (QW) systems where the lead-iodide perovskite layers are the QWs and the organic cation layers separating them are the dielectric barriers.[7,15,16] The optical and electronic properties of 2D RP and DJ perovskites, such as absorbance, bandgap, and photoluminescence (PL) spectra, can be tuned by changing $n, m$ as well as the organic components.[15–19] For example, in the RP perovskites the exciton binding energy ($E_b$) can be modulated by one order of magnitude by increasing $n$ from $n = 1$ (~500 meV) to infinity corresponding to the 3D HOIPs (a few tens of millielectronvolts).[16] These properties have fueled studies on their synthesis,[6,8,15,16] fundamental physical phenomena[7,17] and applications in a variety of optoelectronic devices,[6,20,21] including photovoltaics,[14,22–25] light-emitting diodes (LED),[26–28] photodetectors,[29] and lasers.[30,31]

While numerous studies have now been made on opto-electronic and photonic devices from 2D HOIPs as mentioned above, technology transfer of these proof-of-concept devices to the market will necessitate design and device modelling by use of Maxwell's equations in order to simulate the flow of light in these devices. To achieve this, the accurate knowledge of optical constants and conductivities of the materials is critical. Thus far, a majority of research on the optical properties of RP and DJ perovskites has been focused on the measurement and analysis of PL,[15,16,32] absorption,[18] and Raman spectra.[15] However, there is still a need for a systematic study on the fundamental optical parameters of both of these phases of 2D HOIPs as a function of thickness of the perovskite layers and for large area (>1 mm) single crystals practical for photonic and opto-electronic applications. This study will expand the previous knowledge of the optical constants in 2D HOIPs limited to specific materials and spectral ranges.[33–35]

Here we report a comprehensive and systematic spectroscopic ellipsometry (SE) study of phase pure RP ($n$ = 1–5) and DJ phase ($m$ = 1–4) perovskites. We provide detailed analysis of dielectric function, complex refractive index, complex optical conductivity, absorption coefficient ($\alpha$), electric loss tangent (ELT), energy loss function (ELF), and absorption cross section ($\sigma_{acs}$) of these 2D RP and DJ HOIPs over a broad spectral range of 0.73–3.34 eV (371–1687 nm). We report the inherent optical/dielectric constants and parameters of 2D DJ phase HOIPs mentioned above which to the best of our knowledge have not been reported in prior literature. In addition, we present the previously unreported complex optical conductivity, ELT, ELF of 2D RP phase HOIPs. We also analyze and describe the trends of oscillator strength and damping coefficient of the ground state exciton peaks as a function of perovskite layer thickness and provide a quantitative comparison of extinction and loss-tangent of 2D RP and DJ HOIPs with other known excitonic semiconductors. The ground state exciton carries the maximum oscillator strength in the 2D HOIPs as confirmed again from our dielectric function spectra. Further, 2D HOIPs with small perovskite layers exhibit extraordinarily large loss tangent for large band-gap energies among all known materials suggesting their applicability as gain media in coherent light sources.

## RESULTS AND DISCUSSION

We synthetized phase pure RP ($n$ = 1–5) and DJ ($m$ = 1–4) perovskite crystals with area larger than one millimeter and thickness of a few hundreds of nanometers, whose specific crystal structures aresketched in Figure 1(a).[15,16,36,37] These structures show that the organic barriers have the thicknesses of about 7.1 Å (bilayer BA) and 4.1 Å (monolayer 4AMP) in the RP and DJ phases.[8] In RP perovskites, the thickness of the perovskite layers varies from about 0.64 nm in $n$ = 1 to 3.14 nm in $n$ = 5, and displays a linear dependence in $n$-value; similar observations were obtained for the DJ perovskites. The simulated and experimental out of plane lattice parameters (along the stacking direction) of RP and DJ perovskites can be found in Table S1 and Refs. 15,16. Optical images of the synthesized crystals on glass substrates are depicted in Figure 1(b) and Figure S1. The characteristic sizes of RP and DJ perovskite crystals are up to centimeter-level and millimeter-level, respectively. The x-ray diffraction (XRD) patterns of the RP and DJ perovskites are shown in Figure 1(c,d). We confirmed the structure and phase quality of the crystals by examining the diffraction patterns and comparing them to the ones simulated from the structures in Figure 1(a). As shown in Figure 1(c,d), we assigned Miller indices (*hkl*) to each diffraction peak. Figure 1(e) shows the illumination spot of the ellipsometry measurement on the RP ($n$ = 2) sample, where the green ellipse area marks the shape and size of the measurement spot. The long axis of the elliptical spot is about 1 mm under 70° incidence angle. A pair of focusing and collecting lens attachments are used to obtain a small illumination spot, which ensures that the light beam lies completely within the single crystalline perovskite samples.

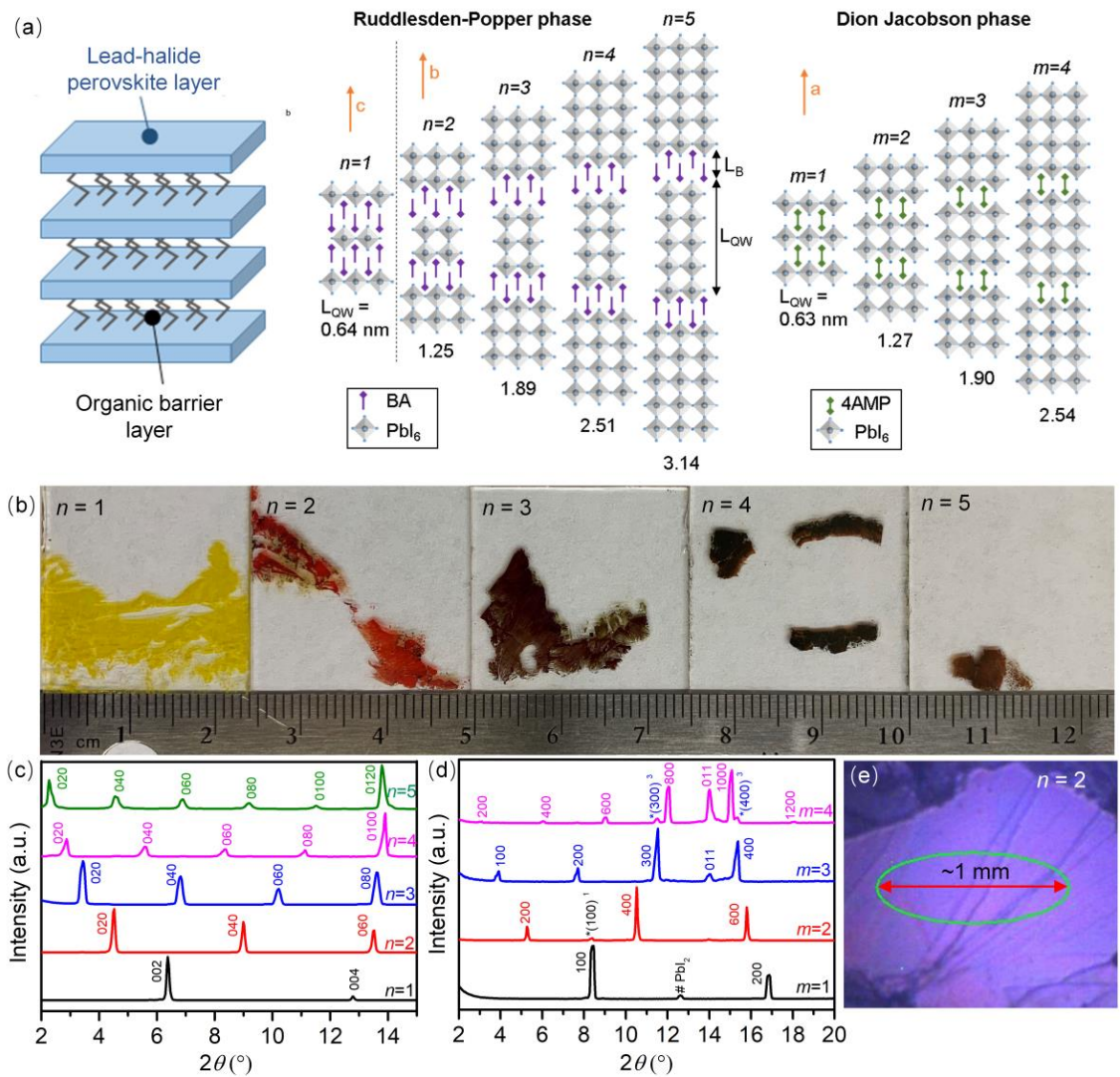

**Figure 1.** (a) Left: Schematics of the QW-like structure showing perovskite layers sandwiched between organic spacing layers. Right: Crystal structure schematics of 2D Ruddlesden-Popper (RP) and Dion-Jacobsen (DJ) hybrid organic inorganic perovskites with varying $n$ and $m$. The $L_{QW}$ denotes the thickness of the inorganic perovskite in each compound. The numerical values denote the distance between the terminal iodide ions of each layer and were determined directly from the refined crystal structures. The orange arrows indicate the direction of the stacking direction in the base (a, b, c). (b) Optical photographs of 2D RP perovskites. (c,d) X-ray diffraction patterns of the 2D RP and DJ perovskite crystals. The diffraction peaks are identified with the Miller ($hkl$) indices derived from the crystal structures. (e) Detection spot of the spectroscopic ellipsometer on the RP ($n = 2$) crystal. The green ellipse indicates the specific measurement position, and the long axis of the elliptical spot is about 1 mm.

The dielectric function spectra and complex refractive indices of the 2D RP and DJ perovskite samples were determined by SE. Our ellipsometer (J.A. Woollam, M-2000) had a detection range from 0.73 eV to 3.34 eV (371–1687 nm). To improve the reliability of the results, we measured every sample at multiple incident illumination angles ranging from 50° to 70° (with respect to the normal of the sample/substrate plane) with a step size of 5°. Upon obtaining the raw ellipsometric parameters $\Psi$ and $\Delta$, where $\tan(\Psi)$ and $\Delta$ refer to the amplitude ratio and phase difference between the reflection coefficient of p-polarization and s-polarization components of the incident light,[38] we used vertical stacking optical models and dielectric function models to derive the geometrical structures and dielectric dispersion functions of the 2D RP and DJ phase perovskites. For the dielectric function model, the Kramers-Kronig constrained B-spline[39,40] fitting is first adopted to embody the dielectric responses of RP and DJ perovskite crystals. Following that a combined Tauc-Lorentz oscillator model[40–43] is used to parameterize the dielectric functions of the 2D RP and DJ perovskite crystals obtained from the B-spline fitting. Finally, the theoretical parameters of the ellipsometric spectra are calculated by the film transfer matrix method.[41] The optical dielectric functions and complex refractive indices of 2D RP and DJ perovskite crystals can be extracted by fitting the measured spectra with the calculated ones using the above models (Figures S2–S4). The optical constant and dielectric function models used yield an excellent goodness of fit (*RMSE* in Table S2) suggesting that they can effectively describe the data. The optical/dielectric constants obtained via the above analysis are isotropic and reflect the overall dielectric response. Isotropic constants are sufficient for the analysis of the oscillator strength and absorption features of these two kinds of HOIPs as has been done below. Besides, such analysis also enables fair comparison with optical/dielectric constants of the existing excitonic materials which are also isotropic. We also extract the out-of-plane dielectric function of 2D RP ($n$ = 1) and DJ ($m$ = 1) phase HOIPs and plot them in the supporting information to provide some guidance on the extent of variation and difference in magnitudes between in and out of plane dielectric response of HOIPs. We note that estimating anisotropic dielectric functions for van der Waals excitonic crystals is a challenge which is limited by available lateral size and thicknesses along arbitrary directions for single crystals. Most of the HOIPs crystal thicknesses as determined by SE are several hundred nanometers as shown in Table S2, meaning that they are suitable for the development of the next-generation photonics and optoelectronic devices.[5,22,26,44] Detailed information about the ellipsometric measurement, fitting and analysis are available in the ellipsometry measurement and analysis section of the supporting information and in prior publications.[45–52]

The real and imaginary parts of the dielectric function spectra ($\varepsilon_r$ and $\varepsilon_i$) of the RP ($n$ = 1–5) and DJ ($m$ = 1–4) perovskites are presented in Figure 2(a,b,e,f), Figure S5, Table S3, and Table S4. Particularly, $\varepsilon_r$ reflects the capacity of a material to propagate the light and $\varepsilon_i$ is correlated to the absorption properties of the material. The dielectric functions presented here are in agreement with prior results.[35,53] We do not observe any negative values in the $\varepsilon_r$ spectra of RP ($n$ = 1, 2) as opposed to those claimed in prior literature.[33] The dash lines in Figure 2(b) indicate that four Tauc-Lorentz oscillators are combined to deconvolute the $\varepsilon_i$ spectrum of RP ($n$ = 2) (see Figure S6 for clearer fitting details). The amplitude reflects the oscillator strength of these feature peaks to some extent. All dielectric function spectra exhibit a pronounced and sharp peak identified to the ground exciton 1$s$ of 2D HOIPs and weaker features (i–iii) at higher energy corresponding to excited exciton states and inter-band transitions, as described in our previous work.[10] As a result of the strong quantum and dielectric confinement effects in the direction normal to the perovskite layers, a significant amount of the total oscillator strength is concentrated in the excitonic features.[9,10,54–56] This effect is particularly strong for RP phase ($n = 1$) crystal since the lead-iodide inorganic quantum well core is of sub-nanometer width (~6.41 Å) coupled with the weak dielectric shielding from the organic barrier layers.[9] The sharp and intense ground-state exciton peaks in $\varepsilon_i$ of RP and DJ perovskites (Figure 2(b,f)) are mainly associated with the transitions at the bandgap that occur between Pb (s-orbitals) – I (p-orbitals) in the valence bands and Pb (p-orbitals) in the conduction bands.[10,16] The transitions occur at the $\Gamma$ point in BA (RP) but at the Brillouin zone edge in 4AMP (DJ).[16]

The broadband complex refractive indices $N$ of the 2D RP and DJ perovskite crystals are plotted in Figure 2(c,d,g,h), including the refractive index $n$ ($n$ denotes the refractive index) and the extinction coefficient $\kappa$. The feature peaks in Figure 2(d,h) are traceable to the same physical origins as the peaks in Figure 2(b,f). Compared to the complex refractive index of the related 3D HOIPs,[57,58] the $n$ and $\kappa$ of 2D RP perovskite crystals reported here exhibit ground exciton peaks with smaller linewidth and larger oscillator strength, which reflects the enhancement of both quantum and dielectric confinement effects in 2D systems. The measurement spectral range of the complex refractive index in this work is also wider than prior reports relevant to the 2D RP perovskites,[5,35,59] which helps us to identify some high-energy characteristic peaks that have not been found before.

By observing $\varepsilon_r$, and $n$ (refractive index), we find that the intensity of these non-energy loss components (the parts associated with the ability of propagate the electromagnetic wave) gradually increases with $n$ and $m$ in the infrared region (about less than 1.5 eV).[60] Since this is sub-gap portion of the spectrum, it is expected that atomic, molecular and bond polarizabilities will mainly contribute to the dielectric response in this frequency/energy range as opposed to any electronic excitations at higher energy. Since the perovskite layers increase in thickness with

- 9 -

increasing *n* and *m*, they are expected to raise the overall dielectric response as well. Further, this phenomenon can also be interpreted by the effective medium approximation (EMA) model $\varepsilon_\infty = (\varepsilon_{QW}L_{QW} + \varepsilon_{B}L_{B})/(L_{QW}+L_{B})$,[9] where, $\varepsilon_\infty$ defines the effective high frequency dielectric constant. For the 2D HOIPs, $\varepsilon_B \approx 2.1$ and $\varepsilon_{QW} \approx 6.5$,[59] $L_B$ = 0.71 nm and 0.41 nm for the 2D RP and DJ perovskites, the value of $L_{QW}$ has been listed in Figure 1(a). The calculated results (Table S2) also suggest that $\varepsilon_\infty$ gradually increases with *n* and *m* (or $L_{QW}$).[10,59,61]

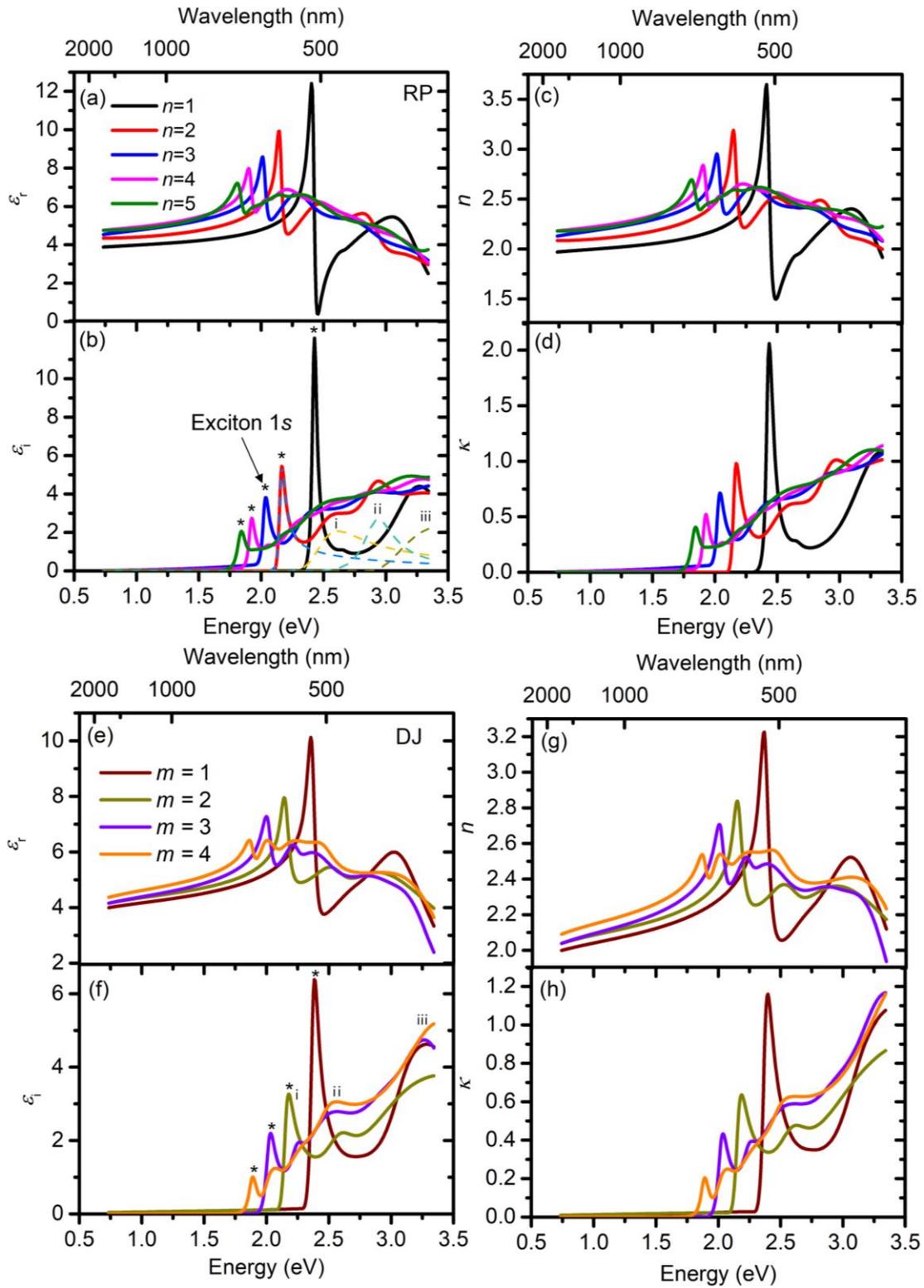

**Figure 2.** Dielectric function spectra and complex refractive indices of 2D RP phase ($n$ = 1–5) and DJ phase ($m$ = 1–4) perovskites.

As illustrated in Figure 2(b,f), the oscillator strength of the ground state exciton peak in the $\varepsilon_i$ spectra exhibits discernible changes with the increasing *n* and *m*. To reveal the specific revolution trends, the amplitude (*Amp*) and damping coefficient ($\Gamma$) of the Tauc-Lorentz oscillator used to describe the ground exciton peak are extracted and plotted in Figure 3(a,b). These values are obtained under the best ellipsometric fitting conditions, and the error bars in Figure 3(a,b) come from the ellipsometry software and with 90% confidence intervals. The amplitude of the ground-state excitonic peak in both the RP and DJ perovskites decreases monotonously with the increasing *n* and *m*. It can be mainly attributed to the decrease in both the quantum and dielectric confinement, concomitant with a reduction in both the binding energy and oscillator strength of the ground exciton state.[10] The damping coefficient $\Gamma$ of the RP perovskites increases with *n* resulting from the gradually enhanced dielectric screening in the perovskite layers. On the other hand, $\Gamma$ in the DJ perovskites attenuates slightly with increasing *m*. We hypothesize that this behavior results from the non-negligible electronic coupling between perovskite layers through van der Waals iodine-iodine interactions in the dielectric barriers, and which decreases for larger perovskite layer thickness (higher *n*).

Figure 2(b,f) and Figure 3(a) show that the strength of the dielectric function of the DJ phase (*m* = 1–4) is weaker than that of RP phase (*n* = 1–4) over the whole measured energy range. Given that the composition and thickness of the perovskite layers are identical in both the RP and DJ perovskites when *n* = *m*, quantum confinement effects are not responsible for the different dielectric responses. Therefore, we interpret the reduction in the dielectric constant and oscillator strength of the ground exciton principally to changes in the organic barrier thickness (0.71 nm in RP versus 0.41 nm in DJ). For example, the oscillator strengths can be calculated using the simple formulation proposed by Ishihara et al.[9] $f_{exc} = \varepsilon_\infty \omega_{LT} \omega_T m_e V_0 / 2\pi e^2$, where $m_e$ and $e$ denote the mess and charge of electron, $V_0$ is the volume of the formula unit[15,16,18] (Table S2), $\omega_{LT}$, $\omega_T$ refers to the frequencies of the longitudinal-transverse exciton splitting and the transverse exciton. The transverse exciton energy is difficult to obtain due to the interference effect and backside reflection in layered 2D RP and DJ perovskite crystals. Thus, the ground state exciton energy[15,16] of 2D RP and DJ perovskites is introduced into the calculation of $f_{exc}$ to replace $\omega_T$,[9] and $\omega_{LT}$ is estimated from Ref. 9. Finally, the oscillator strengths of ground excitons in 2D RP ($n$ = 1–4) and DJ ($m$ = 1–4) perovskite crystals are about 0.40±0.25 and 0.33±0.15 (Figure 3(c)). The $f_{exc}$ of RP ($n$ = 1) is 0.61, which agrees with the previously reported result of 0.7.[9] Considering the simplicity of EMA model parameter $\varepsilon_\infty$, the oscillator strengths presented here are just estimates that are used for the comparative study of 2D RP and DJ perovskite crystals. Nevertheless, the decreasing $f_{exc}$ with $n$ and $m$ matches with the amplitude changing trends shown in Figure 3(a). Additionally, the oscillator strengths of 2D RP are larger than those of 2D DJ in the case of $n = m$. It is mainly due to the larger $V_0$ of 2D RP perovskite crystals. Furthermore, the value of $V_0$ primarily depends on the thickness of organic interlayer which is the larger band-gap or "insulating" barrier $L_B$. A larger $L_B$ will more effectively isolate the excitons in the inorganic perovskite layer, thereby eliminating any interlayer electrostatic interaction and enhancing the excitonic oscillator strength. The image charge effects should also be more pronounced in 2D RP phase HOIPs since the quantum-well/barrier dielectric contrast will be more distinct without any wavefunction overlap or perturbation from the adjacent inorganic layers.[62] Both, the strong 2D spatial confinement and image charge effects therefore play an important role in the larger $f_{ex}$ of 2D RP phase HOIPs.

The absorption coefficient $\alpha$ of the RP and DJ perovskites is calculated from $\alpha = 4\pi\kappa/\lambda$ and plotted in Figure 3(d,e). The ground-state exciton absorption peak observe value between

~$1.2\times10^4$ cm$^{-1}$ and up to ~$5.1\times10^5$ cm$^{-1}$, which is one order of magnitude larger than the exciton features in MAPbI$_3$ (3D perovskites) and GaAs.[63,64] Again, this result reflects the low dimensional character of 2D RP and DJ perovskites where the oscillator strength is largely concentrated in the exciton ground-state.[65] Compared to the relatively flat band-edge absorption features of MAPbI$_3$ and traditional III-V semiconductor GaAs,[66] the step-like $α$ of RP ($n = 1$) strongly suggests 2D nature of the density of state (DOS) akin to GaAs quantum wells.[67] The excitons in 2D RP and DJ perovskites can be regarded as strongly bound Wannier excitons owing to the significant quantum confinement and dielectric confinement effect.[11,62] It is obvious from Figure 3(d,e) that the magnitude of absorption in the DJ phase perovskite crystals is smaller as compared to that of their RP counterparts. We infer the underlying cause of the smaller $α$ of 2D DJ perovskite crystals to a weaker dielectric screening resulting from smaller $L_B$. Further, the perovskite interlayer interaction in RP phases is thoroughly screened by the interdigitated bilayer of organic spacer cations (Figure 1(a)).[8] On the contrary, in the case of DJ phases, the two perovskite layers are connected by only one-layer of organic cations which results in some interlayer interaction albeit weak. This interaction is expected to weaken the excitonic oscillator strength, exciton binding energy, and consequently also absorption strength.

In order to further confirm the decreasing $α$ with the increasing $n$, the absorption spectra around the 1$s$ exciton peaks of 2D RP ($n = 1–4$) are theoretically calculated (Figure 3(f)). The overlap degree of electron and hole wave functions is high in the 2D RP perovskite crystals,[55] but it must be considered together with confinement effects, in a semi-empirical Green function-based solution of the Bethe-Salpeter equation (BSE) for the Wannier exciton.[10] The enhancement of the $E_b$ due to the dielectric confinement effect is mitigated by the vertical extension of the monoelectronic state wavefunctions. From the theoretical viewpoint, multilayered 2D RP and DJ perovskite crystals are very large systems that do not allow a full ab-initio resolution of the BSE, but the semi-empirical Green function method developed in Ref. 10 for $n = 1$–4 RP perovskite crystals is nevertheless able to describe advanced features such as diamagnetic shifts or the systematic deviations from the 2D Rydberg series for $n = 1$–4. This method also allows computing the relative change of optical absorption from $n = 1$ to $n = 4$ for discrete states and also the continuum, but only close to the electronic band gaps (Figure 3(f)). The decrease of the $E_b$ for 1s bound states from $n = 1$ to $n = 4$ is correlated to the decrease of the absorption peaks. The computed absorption curves exhibit oscillations well above the band gaps due to discretization errors in the Fourier transform of the BSE.[10]

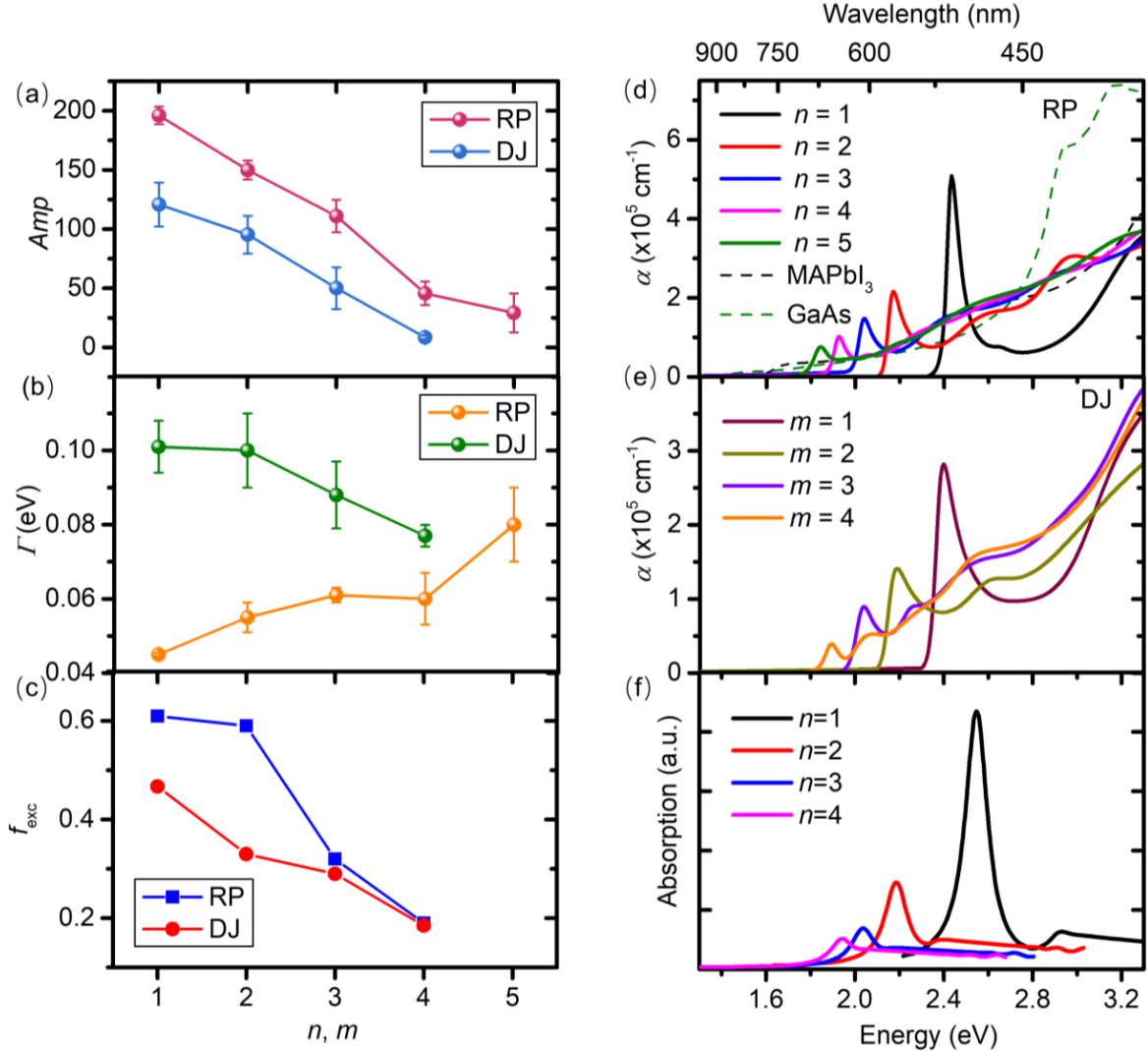

**Figure 3.** (a,b) Amplitude (*Amp*) and damping coefficient ($\Gamma$) of the Tauc-Lorentz oscillator used to describe the ground exciton peak of 2D RP and DJ perovskites. The specific mathematical expression of Tauc-Lorentz oscillator can be found in the supporting information. (c) Ground state exciton oscillator strength of 2D RP and DJ perovskites. (d,e) Absorption coefficients of 2D RP and DJ perovskites, relevant 3D HOIPs MAPbI$_3$, and GaAs. (f) Relative variations of the absorption coefficients $\alpha$ of 2D RP ($n$ = 1–4) phase perovskites deduced from a semi-empirical resolution of the Bethe-Salpeter equation.

To evaluate the expected performances of these materials in optoelectronic/photonic systems, we also derive the complex optical conductivity, electric loss tangent ELT, and energy loss function ELF of 2D RP and DJ perovskite crystals. The complex optical conductivity $\sigma$ of an optoelectronic material is an important parameter, which is essential for the design of photodetectors and solar cells. In general, $\sigma_r$ represents the energy loss component caused by the light-induced conduction current, and $\sigma_i$ characterizes the charge storage capacity of material, which is directly associated with the light-induced displacement current.[48] We calculated the broadband $\sigma$ of 2D RP and DJ perovskite crystals from the dielectric function spectra by the relations: $\sigma_r = \omega\varepsilon_0\varepsilon_i$ and $\sigma_i = \omega\varepsilon_0(\varepsilon_r - 1)$ ($\omega$ and $\varepsilon_0$ stand for the angular frequency of light and the free space permittivity). Figure 4(a–d) presents the real parts $\sigma_r$ and imaginary parts $\sigma_i$ of optical conductivities for these two kinds of HOIPs. These spectra exhibit similar features as the real and imaginary parts of permittivity in Figure 2. To our best knowledge, the $\varepsilon$, $N$, and $\sigma$ of the DJ phase 2D perovskites are systematically reported for the first time, which is expected to provide some parameterized guidance for the structure design and performance optimization of related optoelectronic devices.

The extent of light-trapping in a medium in both the ray optic (bulk) and subwavelength dimensions[68] may be quantified by the ratio of imaginary and real parts of dielectric function, that is, electric loss tangent ELT. Figure 4(e) demonstrates the ELT of 2D RP and DJ perovskite crystals, where the intensity of the of ELT for RP ($n$ = 1) phase perovskites is an order of magnitude larger than those of other materials. The FWHM for EFL ($n$ = 1) is less than 0.05 eV. The intense and sharp electric loss means that the oscillations of electrons and holes driven by the light-induced electric field are closely resonant with the excitation frequency. The sharp line width of the ELT further affirms that the wave functions of electrons and holes in the excitons of $n$ = 1 RP phase crystals are highly overlapping.[55] Such narrow linewidths have also been observed in absorption/extinction spectra of J-aggregate dyes that exhibit coherently coupled transition dipole moments,[69,70] though there is no currently available evidence to suggest coherent coupling of excitonic dipoles in 2D HOIPs. With the increase in $n$, the increased dielectric shielding weakens the overlap of wave functions and oscillator strength in the excitons, which directly gives rise to the larger FWHMs and weaker intensity of ELTs for the RP ($n$ = 2–4) perovskites. Nevertheless, the ELTs of RP ($n$ = 2–4) perovskites are still much larger than those of the 3D RP phase ($n = \infty$) counterpart and the traditional fully inorganic QW material GaAs.[71–73] More importantly, the positions of loss peaks of RP phase perovskites can be tuned flexibly by changing $n$. The ELTs of DJ phase also exhibit the similar optoelectronic characteristics like those of RP phase, while with weaker intensity and broad FWHM.

In order to provide a more comprehensive guidance for the quantitative design of optoelectronic devices, particularly ones entailing maximization of optical absorption or coherent emission, the knowledge of the energy loss function (ELF) (also known as surface energy loss function) and the absorption cross section ($\sigma_{acs}$) is critical. We present the ELF and $\sigma_{acs}$ of the 2D RP and DJ perovskites in Figure 4(f,g). Conceptually, the ELF value at a given energy describe the energy lost by an electron passing through a homogeneous dielectric material. It is a measure or characterization of inelastic scattering processes that the electron undergoes. Typically, it is difficult to determine a momentum dependent energy loss function experimentally, therefore, techniques for deriving it from optical dielectric constants have been adopted. It is calculated by

$$ELF = \mathrm{Im}\left(\frac{-1}{\varepsilon(\omega)}\right) = \frac{\varepsilon_i}{\varepsilon_r^2 + \varepsilon_i^2}. \qquad (1)$$

The sharp energy loss peaks in Figure 4(f) reflect the intense energy dissipation of an incident electron induced by the prominent exciton transitions. The $\sigma_{acs}$ is another important parameter particularly for the design of continuous wave pumped laser and nonlinear optical elements.[74] A high $\sigma_{acs}$ is helpful for achieving continuous wave pumped lasers since it has been established that the two-photon absorption (TPA) cross-section is proportional to the linear absorption cross-section.[75–77] An accurate linear $\sigma_{acs}$ is often equal to the TPA and therefore helps reduce the discrepancy of reported values of TPA.[75,78] Generally, the $\sigma_{acs}$ is given by $\sigma_{acs} = \frac{\alpha}{N} = \alpha \frac{V}{m} \cdot \frac{M}{NA} = \alpha \cdot \frac{M}{\rho \cdot NA}$, where, $N$, $\rho$, NA, and $M$ represent the atomic density (cm$^{-3}$), mass density (g·cm$^{-3}$), Avogadro's constant (6.02214076×10$^{23}$), and relative atomic mass (g), respectively. The mass densities of the 2D RP and DJ perovskites are from previous references.[15,16,18,36,79] Figure 4(g) shows that the $\sigma_{acs}$ of 2D RP and DJ perovskites is significantly greater than that of the 3D HOIPs and traditional in-organic gain medium materials such as GaAs, indicating that the 2D RP and DJ perovskites could have potential applications in coherent light sources.

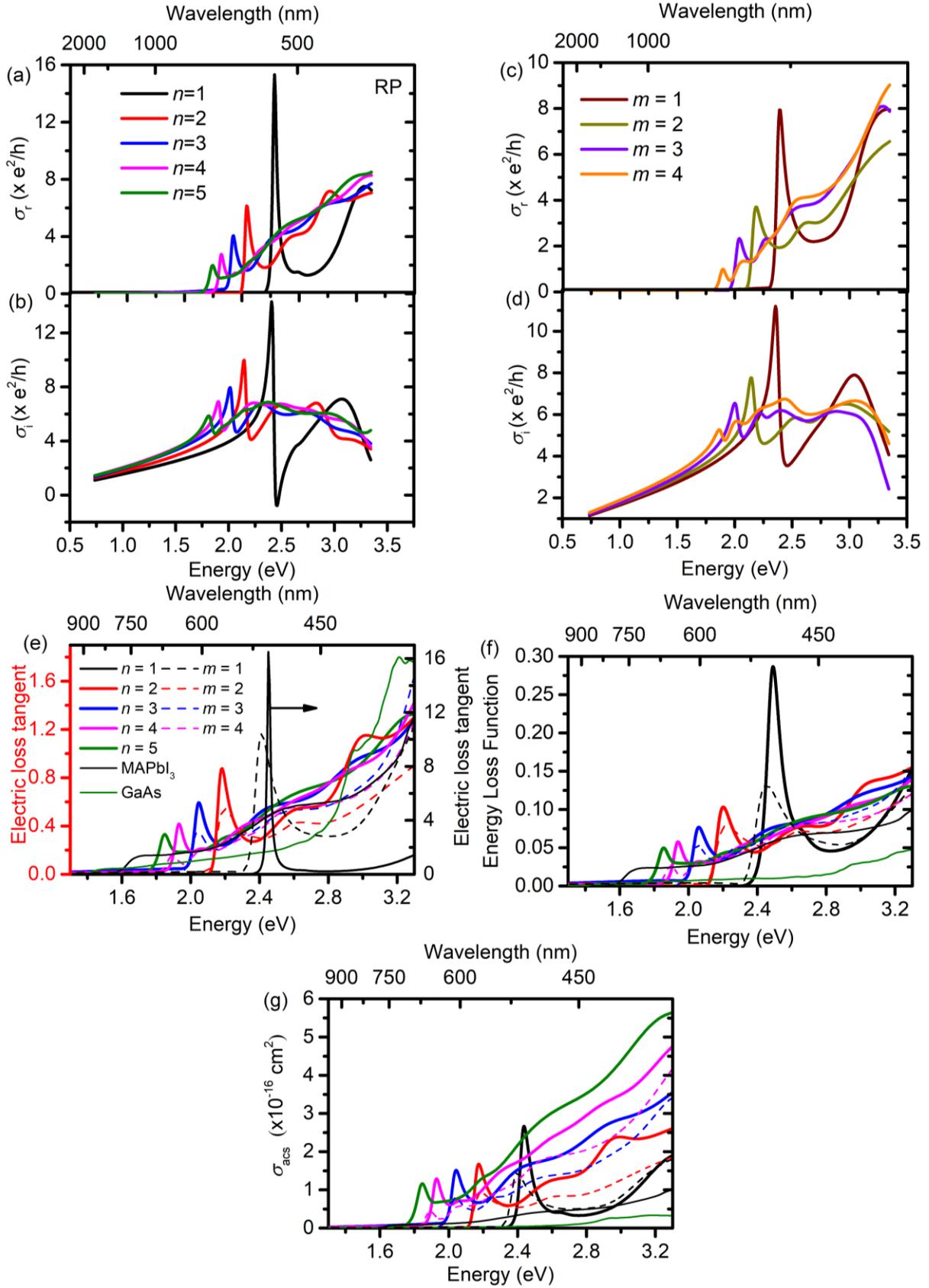

**Figure 4.** (a–d) Complex optical conductivities (reduced by $e^2/h$), (e) electric loss tangent ELT, (f) energy loss function ELF, and (g) absorption cross section spectra $\sigma_{acs}$ of 2D RP and DJ perovskites. The vertical coordinate of ELT for the RP ($n = 1$) phase crystal is on the right.

To compare and evaluate these optical properties and the impact of confinement on excitations in 2D RP and DJ perovskites, we have plotted their extinction coefficient and electric loss tangents as a function of transition energy of the primary excitons as seen in Figure 5. In the same map we have also plotted select well-known excitonic materials for appropriate comparison Figure 5(a).[63,80–87] Broadly speaking, the excitonic semiconductors plotted in Figure 5(a) can be roughly divided into four categories. (I) Bulk layered and non-layered compounds, namely transitional metal dichalcogenides (TMDCs), PbI$_2$, and CdSe.[80–82] (II) Low-dimensional excitonic materials, including monolayer 2D TMDCs (1L-WS$_2$, 1L-MoS$_2$, 1L-WSe$_2$, 1L-MoSe$_2$) and single wall carbon nanotubes (SWCNTs).[83,84] (III) Organic polymers and molecular materials, including dicyanovinyl sexithiophene (DCV6T), fluorine containing polymer Poly(9,9-dioctylfluorene-*alt*-benzothiadiazole), Poly[(9,9-di-*n*-octylfluorenyl-2,7-diyl)-*alt*-(benzo[2,1,3]thiadiazol-4,8-diyl)] abbreviated as F8BT, and the recently developed J-aggregates (J580, J780, and J980, where the numbers following "J" indicates the wavelengths of exciton peaks).[85,86] (IV) Hybrid materials such as 2D and 3D HOIPs (RP, DJ, MAPbI$_3$ and MAPbBr$_3$).[63,64,87] The bulk TMDCs have moderate $\alpha$ (~1.5–3×10$^5$ cm$^{-1}$), while, their $E_{opt}$ (~1.0–1.5 eV for TMDCs) is not entirely in the visible band. Layered bulk PbI$_2$ has an $E_{opt}$ ~2.35 eV while $E_{opt}$ of 3D bonded bulk CdSe is ~1.74 eV, both located in the visible range, while their $\alpha$ at the exciton peak are in the range of ~1.6×10$^5$ cm$^{-1}$ for bulk PbI$_2$ and ~6.6×10$^4$ cm$^{-1}$ for CdSe. Comparatively speaking, the monolayer counterparts of TMDCs exhibit larger $\alpha$ (~1.5–7×10$^5$ cm$^{-1}$) and $E_{opt}$ (~1.5–2.0 eV). Moreover, the 1L TMDCs are direct bandgap semiconductors. These properties make the 1L TMDCs widely studied and appealing for a variety of optoelectronic devices. The $E_{opt}$ of SWCNTs strongly depends on the diameter distribution, purity and synthesis techniques,[84] thus, the values are wide and range from ~0.6 eV to 1.2 eV. The $\alpha$ of SWCNTs is in the same order of magnitude (~$10^4$ cm$^{-1}$) as those of the bulk hybrid perovskites. The large $\alpha$ that J-aggregates present is related to the strong coupling effect between the monomer molecules,[69,70] and the $E_{opt}$ of J-aggregates cover a part of visible range, which promotes some emerging applications.[86] For 2D RP and DJ HOIPs, the $E_{opt}$ of 2D RP HOIPs is larger than that of 2D DJ HOIPs when $n = m$. This is mainly due to the fact that the perovskite layers in 2D RP HOIPs are well isolated by the thicker organic cations (~7.1 Å) and the excitonic transition directly results from the spatial electronic confinement of the separated perovskite layers. On the contrary, the organic spacing between adjacent perovskite layers in the 2D DJ HOIPs is ~4.1 Å, and therefore there is some interlayer interaction between two perovskite layers although weak, which results in a prominent redshift of the exciton position.[8] The $E_{opt}$ of 2D RP and DJ HOIPs covers the entire visible region and have tunable $\alpha$ together with better photo and chemical stability as well as amenable to various types of processing and integration schemes for device making. This combined with their extraordinary optical constants makes them appealing for applications.

Figure 5(b) illustrates the relations between the $E_{opt}$ and the ELT at the first exciton peaks of excitonic semiconductors in Figure 5(a). We can observe that the ELT of 2D and 3D TMDCs are less than 1, except that of 2D WS$_2$ (~3). The ELT of SWCNTs is larger than most of TMDCs, although its $\alpha$ is less than those of TMDCs. This is somewhat counterintuitive but given the extreme 1D confinement in nanotubes and a sp$^2$ bonded graphitic carbon structure, the $\varepsilon_r$ is expected to be smaller than TMDCs resulting in larger ELT. The organic excitonic semiconductors DCVT, F8BT, J780, and J980 remain high ELT comparing to those of most inorganic excitonic materials owing to the same reason as nanotubes but with an even greater degree of confinement. It is worth noting that the ELT of RP phase ($n = 1$) perovskite is exceptionally high (~16.5), which indicates that RP ($n = 1$) can be used as a platform of next-generation LEDs, solar cells, and lasers. With the increase of $n$, ELTs of RP phase gradually decreases to that of 3D RP phase ($n = \infty$). A similar evolution appears for the ELT values of DJ perovskites.

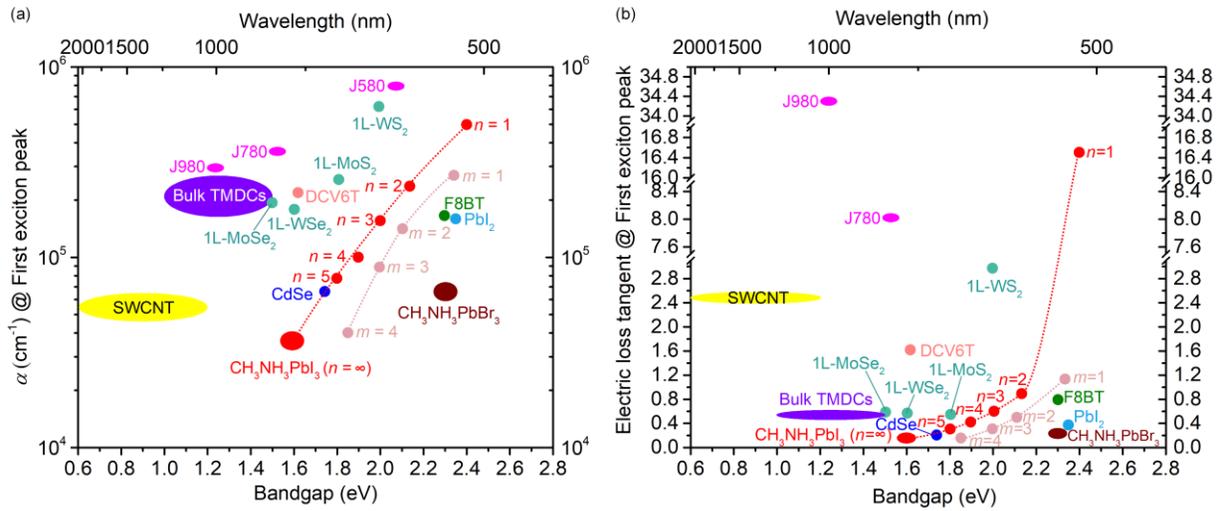

**Figure 5.** (a) Comparison of energy bandgaps (eV) and absorption coefficients (cm$^{-1}$) for a variety of known excitonic semiconductor materials. (b) Comparison of energy bandgaps (eV) and energy loss tangents for the same materials in (a). The dotted lines in (a,b) are just drawn as a guide to the eyes.

## CONCLUSIONS

In summary, the intrinsic optical and dielectric parameters ($\varepsilon$, $N$, $\sigma$, $\alpha$, ELT, ELF, and $\sigma_{acs}$) of 2D RP phase ($n$ = 1–5) and DJ phase ($m$ = 1–4) perovskites are determined by SE over a broadband energy range of 0.73–3.34 eV. Most of the optical/dielectric constants, such as $\sigma$, $\alpha$, ELT, ELF, and $\sigma_{acs}$ are the first to be systematically and completely reported, especially for the 2D DJ phases. A series of excitation features have been observed in each case with some of them having been identified for the first time. The evolution of these excitation features with the increasing QW width are explained under the influence of the quantum confinement effect. An extremely sharp and intense exciton peak is observed in $\alpha$ spectrum of RP phase ($n = 1$), which results from the prominent quantum confinement. With the increase of $n$ and $m$, the ground state exciton oscillator strength of RP and DJ perovskites gradually decreases, which is explained with the increasing dielectric screening from the perovskite layers. We observe that the 2D DJ perovskites exhibit lower $\alpha$, and dielectric permittivity constants in and above the energy range of electronic transitions as a compared to 2D RP phase perovskites which we attribute to the thicker organic barrier thickness in RP perovskites. We also perform a systematic comparison of absorption coefficient and electric loss tangents among select well-known excitonic materials and determine that 2D RP phase ($n = 1$) has among the highest values of $\alpha$ and ELT in the green part of the spectrum. Our results will be critical for the design and optimization of optoelectronic and photonic devices comprising of RP and DJ perovskite active layers and are also expected to advance the understanding of the fundamental electronic structures of phase pure 2D HOIPs.

## METHODS

**Synthesis of $(BA)_2(MA)_{n-1}Pb_nI_{3n+1}$ solution:** RP perovskites solution were synthesized by adopting the previously reported procedure, using 0.4 times the scale.[15]

**Synthesis of $(4AMP)(MA)_{m-1}Pb_mI_{3m+1}$ solution:** DJ perovskites solution were synthesized by adopting the previously reported procedure, using 0.5 times the scale.[16]

**Growth of 2D RP and DJ HOIPs:** Glass was used as the substrate for the 2D perovskite growth. Glass substrates were cut into 2.5 ×2.5 cm squares, cleaned in soap water, isopropanol, acetone by ultrasonication for 20 min each; then dried by an argon gun. The substrates were transferred into a plasma cleaner, cleaned for 10 mins. 10 μL of the diluted solution was dropped onto the glass surface, another glass was put on top to fully cover the bottom glass and dried overnight at 60 °C. Thin sheets of thin HOIPs grew spontaneously as the solvent evaporated. After drying, the largest crystals were taken to further characterization. The whole growth process was carried out in ambient conditions.[88] The nature and quality of the sample were verified by x-ray diffractions using a Rigaku SmartLab x-ray diffractometer.

**Characterization of 2D RP and DJ HOIPs:** The XRD diffraction patterns of 2D perovskites were measured by powder X-ray diffraction system (PXRD) (Siemens D5000) with Cu(Ka) radiation (l = 1.5406 Å) at 0.05 per step with a holding time of 5s per step under the operation conditions of 40 kV and 35 mA. A commercial spectroscopic ellipsometer (M-2000 type spectroscopic ellipsometer, J.A. Woollam Company) was used to investigate the basic optical and dielectric parameters of perovskite specimens. The illumination spot diameter of the ellipsometer can be reduced to ~1 mm by using a pair of focusing probes. Multiangle-incidence measurements were performed from 50° to 70° with a step size of 5° to obtain accurate fitting parameters.

## ASSOCIATED CONTENT

**Supporting Information**

The Supporting Information is available free of charge on the ACS Publications website at DOI: xx.xxxx/acs.nano.xxxxxxx. Optical photographs of 2D DJ perovskites; Ellipsometric measurement and analysis of 2D RP and DJ perovskites; Anisotropic dielectric functions of 2D RP ($n$ = 1) and DJ ($m$ = 1) perovskites; Tabulated optical and dielectric parameters of 2D RP ($n$ = 1–5) and DJ ($m$ = 1–4) phase perovskites. (PDF)

# AUTHOR INFORMATION

**Corresponding Author**

*E-mail: shyliu@hust.edu.cn
*E-mail: adm4@rice.edu
*E-mail: dmj@seas.upenn.edu

**Author Contributions**
#These authors contributed equally to this work.

**ORCID**
Baokun Song: 0000-0002-8184-5616
Jin Hou:
Haonan Wang: 0000-0003-2047-5380
Siraj Sidhik: 0000-0002-2097-2830
Jinshui Miao: 0000-0002-7571-2454
Honggang Gu: 0000-0001-8812-1621
Huiqin Zhang:
Shiyuan Liu: 0000-0002-0756-1439
Zahra Fakhraai: 0000-0002-0597-9882
Jacky Even: 0000-0002-4607-3390
Jean-Christophe Blancon: 0000-0002-3833-5792
Aditya D. Mohite: 0000-0001-9435-0201
Deep M. Jariwala: 0000-0002-3570-8768

**Notes**

The authors declare no competing financial interest.

# ACKNOWLEDGEMENTS

The work at Penn was primarily funded by the U.S. Army Research Office under contract number W911NF-19-1-0109. D.J. and Z. F. also acknowledges support from University of Pennsylvania Materials Research Science and Engineering Center (MRSEC) (DMR-1720530). Baokun Song acknowledges the support from China scholarship council (CSC). Baokun Song, Honggang Gu and Shiyuan Liu acknowledge the National Natural Science Foundation of China (51525502 and 51775217). The work at Rice University was supported by the DOE-EERE 2022-1652 program. J.E. acknowledges the financial support from the Institut Universitaire de France. The authors acknowledge assistance from Surendra B. Anantharaman of the complex refractive indices of J580, J780, and J980, and we also would like to acknowledge the constructive suggestions given by Prof. Artur R. Davoyan.

**Graphical Table of Contents**

Determination of Dielectric Functions and Exciton Oscillator Strength of Two-Dimensional Hybrid Perovskites

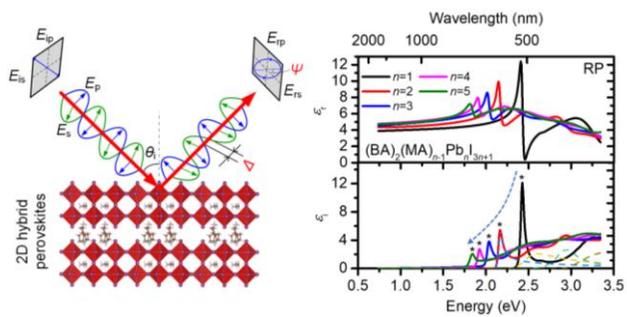

Supporting Information

# Determination of Dielectric Functions and Exciton Oscillator Strength of Two-Dimensional Hybrid Perovskites

Baokun Song,[†,‡,#] Jin Hou,[$,&,#] Haonan Wang,[§] Siraj Sidhik,[$] Jinshui Miao,[†] Honggang Gu,[‡] Huiqin Zhang,[†] Shiyuan Liu,[‡,*] Zahra Fakhraai,[§] Jacky Even,[∥] Jean-Christophe Blancon,[$] Aditya D. Mohite,[$,&,*] Deep Jariwala[†,*]

[†]Department of Electrical and Systems Engineering, University of Pennsylvania, Philadelphia, PA 19104, USA

[‡]School of Mechanical Science and Engineering, Huazhong University of Science and Technology, Wuhan 430074, P. R. China

[$]Department of Chemical and Biomolecular Engineering Rice University, Houston, TX 77005, USA

[&]Department of Material Science and Nanoengineering Rice University, Houston, TX 77005, USA

[§]Department of Chemistry, University of Pennsylvania, Philadelphia, PA 19104, USA

[∥]Univ Rennes, INSA Rennes, CNRS, Institute FOTON, UMR 6082, Rennes F-35000, France

*E-mail: shyliu@hust.edu.cn

*E-mail: adm4@rice.edu

*E-mail: dmj@seas.upenn.edu

**Optical photograph of 2D DJ (*m* = 1–4) perovskites**

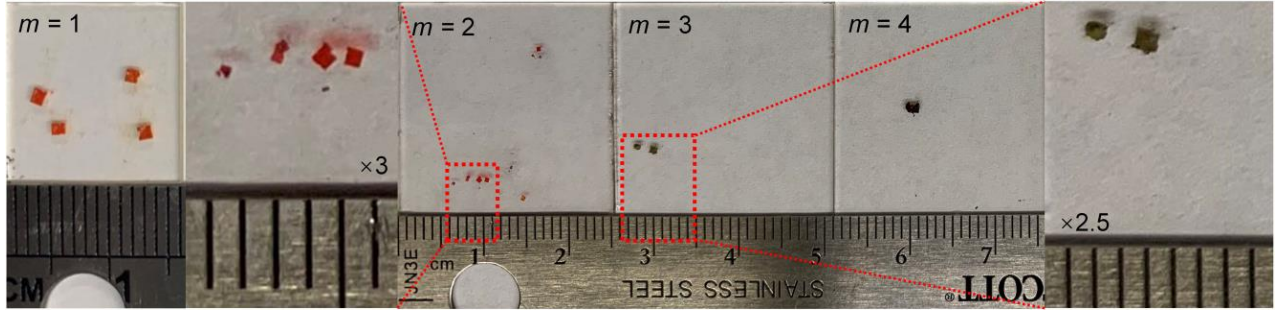

**Figure S1** Optical photographs of 2D DJ perovskites. The red dotted box regions for *m* = 2 and *m* = 3 DJ perovskites are magnified 3 times and 2.5 times, respectively, for visual clarity.

**Ellipsometry measurement and analysis**

The spectroscopic ellipsometry is a non-contact, non-invasive optical characterization technique, which can extract the geometrical and inherent optical/dielectric parameters of materials by analyzing the polarization state changes of polarized light after interacting with the samples.[1] The changes of polarization state are generally expressed as a pair of ellipsometric parameters [$\Psi(\omega)$, $\Delta(\omega)$], where $\tan(\Psi)$ and $\Delta$ refer to the amplitude ratio and phase difference of the reflection coefficient of *p* and *s* polarized light ($r_p/r_s = \tan(\Psi)e^{i\Delta}$), and $\omega$ denotes the angular frequency of light.[2] The ellipsometer we used in the experiment are M-2000 type variable angle spectroscopic ellipsometer (VASE). The applicable energy region of the ellipsometer is 0.73–6.42 eV (371–1687 nm), and its incidence angle can be adjusted from 45° to 80° with 0.1 ° resolution. To obtain the reliable intrinsic optical and dielectric parameters of 2-dimensional (2D) RP perovskites, multiangle measurement mode (incidence angle: 50°, 55° 60°, 65°, 70°) is selected to carry out the ellipsometric characterization.

Ellipsometry is a model-based technique. To analyze the measured ellipsometric spectra, two models need to be constructed first, namely the optical model and the dielectric function model. The optical model describes the geometric structure of the sample, and the dielectric function model gives the dielectric dispersion of the material which would be involved in the optical model. The theoretical ellipsometric spectra of the sample can be calculated by introducing the interference theory in multilayer films into the above two models.[1] By fitting the measured ellipsometric spectra with the theoretical calculated ones, the basic optical and dielectric parameters (such as the complex refractive index and the dielectric function) and geometrical parameters (such as the thickness) of the sample can be obtained simultaneously.

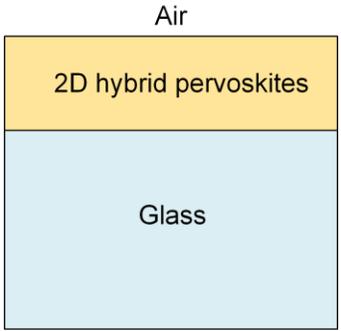

**Figure S2.** Optical model of the 2D hybrid perovskites film on glass substrates.

The optical model for the 2D perovskite samples is a vertical stacking multilayer structure, including the ambient air, the 2D perovskite, and the glass substrate (Figure S2). For the dielectric model, we first adopt the Kramers-Kronig constrained B-spline to embody the dielectric dispersions of these two kinds of 2D perovskites due to lack of transcendental physical models. Then, a generalized oscillator model combining several Tauc-Lorentz oscillators is used to parameterize the dielectric functions of the 2D perovskites obtained by the B-spline fitting.[1,3,4] The specific form of combined Tauc-Lorentz oscillator is as follows

$$\varepsilon(E) = \sum_{i}^{m} \varepsilon_{\text{Tauc-Lorentz}}^{i}(Amp_i, \Gamma_i, E_{0i}, E_{gi}; E) \tag{S1a}$$

$$\varepsilon_{\text{Tauc-Lorentz}}^{i}(E)=\begin{cases} \dfrac{Amp_i E_{0i} \Gamma_i (E-E_{gi})}{E(E_{0i}^2 - E^2)^2 + \Gamma_i^2 E^3} & (E>E_{gi}) \\ 0 & (E \leq E_{gi}) \end{cases}. \quad (S1b)$$

Where, $Amp_i$, $\Gamma_i$, $E_{0i}$, $E_{gi}$ denote the amplitude, damping coefficient, central energy, and bandgap of the $i$th Tuac-Lorentz oscillator.

On the basic of the above two models, the theoretical ellipsometric spectra of the 2D perovskites can be calculated. Figure S3 and Figure S4 illustrates the measured and best-fitting ellipsometric spectra of the 2D RP and DJ perovskites. Obviously, the theoretical ellipsometric spectra are in good agreement with the experimental ellipsometric spectra, indicating that the models established by us are reasonable. Further information about the basic principles of ellipsometry and more application examples of ellipsometry are available in our previous research works.[2,5,6] Moreover, the dielectric function spectra of 2D RP and DJ perovskites extracted based on the B-spline model are also plotted in Figure S5, where the local fluctuations in the dielectric function spectra can be attributed to some ambient noise.

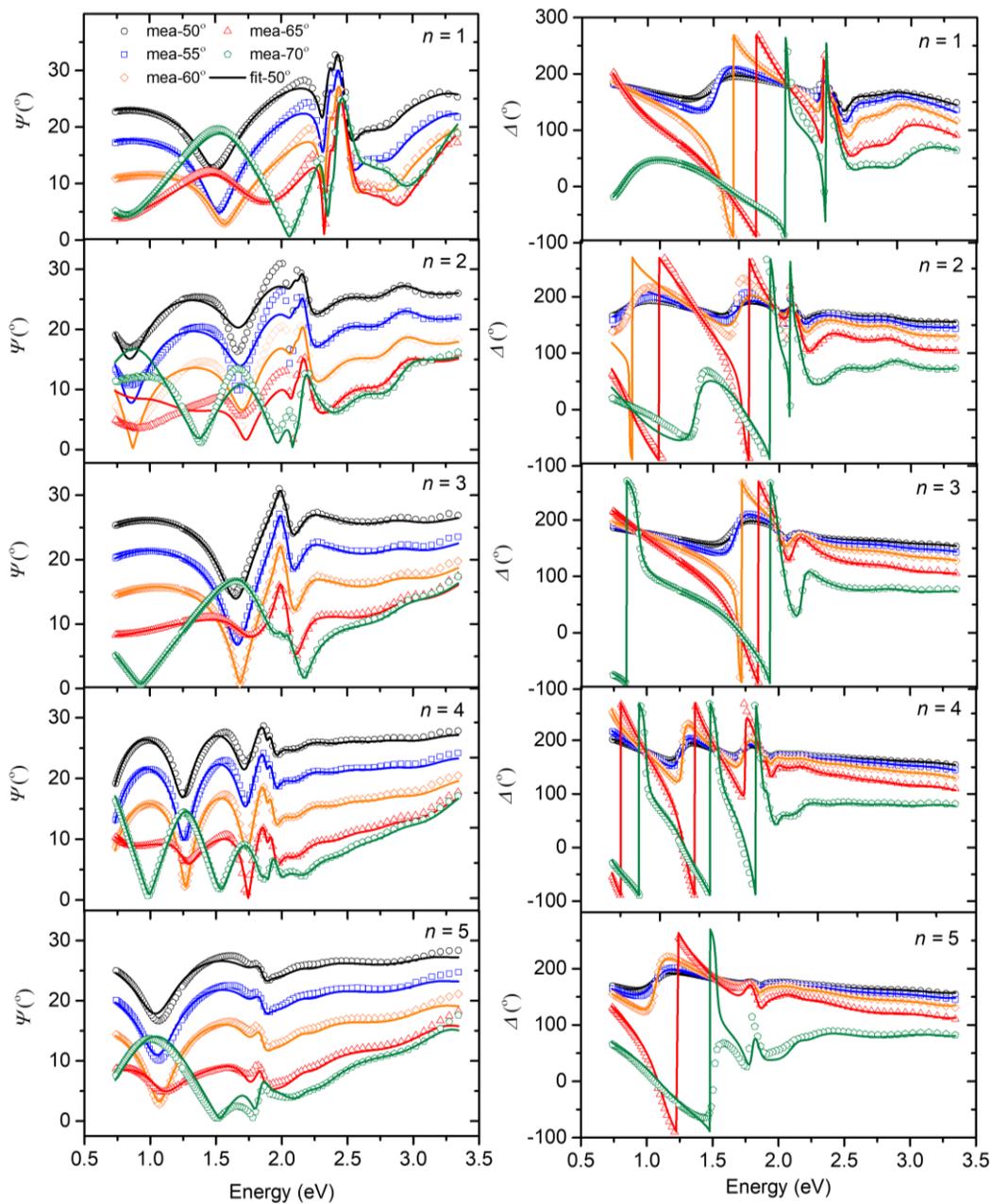

**Figure S3.** Measured and best-fitting ellipsometric parameter spectra of 2D RP perovskites.

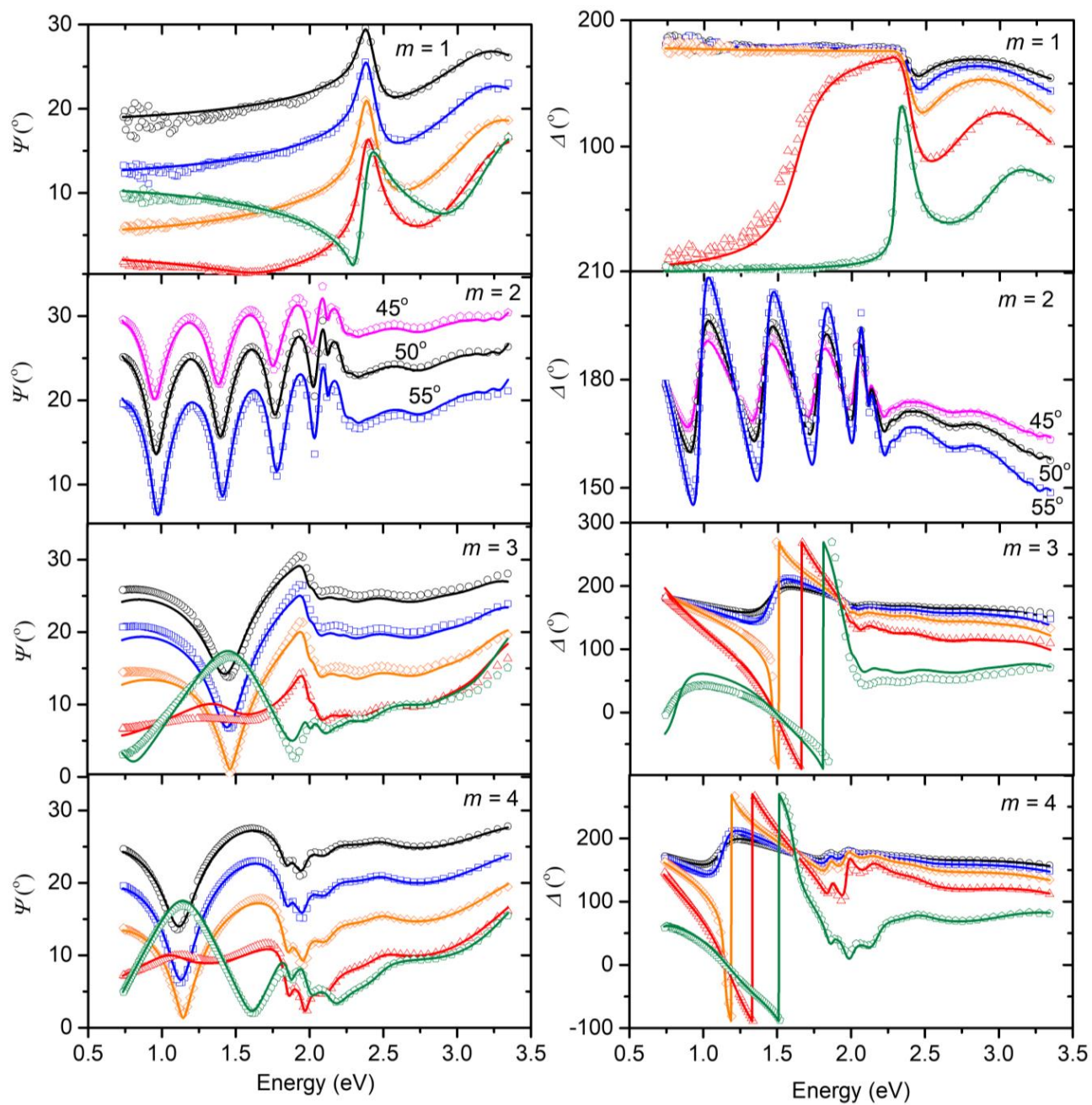

**Figure S4.** Measured and best-fitting ellipsometric spectra of 2D DJ perovskites. In consideration of the smaller area of DJ ($m = 2$), the ellipsometric spectra at 45°, 50°, and 55° incident angles are fitted.

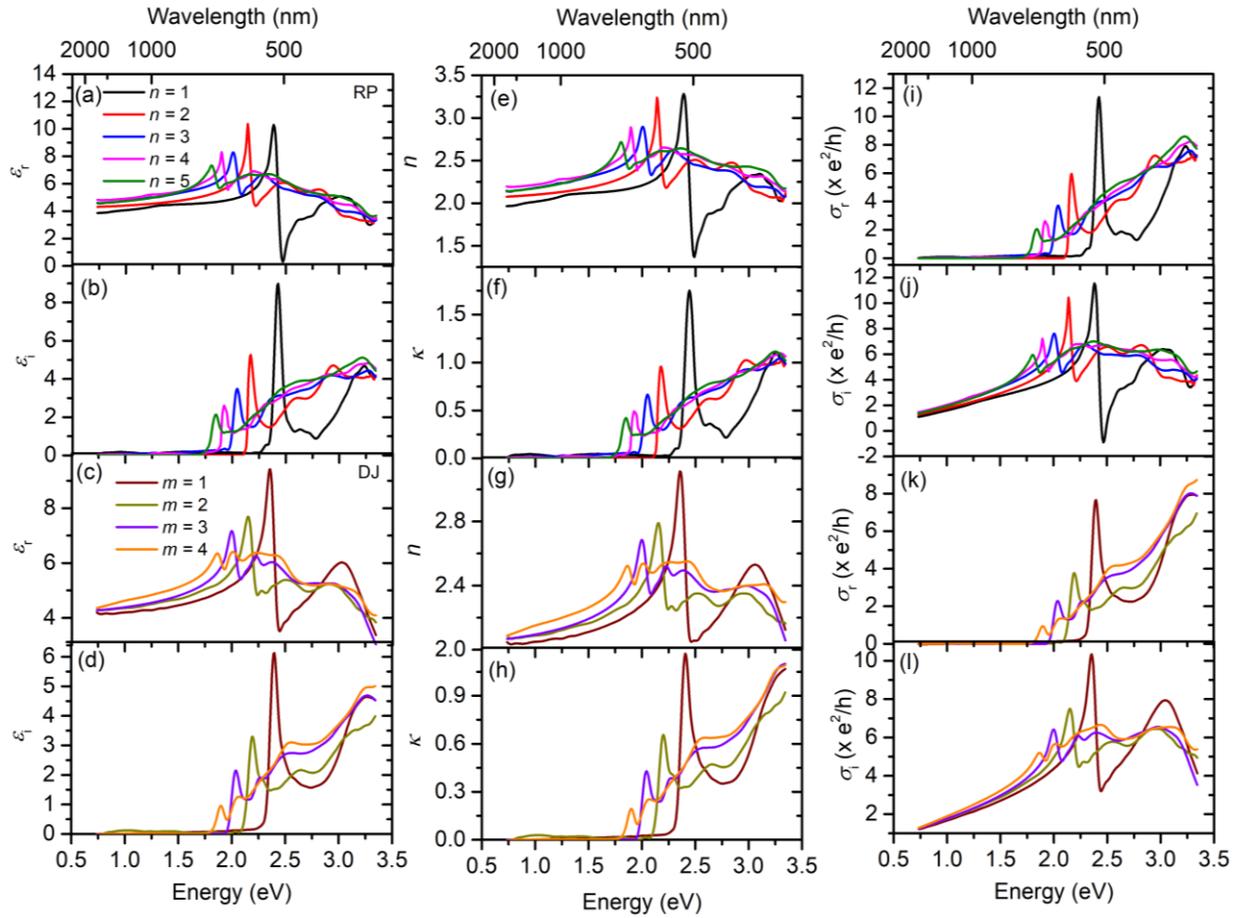

**Figure S5.** Dielectric functions, complex refractive indices, and complex optical conductivities (reduced by $e^2/h$) of 2D RP and DJ phase perovskites obtained from B-spline fitting.

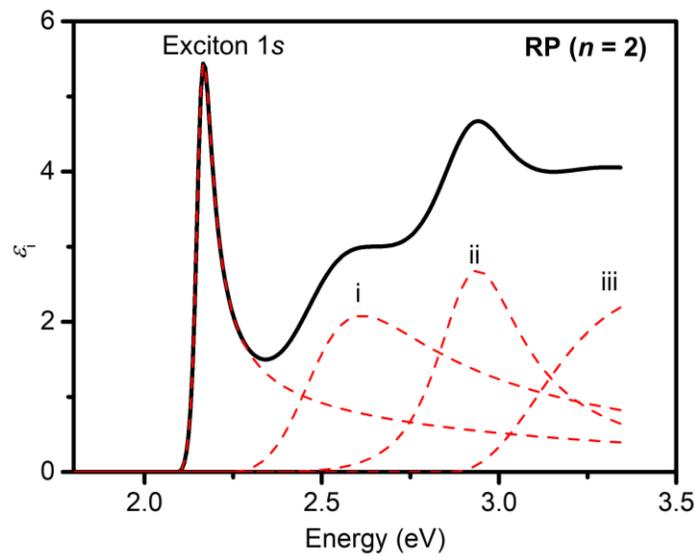

**Figure S6.** Fitting the imaginary part of dielectric function of 2D RP ($n = 2$) perovskite via combined Tauc-Lorentz oscillators.

To obtain the out-of-plane dielectric functions of 2D RP and DJ phase HOIPs, we fit the multiangle Mueller matrix (MM) spectra (Figure S7) using an anisotropic dielectric function model. The 2D HOIPs thin-film crystal can be regarded as a uniaxial crystal with the following dielectric tensor

$$\varepsilon = \begin{bmatrix} \varepsilon_o & & \\ & \varepsilon_o & \\ & & \varepsilon_e \end{bmatrix}. \qquad (S2)$$

Where, $\varepsilon_o$ and $\varepsilon_e$ refer to the ordinary (in-plane) and extraordinary (out-of-plane) dielectric functions. The optical model used in the anisotropic ellipsometric analysis is the same as Figure S2, and the overall dielectric response of the 2D HOIPs are also first described by B-splines. Then, the B-spline model is converted to a uniaxial anisotropy model and used to fit the MM spectra of 2D HOIPs. Next, the ordinary/extraordinary dielectric functions extracted from the anisotropic fitting based on the B-splines are parametrized by two sets of combined Tauc-Lorentz oscillators. The regression analysis processes are similar with those of the isotropic analysis discussed earlier. The measured and best-fits of MM spectra of 2D RP ($n = 1$) and DJ ($m = 1$) phase perovskites are presented in Figure S7. The near-zero off-diagonal elements ($m_{13}$, $m_{14}$, $m_{23}$, $m_{24}$, $m_{31}$, $m_{32}$) indicate that there is no obvious cross polarization between p- and s-polarized light. The information associated with the out-of-plane dielectric functions is enclosed into the diagonal elements ($m_{12}$, $m_{21}$, $m_{22}$, $m_{33}$, $m_{34}$). As shown in Figure S7, the theoretically calculated and measured MM spectra exhibit good degree of matching ($RMSE = 12.33$ for RP ($n =1$) and 18.40 for DJ ($m = 1$)), meaning that the anisotropic dielectric model can effectively embody the polarized dielectric responses of the 2D HOIPs.[5] Figure S8 presents the ordinary and extraordinary dielectric functions of RP ($n = 1$) and DJ ($m = 1$) HOIPs. We interpret that the differences between $\varepsilon_{iso}$ (Figure 2) and $\varepsilon_o$ are mainly due to the underestimate of thickness in the isotropic analysis and the different fitting goodness as different models are adopted for fitting the raw ellipsometric spectra.

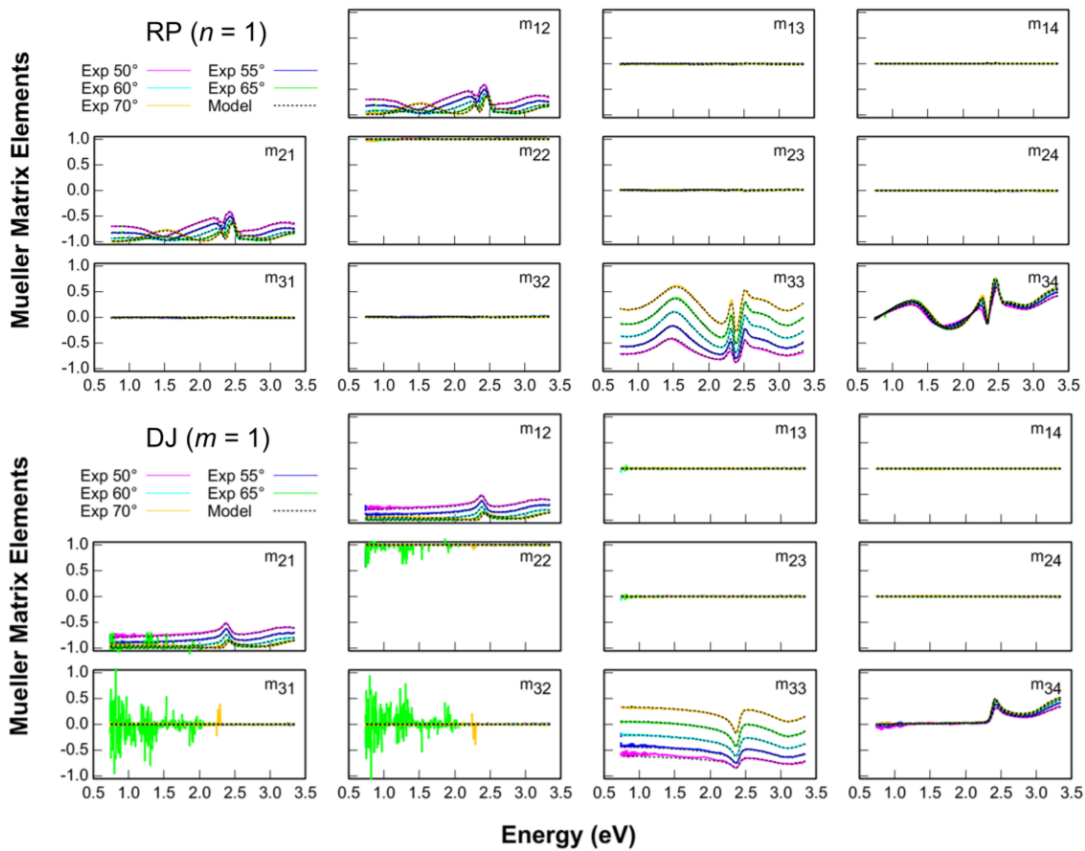

**Figure S7.** Measured and best-fits of Mueller matrix spectra of 2D RP ($n$ = 1) and DJ ($m$ = 1) phase perovskites.

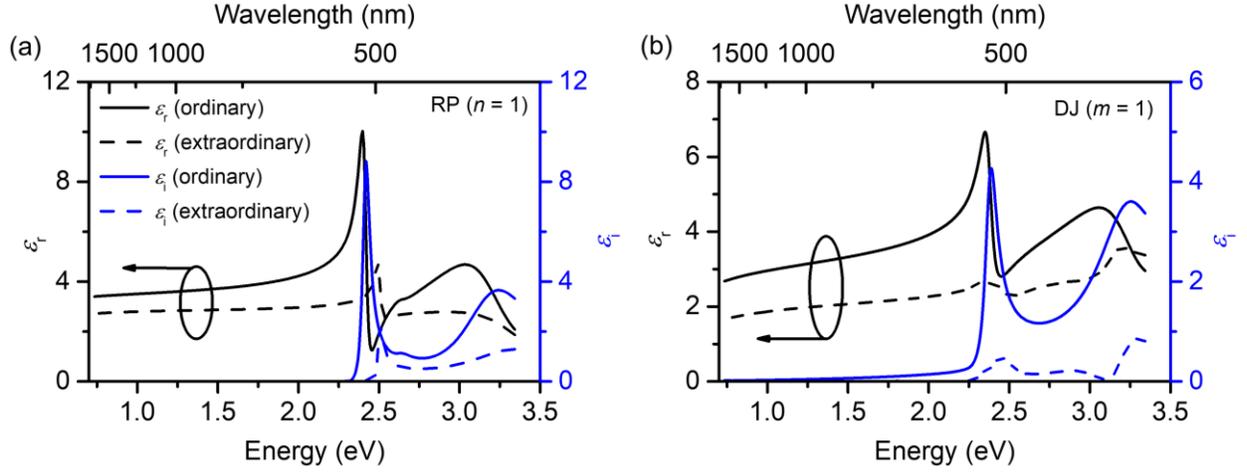

**Figure S8.** Ordinary and extraordinary dielectric functions of 2D RP ($n = 1$) and DJ ($m = 1$) phase perovskites.

**Table S1.** Out-of-plane lattice parameters (along the stacking direction) between experiment and simulation for 2D RP and DJ phase perovskites

| No. | Lattice parameter along stacking axis (Å) | Lattice parameter previous work (Å)[7,8] | Relative difference (%) |
|---|---|---|---|
| $n = 1$ | 27.67744 | 28.014 | −1.2014 |
| $n = 2$ | 39.2354 | 39.347 | −0.28364 |
| $n = 3$ | 51.76406 | 51.959 | −0.37517 |
| $n = 4$ | 63.51412 | 64.383 | −1.34954 |
| $n = 5$ | 77.20004 | 76.613 | 0.766242 |
| $m = 1$ | 10.48131 | 10.4999 | −0.17702 |
| $m = 2$ | 16.79971 | 16.80225 | −0.01509 |
| $m = 3$ | 22.88126 | 23.1333 | −1.08949 |
| $m = 4$ | 29.18833 | 29.457 | −0.91206 |

**Table S2.** Tauc-Lorentz parameters of ground state exciton peaks in $\varepsilon_i$, optical bandgaps $E_{opt}$, thicknesses, *RMSE*, the effective high-frequency dielectric function $\varepsilon_\infty$, and the volume of the formula unit of 2D RP and DJ phase perovskites

| No. | *Amp* | $\Gamma$ (eV) | $E_0$ (eV) | Thicknesses (nm) | RMSE from B-Spline | RMSE from Gen-Osc | $\varepsilon_\infty$ from EMA[a] | $V_0$ (cm$^3$) |
|---|---|---|---|---|---|---|---|---|
| $n = 1$ | 196.06±7.37 | 0.045±0.0005 | 2.428±0.00311 | 216.94±0.135 | 16.1 | 16.77 | 4.19 | 532.99 |
| $n = 2$ | 149.82±8.01 | 0.055±0.004 | 2.165±0.00224 | 351.41±0.470 | 34.5 | 31.7 | 4.91 | 779.68 |
| $n = 3$ | 110.92±13.71 | 0.061±0.002 | 2.037±0.002313 | 168.08±0.314 | 29.7 | 29.7 | 5.30 | 1029.51 |
| $n = 4$ | 45.63±9.87 | 0.060±0.007 | 1.924±0.00312 | 464.61±0.932 | 37.8 | 38.1 | 5.53 | 1276.24 |
| $n = 5$ | 29.18±16.49 | 0.08±0.01 | 1.842±0.008016 | 283.78±0.496 | 38.8 | 38.8 | 5.68 | 1531.29 |
| $m = 1$ | 120.75±18.48 | 0.11±0.007 | 2.377±0.002 | bulk | 11.86 | 10.89 | 4.77 | 412.51 |
| $m = 2$ | 95.31±15.96 | 0.1±0.01 | 2.183±0.004014 | 654.86±0.681 | 14.62 | 12.26 | 5.43 | 656.88 |
| *m* = 3 | 41.69±17.7 | 0.088±0.009 | 2.035±0.02312 | 207.13±0.725 | 20.6 | 21.5 | 5.71 | 903.06 |
| *m* = 4 | 8.45±1.90 | 0.077±0.0029 | 1.891±0.002033 | 272.43±0.100 | 9.19 | 9.31 | 5.89 | 1155.65 |

[a]The effective high-frequency dielectric function is calculated by the effective medium approximation (EMA) model.

**Table S3.** Dielectric functions and absorption coefficients of 2D RP perovskites

| Wavelength (nm) | *n* = 1 | | | *n* = 2 | | | *n* = 3 | | | *n* = 4 | | | *n* = 5 | | |
|---|---|---|---|---|---|---|---|---|---|---|---|---|---|---|---|
| | $\varepsilon_r$ | $\varepsilon_i$ | $\alpha$ | $\varepsilon_r$ | $\varepsilon_i$ | $\alpha$ | $\varepsilon_r$ | $\varepsilon_i$ | $\alpha$ | $\varepsilon_r$ | $\varepsilon_i$ | $\alpha$ | $\varepsilon_r$ | $\varepsilon_i$ | $\alpha$ |
| 370.91 | 2.49 | 4.16 | 367445.81 | 2.96 | 4.05 | 343664.03 | 3.19 | 4.42 | 360580.11 | 3.06 | 4.75 | 385859.03 | 3.75 | 4.88 | 371544.56 |
| 372.49 | 2.66 | 4.24 | 365013.23 | 3.01 | 4.06 | 340692.46 | 3.26 | 4.40 | 355481.87 | 3.16 | 4.77 | 381934.06 | 3.71 | 4.87 | 370427.48 |
| 374.08 | 2.84 | 4.30 | 361475.96 | 3.06 | 4.06 | 337657.13 | 3.32 | 4.38 | 350449.37 | 3.26 | 4.78 | 377708.81 | 3.69 | 4.87 | 369467.76 |
| 375.66 | 3.03 | 4.35 | 356821.71 | 3.11 | 4.06 | 334572.70 | 3.39 | 4.36 | 345507.70 | 3.35 | 4.78 | 373213.67 | 3.68 | 4.88 | 368489.63 |
| 377.24 | 3.22 | 4.39 | 351058.80 | 3.16 | 4.06 | 331456.41 | 3.45 | 4.33 | 340679.32 | 3.45 | 4.78 | 368481.29 | 3.69 | 4.89 | 367373.02 |
| 378.82 | 3.42 | 4.40 | 344218.98 | 3.21 | 4.05 | 328327.51 | 3.50 | 4.31 | 335983.99 | 3.54 | 4.77 | 363548.43 | 3.71 | 4.90 | 366036.67 |
| 380.40 | 3.62 | 4.40 | 336357.38 | 3.25 | 4.05 | 325208.33 | 3.55 | 4.28 | 331439.24 | 3.63 | 4.76 | 358454.30 | 3.73 | 4.91 | 364428.53 |
| 381.99 | 3.83 | 4.37 | 327551.76 | 3.29 | 4.04 | 322124.02 | 3.60 | 4.26 | 327060.27 | 3.72 | 4.74 | 353239.66 | 3.77 | 4.92 | 362519.35 |
| 383.57 | 4.02 | 4.32 | 317902.09 | 3.33 | 4.03 | 319102.80 | 3.64 | 4.23 | 322859.91 | 3.81 | 4.72 | 347947.12 | 3.81 | 4.93 | 360295.96 |
| 385.15 | 4.22 | 4.26 | 307524.76 | 3.37 | 4.02 | 316174.17 | 3.68 | 4.21 | 318848.61 | 3.89 | 4.69 | 342619.33 | 3.86 | 4.93 | 357757.28 |
| 386.74 | 4.40 | 4.18 | 296550.17 | 3.40 | 4.02 | 313372.03 | 3.72 | 4.18 | 315034.72 | 3.96 | 4.66 | 337297.60 | 3.91 | 4.94 | 354912.30 |
| 388.32 | 4.57 | 4.08 | 285116.40 | 3.43 | 4.01 | 310730.77 | 3.75 | 4.16 | 311423.94 | 4.03 | 4.63 | 332023.64 | 3.97 | 4.93 | 351777.11 |
| 389.90 | 4.73 | 3.96 | 273365.19 | 3.46 | 4.00 | 308287.24 | 3.78 | 4.14 | 308020.05 | 4.09 | 4.59 | 326836.33 | 4.03 | 4.93 | 348373.10 |
| 391.49 | 4.87 | 3.83 | 261434.75 | 3.49 | 4.00 | 306078.04 | 3.81 | 4.12 | 304824.24 | 4.15 | 4.55 | 321771.03 | 4.09 | 4.92 | 344725.51 |
| 393.07 | 5.00 | 3.70 | 249457.38 | 3.51 | 4.00 | 304140.00 | 3.84 | 4.11 | 301835.61 | 4.21 | 4.51 | 316860.83 | 4.15 | 4.90 | 340862.56 |
| 394.65 | 5.11 | 3.55 | 237553.18 | 3.53 | 4.00 | 302508.19 | 3.87 | 4.09 | 299050.51 | 4.25 | 4.48 | 312133.91 | 4.21 | 4.89 | 336814.36 |
| 396.24 | 5.21 | 3.40 | 225829.68 | 3.54 | 4.00 | 301213.43 | 3.89 | 4.08 | 296463.26 | 4.30 | 4.44 | 307614.03 | 4.28 | 4.86 | 332611.28 |
| 397.82 | 5.28 | 3.25 | 214378.88 | 3.56 | 4.02 | 300281.73 | 3.92 | 4.07 | 294065.51 | 4.34 | 4.40 | 303320.19 | 4.34 | 4.84 | 328283.62 |
| 399.40 | 5.34 | 3.10 | 203276.20 | 3.57 | 4.03 | 299729.55 | 3.94 | 4.06 | 291846.54 | 4.37 | 4.37 | 299265.78 | 4.40 | 4.81 | 323862.15 |
| 400.99 | 5.39 | 2.95 | 192582.42 | 3.58 | 4.05 | 299562.11 | 3.96 | 4.06 | 289792.01 | 4.41 | 4.34 | 295458.96 | 4.45 | 4.78 | 319376.20 |
| 402.57 | 5.42 | 2.81 | 182341.93 | 3.60 | 4.08 | 299768.77 | 3.99 | 4.06 | 287886.40 | 4.44 | 4.31 | 291901.61 | 4.51 | 4.74 | 314853.19 |
| 404.16 | 5.45 | 2.66 | 172587.21 | 3.62 | 4.12 | 300320.86 | 4.01 | 4.06 | 286110.20 | 4.46 | 4.28 | 288591.72 | 4.56 | 4.70 | 310319.38 |
| 405.74 | 5.45 | 2.53 | 163337.51 | 3.64 | 4.17 | 301165.74 | 4.04 | 4.06 | 284442.43 | 4.49 | 4.26 | 285520.24 | 4.61 | 4.67 | 305799.06 |
| 407.33 | 5.45 | 2.39 | 154602.47 | 3.67 | 4.22 | 302224.30 | 4.07 | 4.06 | 282858.58 | 4.51 | 4.24 | 282674.09 | 4.66 | 4.63 | 301314.20 |
| 408.91 | 5.44 | 2.27 | 146382.52 | 3.71 | 4.28 | 303388.87 | 4.10 | 4.07 | 281332.69 | 4.53 | 4.22 | 280035.21 | 4.70 | 4.59 | 296884.58 |
| 410.50 | 5.43 | 2.15 | 138672.64 | 3.76 | 4.34 | 304521.67 | 4.13 | 4.07 | 279836.54 | 4.56 | 4.21 | 277582.53 | 4.74 | 4.54 | 292527.63 |
| 412.08 | 5.40 | 2.04 | 131460.88 | 3.83 | 4.40 | 305456.69 | 4.17 | 4.08 | 278341.52 | 4.58 | 4.19 | 275290.50 | 4.78 | 4.50 | 288259.08 |
| 413.67 | 5.37 | 1.93 | 124732.33 | 3.91 | 4.47 | 306005.53 | 4.20 | 4.09 | 276817.37 | 4.61 | 4.18 | 273131.30 | 4.81 | 4.46 | 284091.70 |
| 415.25 | 5.34 | 1.83 | 118469.14 | 4.00 | 4.53 | 305966.41 | 4.24 | 4.10 | 275234.65 | 4.63 | 4.18 | 271076.34 | 4.84 | 4.42 | 280036.34 |
| 416.84 | 5.30 | 1.74 | 112651.48 | 4.11 | 4.59 | 305139.75 | 4.29 | 4.10 | 273564.69 | 4.66 | 4.17 | 269095.09 | 4.87 | 4.38 | 276102.16 |
| 418.42 | 5.26 | 1.66 | 107257.45 | 4.24 | 4.63 | 303345.77 | 4.33 | 4.11 | 271781.65 | 4.69 | 4.16 | 267158.11 | 4.90 | 4.34 | 272295.51 |
| 420.01 | 5.21 | 1.58 | 102266.81 | 4.38 | 4.66 | 300444.95 | 4.38 | 4.11 | 269861.97 | 4.72 | 4.16 | 265237.33 | 4.93 | 4.30 | 268622.41 |
| 421.59 | 5.17 | 1.50 | 97657.01 | 4.53 | 4.67 | 296359.16 | 4.43 | 4.11 | 267786.93 | 4.75 | 4.16 | 263306.07 | 4.95 | 4.26 | 265085.54 |
| 423.18 | 5.12 | 1.44 | 93407.80 | 4.69 | 4.66 | 291086.76 | 4.49 | 4.11 | 265542.72 | 4.79 | 4.15 | 261340.86 | 4.97 | 4.22 | 261687.40 |
| 424.77 | 5.07 | 1.37 | 89498.55 | 4.85 | 4.62 | 284708.99 | 4.54 | 4.11 | 263120.19 | 4.82 | 4.15 | 259321.87 | 4.99 | 4.19 | 258428.13 |
| 426.35 | 5.02 | 1.32 | 85910.17 | 5.00 | 4.57 | 277387.72 | 4.60 | 4.10 | 260517.58 | 4.86 | 4.14 | 257233.56 | 5.00 | 4.16 | 255307.43 |
| 427.94 | 4.96 | 1.26 | 82624.11 | 5.14 | 4.49 | 269348.07 | 4.65 | 4.09 | 257738.06 | 4.90 | 4.13 | 255064.97 | 5.02 | 4.12 | 252323.46 |
| 429.52 | 4.91 | 1.22 | 79623.83 | 5.26 | 4.39 | 260854.80 | 4.71 | 4.07 | 254791.10 | 4.94 | 4.12 | 252809.27 | 5.04 | 4.09 | 249473.45 |
| 431.11 | 4.86 | 1.17 | 76893.40 | 5.37 | 4.28 | 252183.87 | 4.76 | 4.05 | 251691.55 | 4.98 | 4.11 | 250464.96 | 5.05 | 4.07 | 246754.03 |
| 432.70 | 4.80 | 1.13 | 74418.50 | 5.46 | 4.16 | 243570.64 | 4.82 | 4.03 | 248458.39 | 5.02 | 4.10 | 248034.09 | 5.06 | 4.04 | 244160.48 |
| 434.28 | 4.75 | 1.09 | 72186.49 | 5.53 | 4.04 | 235097.22 | 4.87 | 4.00 | 245114.22 | 5.06 | 4.08 | 245523.38 | 5.08 | 4.01 | 241688.22 |
| 435.87 | 4.69 | 1.06 | 70186.07 | 5.58 | 3.92 | 226955.83 | 4.92 | 3.98 | 241684.74 | 5.10 | 4.06 | 242942.61 | 5.09 | 3.99 | 239331.02 |
| 437.46 | 4.64 | 1.03 | 68407.26 | 5.61 | 3.80 | 219308.67 | 4.97 | 3.94 | 238196.51 | 5.14 | 4.04 | 240304.71 | 5.10 | 3.97 | 237083.96 |
| 439.05 | 4.59 | 1.01 | 66842.22 | 5.63 | 3.69 | 212260.28 | 5.01 | 3.91 | 234676.21 | 5.18 | 4.02 | 237624.57 | 5.11 | 3.95 | 234940.10 |
| 440.63 | 4.53 | 0.98 | 65484.65 | 5.63 | 3.58 | 205867.77 | 5.06 | 3.87 | 231150.40 | 5.22 | 4.00 | 234918.04 | 5.12 | 3.93 | 232892.62 |
| 442.22 | 4.48 | 0.96 | 64330.63 | 5.63 | 3.49 | 200149.73 | 5.09 | 3.84 | 227644.21 | 5.26 | 3.98 | 232202.23 | 5.13 | 3.91 | 230935.42 |
| 443.81 | 4.42 | 0.95 | 63377.37 | 5.62 | 3.40 | 195096.57 | 5.13 | 3.80 | 224179.99 | 5.29 | 3.95 | 229493.86 | 5.15 | 3.90 | 229060.64 |
| 445.39 | 4.36 | 0.93 | 62625.85 | 5.60 | 3.33 | 190678.65 | 5.16 | 3.76 | 220778.51 | 5.33 | 3.93 | 226809.27 | 5.16 | 3.88 | 227261.54 |
| 446.98 | 4.31 | 0.92 | 62079.25 | 5.58 | 3.26 | 186852.59 | 5.19 | 3.72 | 217457.57 | 5.36 | 3.90 | 224163.84 | 5.17 | 3.87 | 225531.02 |
| 448.57 | 4.25 | 0.91 | 61744.40 | 5.55 | 3.20 | 183566.62 | 5.22 | 3.68 | 214232.22 | 5.39 | 3.88 | 221571.13 | 5.19 | 3.86 | 223861.86 |
| 450.16 | 4.20 | 0.91 | 61633.25 | 5.53 | 3.16 | 180765.47 | 5.24 | 3.65 | 211114.38 | 5.41 | 3.85 | 219042.94 | 5.20 | 3.85 | 222246.52 |
| 451.75 | 4.14 | 0.91 | 61761.25 | 5.50 | 3.12 | 178392.55 | 5.26 | 3.61 | 208113.47 | 5.44 | 3.83 | 216588.31 | 5.22 | 3.84 | 220678.30 |
| 453.33 | 4.08 | 0.91 | 62151.14 | 5.48 | 3.09 | 176391.36 | 5.28 | 3.57 | 205236.72 | 5.46 | 3.80 | 214214.66 | 5.23 | 3.83 | 219149.91 |
| 454.92 | 4.02 | 0.92 | 62830.56 | 5.46 | 3.06 | 174708.26 | 5.29 | 3.54 | 202488.89 | 5.49 | 3.78 | 211926.34 | 5.25 | 3.82 | 217654.83 |
| 456.51 | 3.97 | 0.93 | 63832.51 | 5.45 | 3.04 | 173290.86 | 5.31 | 3.51 | 199872.09 | 5.51 | 3.76 | 209724.86 | 5.27 | 3.81 | 216186.29 |
| 458.10 | 3.91 | 0.95 | 65187.70 | 5.43 | 3.03 | 172090.68 | 5.32 | 3.48 | 197387.86 | 5.53 | 3.74 | 207609.98 | 5.29 | 3.80 | 214737.54 |
| 459.69 | 3.86 | 0.97 | 66909.66 | 5.42 | 3.02 | 171061.11 | 5.33 | 3.45 | 195035.25 | 5.55 | 3.72 | 205577.98 | 5.31 | 3.80 | 213302.28 |
| 461.27 | 3.81 | 1.00 | 68964.06 | 5.42 | 3.01 | 170158.96 | 5.34 | 3.42 | 192812.62 | 5.58 | 3.70 | 203622.79 | 5.33 | 3.79 | 211874.33 |
| 462.86 | 3.77 | 1.03 | 71218.07 | 5.42 | 3.01 | 169343.71 | 5.35 | 3.40 | 190717.19 | 5.60 | 3.69 | 201736.78 | 5.36 | 3.78 | 210447.60 |
| 464.45 | 3.73 | 1.06 | 73399.50 | 5.42 | 3.00 | 168576.65 | 5.36 | 3.37 | 188744.71 | 5.62 | 3.67 | 199908.42 | 5.38 | 3.78 | 209016.40 |
| 466.04 | 3.70 | 1.08 | 75137.62 | 5.43 | 3.00 | 167822.13 | 5.36 | 3.35 | 186890.88 | 5.64 | 3.65 | 198126.55 | 5.40 | 3.77 | 207576.12 |
| 467.63 | 3.68 | 1.10 | 76153.40 | 5.44 | 3.00 | 167045.55 | 5.37 | 3.33 | 185151.16 | 5.67 | 3.64 | 196376.33 | 5.43 | 3.76 | 206121.64 |
| 469.22 | 3.65 | 1.10 | 76485.23 | 5.46 | 3.00 | 166215.50 | 5.38 | 3.31 | 183519.28 | 5.69 | 3.63 | 194643.24 | 5.46 | 3.76 | 204648.53 |
| 470.81 | 3.61 | 1.10 | 76491.86 | 5.48 | 3.00 | 165302.20 | 5.38 | 3.30 | 181989.74 | 5.72 | 3.61 | 192911.20 | 5.49 | 3.75 | 203153.10 |
| 472.40 | 3.56 | 1.10 | 76618.66 | 5.51 | 3.00 | 164277.66 | 5.39 | 3.28 | 180556.33 | 5.75 | 3.60 | 191164.66 | 5.52 | 3.74 | 201630.95 |
| 473.99 | 3.49 | 1.10 | 77192.39 | 5.54 | 2.99 | 163116.01 | 5.40 | 3.27 | 179212.58 | 5.78 | 3.58 | 189387.92 | 5.55 | 3.73 | 200079.20 |
| 475.58 | 3.42 | 1.11 | 78384.53 | 5.57 | 2.99 | 161794.82 | 5.41 | 3.26 | 177952.00 | 5.81 | 3.57 | 187567.25 | 5.58 | 3.72 | 198494.16 |
| 477.16 | 3.34 | 1.13 | 80266.88 | 5.61 | 2.98 | 160293.05 | 5.41 | 3.25 | 176767.75 | 5.84 | 3.55 | 185689.95 | 5.61 | 3.71 | 196873.05 |
| 478.75 | 3.25 | 1.16 | 82875.10 | 5.65 | 2.96 | 158593.40 | 5.42 | 3.24 | 175652.54 | 5.87 | 3.53 | 183746.39 | 5.64 | 3.70 | 195213.64 |
| 480.34 | 3.15 | 1.19 | 86247.45 | 5.69 | 2.95 | 156682.03 | 5.43 | 3.24 | 174599.58 | 5.90 | 3.51 | 181730.07 | 5.68 | 3.69 | 193513.08 |
| 481.93 | 3.05 | 1.24 | 90446.54 | 5.73 | 2.92 | 154548.57 | 5.44 | 3.23 | 173601.24 | 5.94 | 3.49 | 179637.32 | 5.71 | 3.68 | 191769.71 |
| 483.52 | 2.94 | 1.29 | 95570.23 | 5.78 | 2.90 | 152187.11 | 5.46 | 3.23 | 172650.21 | 5.97 | 3.46 | 177469.38 | 5.75 | 3.66 | 189981.34 |
| 485.11 | 2.83 | 1.36 | 101760.76 | 5.82 | 2.87 | 149596.12 | 5.47 | 3.22 | 171738.86 | 6.00 | 3.44 | 175230.72 | 5.78 | 3.65 | 188146.97 |
| 486.70 | 2.70 | 1.44 | 109214.68 | 5.87 | 2.83 | 146779.30 | 5.49 | 3.22 | 170859.23 | 6.03 | 3.41 | 172930.46 | 5.82 | 3.63 | 186264.88 |
| 488.29 | 2.56 | 1.53 | 118196.85 | 5.91 | 2.79 | 143744.81 | 5.50 | 3.22 | 170003.05 | 6.06 | 3.38 | 170580.04 | 5.86 | 3.62 | 184333.76 |
| 489.88 | 2.42 | 1.64 | 129052.30 | 5.95 | 2.74 | 140506.59 | 5.52 | 3.22 | 169161.68 | 6.09 | 3.35 | 168195.12 | 5.89 | 3.60 | 182352.85 |
| 491.47 | 2.25 | 1.78 | 142227.36 | 5.99 | 2.69 | 137083.43 | 5.54 | 3.22 | 168326.95 | 6.12 | 3.32 | 165792.57 | 5.93 | 3.58 | 180321.26 |
| 493.06 | 2.07 | 1.95 | 158276.91 | 6.02 | 2.63 | 133499.11 | 5.56 | 3.22 | 167489.45 | 6.14 | 3.28 | 163391.25 | 5.97 | 3.55 | 178238.31 |
| 494.65 | 1.87 | 2.15 | 177862.36 | 6.06 | 2.57 | 129781.03 | 5.59 | 3.22 | 166640.23 | 6.16 | 3.25 | 161009.61 | 6.01 | 3.53 | 176103.36 |
| 496.24 | 1.65 | 2.41 | 201707.71 | 6.08 | 2.50 | 125961.44 | 5.61 | 3.22 | 165769.37 | 6.18 | 3.21 | 158666.28 | 6.05 | 3.51 | 173916.27 |
| 497.83 | 1.41 | 2.73 | 230506.13 | 6.10 | 2.43 | 122074.35 | 5.64 | 3.22 | 164867.20 | 6.20 | 3.18 | 156378.05 | 6.08 | 3.48 | 171677.00 |
| 499.43 | 1.14 | 3.16 | 264765.41 | 6.12 | 2.36 | 118156.42 | 5.67 | 3.22 | 163923.49 | 6.22 | 3.15 | 154159.75 | 6.12 | 3.45 | 169385.58 |
| 501.02 | 0.86 | 3.71 | 304640.66 | 6.13 | 2.29 | 114245.05 | 5.70 | 3.22 | 162928.05 | 6.23 | 3.11 | 152018.72 | 6.16 | 3.42 | 167042.21 |
| 502.61 | 0.59 | 4.46 | 349743.31 | 6.13 | 2.22 | 110377.84 | 5.74 | 3.21 | 161869.78 | 6.24 | 3.08 | 149958.14 | 6.20 | 3.39 | 164647.80 |
| 504.20 | 0.39 | 5.50 | 398770.16 | 6.12 | 2.15 | 106591.19 | 5.77 | 3.21 | 160738.54 | 6.25 | 3.05 | 147986.67 | 6.23 | 3.36 | 162203.54 |
| 505.79 | 0.42 | 6.92 | 448413.54 | 6.11 | 2.08 | 102919.70 | 5.81 | 3.20 | 159523.11 | 6.26 | 3.02 | 146108.72 | 6.27 | 3.32 | 159710.93 |
| 507.38 | 1.00 | 8.80 | 490687.03 | 6.10 | 2.01 | 99395.33 | 5.85 | 3.20 | 158212.79 | 6.27 | 3.00 | 144326.05 | 6.30 | 3.29 | 157172.32 |
| 508.97 | 2.73 | 10.88 | 508539.49 | 6.07 | 1.94 | 96047.31 | 5.89 | 3.19 | 156797.33 | 6.28 | 2.97 | 142637.97 | 6.34 | 3.25 | 154590.31 |
| 510.56 | 6.00 | 12.10 | 476906.40 | 6.04 | 1.88 | 92900.55 | 5.94 | 3.18 | 155266.61 | 6.29 | 2.95 | 141039.88 | 6.37 | 3.21 | 151968.73 |
| 512.15 | 9.78 | 11.08 | 388017.57 | 6.01 | 1.82 | 89976.98 | 5.99 | 3.16 | 153611.18 | 6.30 | 2.93 | 139526.54 | 6.40 | 3.17 | 149311.66 |
| 513.74 | 12.02 | 8.26 | 277003.19 | 5.97 | 1.76 | 87294.33 | 6.03 | 3.14 | 151822.12 | 6.31 | 2.91 | 138089.31 | 6.43 | 3.12 | 146624.29 |
| 515.33 | 12.41 | 5.44 | 184269.75 | 5.93 | 1.71 | 84866.71 | 6.08 | 3.12 | 149892.18 | 6.32 | 2.89 | 136719.22 | 6.46 | 3.08 | 143913.01 |
| 516.93 | 11.84 | 3.44 | 120345.04 | 5.88 | 1.67 | 82704.93 | 6.13 | 3.10 | 147814.90 | 6.33 | 2.87 | 135404.98 | 6.48 | 3.03 | 141184.39 |
| 518.52 | 11.03 | 2.18 | 79046.30 | 5.83 | 1.63 | 80816.68 | 6.18 | 3.07 | 145585.54 | 6.34 | 2.85 | 134134.88 | 6.51 | 2.99 | 138446.09 |
| 520.11 | 10.24 | 1.39 | 52538.71 | 5.78 | 1.59 | 79206.73 | 6.23 | 3.04 | 143201.65 | 6.35 | 2.84 | 132896.72 | 6.53 | 2.94 | 135706.89 |
| 521.70 | 9.56 | 0.91 | 35279.07 | 5.73 | 1.56 | 77878.28 | 6.28 | 3.01 | 140662.46 | 6.37 | 2.82 | 131677.16 | 6.55 | 2.89 | 132975.95 |
| 523.29 | 8.99 | 0.60 | 23825.10 | 5.67 | 1.54 | 76832.12 | 6.33 | 2.97 | 137970.30 | 6.38 | 2.81 | 130463.02 | 6.56 | 2.84 | 130262.26 |
| 524.88 | 8.52 | 0.39 | 16089.97 | 5.61 | 1.52 | 76068.37 | 6.38 | 2.92 | 135129.35 | 6.40 | 2.79 | 129241.47 | 6.58 | 2.79 | 127576.59 |

| | | | | | | | | | | | | | | | | |
|---|---|---|---|---|---|---|---|---|---|---|---|---|---|---|---|---|
| 526.47 | 8.12 | 0.26 | 10795.90 | 5.56 | 1.51 | 75583.49 | 6.42 | 2.87 | 132147.53 | 6.42 | 2.78 | 127999.35 | 6.59 | 2.74 | 124928.62 |
| 528.07 | 7.79 | 0.17 | 7142.66 | 5.50 | 1.50 | 75375.43 | 6.46 | 2.82 | 129034.95 | 6.44 | 2.76 | 126724.03 | 6.60 | 2.69 | 122328.25 |
| 529.66 | 7.50 | 0.11 | 4616.97 | 5.44 | 1.50 | 75443.78 | 6.50 | 2.76 | 125804.57 | 6.46 | 2.74 | 125403.61 | 6.60 | 2.64 | 119785.19 |
| 531.25 | 7.26 | 0.07 | 2879.21 | 5.38 | 1.50 | 75788.98 | 6.54 | 2.70 | 122472.71 | 6.49 | 2.73 | 124027.51 | 6.61 | 2.60 | 117309.20 |
| 532.84 | 7.05 | 0.04 | 1700.85 | 5.32 | 1.51 | 76412.76 | 6.57 | 2.64 | 119057.70 | 6.51 | 2.71 | 122585.35 | 6.61 | 2.55 | 114909.10 |
| 534.43 | 6.87 | 0.02 | 925.49 | 5.27 | 1.52 | 77318.02 | 6.60 | 2.57 | 115580.25 | 6.54 | 2.69 | 121068.52 | 6.61 | 2.50 | 112592.40 |
| 536.03 | 6.71 | 0.01 | 444.26 | 5.21 | 1.54 | 78510.72 | 6.62 | 2.50 | 112063.19 | 6.57 | 2.66 | 119469.25 | 6.61 | 2.46 | 110366.81 |
| 537.62 | 6.57 | 0.00 | 179.28 | 5.15 | 1.57 | 79998.11 | 6.64 | 2.43 | 108530.50 | 6.60 | 2.64 | 117782.23 | 6.60 | 2.42 | 108238.56 |
| 539.21 | 6.44 | 0.00 | 74.58 | 5.09 | 1.60 | 81791.76 | 6.65 | 2.36 | 105006.96 | 6.63 | 2.61 | 116002.56 | 6.60 | 2.38 | 106212.53 |
| 540.80 | 6.33 | 0.00 | 65.06 | 5.04 | 1.64 | 83906.70 | 6.66 | 2.29 | 101517.21 | 6.65 | 2.58 | 114128.38 | 6.59 | 2.34 | 104293.28 |
| 542.39 | 6.23 | 0.00 | 61.40 | 4.98 | 1.69 | 86362.05 | 6.66 | 2.21 | 98086.85 | 6.68 | 2.55 | 112159.06 | 6.58 | 2.30 | 102483.73 |
| 543.99 | 6.14 | 0.00 | 57.98 | 4.93 | 1.74 | 89182.55 | 6.65 | 2.14 | 94739.15 | 6.71 | 2.51 | 110096.14 | 6.57 | 2.27 | 100785.02 |
| 545.58 | 6.06 | 0.00 | 54.82 | 4.88 | 1.80 | 92399.34 | 6.64 | 2.07 | 91497.13 | 6.74 | 2.47 | 107942.99 | 6.57 | 2.24 | 99197.56 |
| 547.17 | 5.99 | 0.00 | 51.67 | 4.82 | 1.87 | 96050.54 | 6.63 | 2.00 | 88381.71 | 6.76 | 2.43 | 105705.24 | 6.56 | 2.21 | 97719.25 |
| 548.76 | 5.92 | 0.00 | 48.78 | 4.77 | 1.95 | 100182.71 | 6.61 | 1.94 | 85411.44 | 6.79 | 2.39 | 103390.01 | 6.55 | 2.18 | 96346.37 |
| 550.36 | 5.86 | 0.00 | 45.89 | 4.73 | 2.04 | 104851.99 | 6.58 | 1.87 | 82603.57 | 6.81 | 2.34 | 101006.43 | 6.54 | 2.16 | 95072.77 |
| 551.95 | 5.80 | 0.00 | 43.49 | 4.68 | 2.14 | 110124.26 | 6.55 | 1.81 | 79972.17 | 6.83 | 2.29 | 98565.08 | 6.53 | 2.13 | 93888.67 |
| 553.54 | 5.75 | 0.00 | 40.86 | 4.64 | 2.26 | 116051.14 | 6.52 | 1.76 | 77529.14 | 6.85 | 2.24 | 96078.09 | 6.52 | 2.11 | 92781.57 |
| 555.13 | 5.70 | 0.00 | 38.71 | 4.61 | 2.40 | 122704.77 | 6.48 | 1.71 | 75284.57 | 6.86 | 2.19 | 93558.93 | 6.52 | 2.10 | 91735.09 |
| 556.73 | 5.65 | 0.00 | 36.34 | 4.58 | 2.56 | 130196.42 | 6.44 | 1.66 | 73246.11 | 6.87 | 2.14 | 91021.73 | 6.52 | 2.08 | 90728.97 |
| 558.32 | 5.60 | 0.00 | 34.44 | 4.56 | 2.74 | 138638.74 | 6.39 | 1.62 | 71419.10 | 6.88 | 2.09 | 88481.16 | 6.52 | 2.06 | 89738.65 |
| 559.91 | 5.56 | 0.00 | 32.32 | 4.56 | 2.95 | 148131.62 | 6.34 | 1.58 | 69808.28 | 6.89 | 2.03 | 85952.12 | 6.52 | 2.04 | 88736.68 |
| 561.50 | 5.52 | 0.00 | 30.66 | 4.58 | 3.20 | 158730.83 | 6.30 | 1.55 | 68415.85 | 6.89 | 1.98 | 83450.41 | 6.53 | 2.03 | 87693.41 |
| 563.10 | 5.48 | 0.00 | 28.79 | 4.64 | 3.49 | 170389.08 | 6.25 | 1.52 | 67244.45 | 6.88 | 1.92 | 80990.77 | 6.54 | 2.01 | 86579.50 |
| 564.69 | 5.45 | 0.00 | 27.15 | 4.76 | 3.83 | 182847.34 | 6.19 | 1.49 | 66294.58 | 6.88 | 1.87 | 78587.69 | 6.55 | 1.98 | 85368.13 |
| 566.28 | 5.41 | 0.00 | 25.74 | 4.96 | 4.22 | 195451.00 | 6.14 | 1.47 | 65567.53 | 6.87 | 1.82 | 76255.60 | 6.56 | 1.96 | 84038.43 |
| 567.88 | 5.38 | 0.00 | 24.12 | 5.30 | 4.64 | 206868.48 | 6.09 | 1.46 | 65064.30 | 6.85 | 1.77 | 74007.64 | 6.57 | 1.93 | 82578.55 |
| 569.47 | 5.35 | 0.00 | 22.73 | 5.82 | 5.06 | 214746.18 | 6.03 | 1.45 | 64786.13 | 6.84 | 1.72 | 71856.13 | 6.58 | 1.90 | 80987.40 |
| 571.06 | 5.32 | 0.00 | 21.57 | 6.59 | 5.38 | 215537.15 | 5.98 | 1.45 | 64735.43 | 6.82 | 1.67 | 69812.50 | 6.59 | 1.87 | 79275.22 |
| 572.65 | 5.29 | 0.00 | 20.19 | 7.58 | 5.44 | 205172.72 | 5.93 | 1.45 | 64915.47 | 6.79 | 1.62 | 67887.36 | 6.60 | 1.83 | 77462.89 |
| 574.25 | 5.26 | 0.00 | 19.04 | 8.65 | 5.06 | 181395.09 | 5.87 | 1.46 | 65331.50 | 6.77 | 1.58 | 66089.53 | 6.61 | 1.79 | 75578.75 |
| 575.84 | 5.24 | 0.00 | 17.89 | 9.50 | 4.25 | 146821.07 | 5.82 | 1.47 | 65990.28 | 6.74 | 1.54 | 64427.55 | 6.61 | 1.75 | 73655.93 |
| 577.43 | 5.21 | 0.00 | 16.97 | 9.92 | 3.19 | 108805.23 | 5.76 | 1.49 | 66901.35 | 6.71 | 1.51 | 62908.37 | 6.61 | 1.71 | 71727.84 |
| 579.03 | 5.19 | 0.00 | 15.84 | 9.91 | 2.18 | 74797.83 | 5.71 | 1.51 | 68077.62 | 6.68 | 1.47 | 61536.89 | 6.60 | 1.67 | 69825.98 |
| 580.62 | 5.17 | 0.00 | 14.93 | 9.61 | 1.39 | 48450.60 | 5.66 | 1.54 | 69534.63 | 6.65 | 1.45 | 60315.11 | 6.59 | 1.63 | 67976.54 |
| 582.21 | 5.14 | 0.00 | 14.03 | 9.21 | 0.84 | 29841.18 | 5.61 | 1.58 | 71293.16 | 6.62 | 1.42 | 59246.36 | 6.58 | 1.59 | 66200.24 |
| 583.80 | 5.12 | 0.00 | 13.13 | 8.79 | 0.48 | 17444.25 | 5.56 | 1.62 | 73378.54 | 6.59 | 1.40 | 58333.25 | 6.57 | 1.55 | 64512.20 |
| 585.40 | 5.10 | 0.00 | 12.45 | 8.41 | 0.26 | 9532.35 | 5.51 | 1.68 | 75821.13 | 6.55 | 1.38 | 57576.01 | 6.55 | 1.51 | 62917.70 |
| 586.99 | 5.08 | 0.00 | 11.56 | 8.07 | 0.12 | 4697.37 | 5.46 | 1.74 | 78658.52 | 6.52 | 1.37 | 56972.32 | 6.54 | 1.48 | 61418.14 |
| 588.58 | 5.06 | 0.00 | 10.89 | 7.79 | 0.05 | 1925.78 | 5.42 | 1.81 | 81934.19 | 6.49 | 1.36 | 56517.05 | 6.52 | 1.44 | 60015.49 |
| 590.18 | 5.04 | 0.00 | 10.22 | 7.55 | 0.01 | 522.73 | 5.38 | 1.89 | 85698.14 | 6.46 | 1.35 | 56200.76 | 6.50 | 1.41 | 58708.80 |
| 591.77 | 5.03 | 0.00 | 9.56 | 7.35 | 0.00 | 23.15 | 5.35 | 1.99 | 89977.98 | 6.44 | 1.35 | 56008.89 | 6.48 | 1.39 | 57495.56 |
| 593.36 | 5.01 | 0.00 | 8.89 | 7.19 | 0.00 | 0.00 | 5.32 | 2.10 | 94821.66 | 6.41 | 1.34 | 55919.27 | 6.46 | 1.36 | 56371.63 |
| 594.96 | 4.99 | 0.00 | 8.45 | 7.05 | 0.00 | 0.00 | 5.30 | 2.23 | 100293.79 | 6.39 | 1.35 | 55902.60 | 6.43 | 1.34 | 55332.32 |
| 596.55 | 4.98 | 0.00 | 7.79 | 6.94 | 0.00 | 0.00 | 5.30 | 2.38 | 106441.70 | 6.38 | 1.35 | 55919.49 | 6.41 | 1.31 | 54372.26 |
| 598.14 | 4.96 | 0.00 | 7.35 | 6.84 | 0.00 | 0.00 | 5.32 | 2.55 | 113270.96 | 6.37 | 1.35 | 55921.39 | 6.39 | 1.29 | 53486.45 |
| 599.74 | 4.94 | 0.00 | 6.91 | 6.74 | 0.00 | 0.00 | 5.37 | 2.75 | 120700.89 | 6.36 | 1.35 | 55853.48 | 6.37 | 1.27 | 52669.87 |
| 601.33 | 4.93 | 0.00 | 6.27 | 6.66 | 0.00 | 0.00 | 5.46 | 2.97 | 128494.75 | 6.36 | 1.35 | 55657.84 | 6.35 | 1.26 | 51917.79 |
| 602.92 | 4.91 | 0.00 | 5.84 | 6.59 | 0.00 | 0.00 | 5.61 | 3.21 | 136157.04 | 6.36 | 1.34 | 55282.96 | 6.33 | 1.24 | 51225.58 |
| 604.52 | 4.90 | 0.00 | 5.40 | 6.52 | 0.00 | 0.00 | 5.85 | 3.45 | 142810.33 | 6.36 | 1.33 | 54694.66 | 6.30 | 1.23 | 50588.73 |
| 606.11 | 4.89 | 0.00 | 5.18 | 6.45 | 0.00 | 0.00 | 6.19 | 3.67 | 147119.32 | 6.36 | 1.32 | 53885.99 | 6.28 | 1.21 | 50003.99 |
| 607.70 | 4.87 | 0.00 | 4.76 | 6.39 | 0.00 | 0.00 | 6.63 | 3.81 | 147395.73 | 6.35 | 1.30 | 52884.70 | 6.26 | 1.20 | 49467.18 |
| 609.30 | 4.86 | 0.00 | 4.33 | 6.34 | 0.00 | 0.00 | 7.17 | 3.81 | 142083.48 | 6.34 | 1.27 | 51754.15 | 6.24 | 1.19 | 48975.02 |
| 610.89 | 4.85 | 0.00 | 4.11 | 6.29 | 0.00 | 0.00 | 7.71 | 3.62 | 130618.18 | 6.32 | 1.24 | 50583.65 | 6.23 | 1.18 | 48524.54 |
| 612.48 | 4.84 | 0.00 | 3.69 | 6.24 | 0.00 | 0.00 | 8.17 | 3.24 | 114141.10 | 6.30 | 1.22 | 49476.26 | 6.21 | 1.17 | 48112.49 |
| 614.08 | 4.82 | 0.00 | 3.48 | 6.19 | 0.00 | 0.00 | 8.47 | 2.74 | 95245.75 | 6.26 | 1.19 | 48533.41 | 6.19 | 1.17 | 47736.14 |
| 615.67 | 4.81 | 0.00 | 3.27 | 6.15 | 0.00 | 0.00 | 8.59 | 2.22 | 76766.86 | 6.22 | 1.17 | 47844.07 | 6.17 | 1.16 | 47392.78 |
| 617.26 | 4.80 | 0.00 | 2.85 | 6.11 | 0.00 | 0.00 | 8.55 | 1.75 | 60635.76 | 6.18 | 1.16 | 47479.71 | 6.16 | 1.15 | 47079.27 |
| 618.86 | 4.79 | 0.00 | 2.64 | 6.07 | 0.00 | 0.00 | 8.42 | 1.36 | 47600.92 | 6.12 | 1.16 | 47494.72 | 6.14 | 1.15 | 46793.16 |
| 620.45 | 4.78 | 0.00 | 2.43 | 6.04 | 0.00 | 0.00 | 8.24 | 1.07 | 37578.31 | 6.07 | 1.17 | 47930.33 | 6.12 | 1.14 | 46531.62 |
| 622.04 | 4.77 | 0.00 | 2.22 | 6.00 | 0.00 | 0.00 | 8.05 | 0.85 | 30102.38 | 6.01 | 1.19 | 48820.28 | 6.11 | 1.14 | 46291.43 |
| 623.64 | 4.76 | 0.00 | 2.02 | 5.97 | 0.00 | 0.00 | 7.86 | 0.69 | 24627.25 | 5.95 | 1.22 | 50196.72 | 6.09 | 1.13 | 46069.78 |
| 625.23 | 4.75 | 0.00 | 1.81 | 5.94 | 0.00 | 0.00 | 7.68 | 0.57 | 20664.12 | 5.90 | 1.27 | 52093.73 | 6.08 | 1.13 | 45863.12 |
| 626.83 | 4.74 | 0.00 | 1.60 | 5.91 | 0.00 | 0.00 | 7.52 | 0.49 | 17820.76 | 5.84 | 1.32 | 54551.86 | 6.07 | 1.13 | 45668.55 |
| 628.42 | 4.73 | 0.00 | 1.40 | 5.88 | 0.00 | 0.00 | 7.38 | 0.43 | 15799.09 | 5.79 | 1.40 | 57616.84 | 6.05 | 1.12 | 45482.17 |
| 630.01 | 4.72 | 0.00 | 1.40 | 5.85 | 0.00 | 0.00 | 7.25 | 0.39 | 14378.85 | 5.75 | 1.49 | 61338.29 | 6.04 | 1.12 | 45300.91 |
| 631.61 | 4.71 | 0.00 | 1.19 | 5.82 | 0.00 | 0.00 | 7.13 | 0.36 | 13398.51 | 5.72 | 1.60 | 65759.80 | 6.03 | 1.12 | 45121.37 |
| 633.20 | 4.70 | 0.00 | 0.99 | 5.80 | 0.00 | 0.00 | 7.03 | 0.34 | 12740.64 | 5.70 | 1.73 | 70900.90 | 6.02 | 1.12 | 44940.74 |
| 634.79 | 4.69 | 0.00 | 0.99 | 5.77 | 0.00 | 0.00 | 6.94 | 0.33 | 12319.67 | 5.71 | 1.88 | 76720.93 | 6.01 | 1.11 | 44757.26 |
| 636.39 | 4.68 | 0.00 | 0.79 | 5.75 | 0.00 | 0.00 | 6.86 | 0.32 | 12072.60 | 5.76 | 2.05 | 83054.07 | 5.99 | 1.11 | 44570.56 |
| 637.98 | 4.67 | 0.00 | 0.79 | 5.72 | 0.00 | 0.00 | 6.79 | 0.32 | 11952.82 | 5.85 | 2.24 | 89546.42 | 5.98 | 1.11 | 44382.81 |
| 639.57 | 4.67 | 0.00 | 0.59 | 5.70 | 0.00 | 0.00 | 6.73 | 0.31 | 11872.95 | 6.00 | 2.43 | 95551.61 | 5.97 | 1.10 | 44199.75 |
| 641.17 | 4.66 | 0.00 | 0.59 | 5.68 | 0.00 | 0.00 | 6.67 | 0.31 | 11789.15 | 6.23 | 2.60 | 100073.61 | 5.95 | 1.10 | 44031.49 |
| 642.76 | 4.65 | 0.00 | 0.39 | 5.66 | 0.00 | 0.00 | 6.62 | 0.31 | 11702.64 | 6.54 | 2.72 | 101867.63 | 5.93 | 1.10 | 43895.38 |
| 644.35 | 4.64 | 0.00 | 0.39 | 5.64 | 0.00 | 0.00 | 6.57 | 0.31 | 11614.22 | 6.90 | 2.74 | 99808.45 | 5.91 | 1.10 | 43815.89 |
| 645.95 | 4.63 | 0.00 | 0.39 | 5.62 | 0.00 | 0.00 | 6.53 | 0.30 | 11524.48 | 7.28 | 2.63 | 93470.13 | 5.89 | 1.10 | 43827.05 |
| 647.54 | 4.63 | 0.00 | 0.19 | 5.60 | 0.00 | 0.00 | 6.49 | 0.30 | 11433.64 | 7.61 | 2.40 | 83534.48 | 5.86 | 1.10 | 43972.48 |
| 649.13 | 4.62 | 0.00 | 0.19 | 5.58 | 0.00 | 0.00 | 6.45 | 0.30 | 11342.08 | 7.84 | 2.09 | 71588.55 | 5.83 | 1.11 | 44305.34 |
| 650.73 | 4.61 | 0.00 | 0.19 | 5.56 | 0.00 | 0.00 | 6.41 | 0.30 | 11250.00 | 7.96 | 1.75 | 59403.16 | 5.80 | 1.12 | 44885.30 |
| 652.32 | 4.61 | 0.00 | 0.19 | 5.54 | 0.00 | 0.00 | 6.38 | 0.29 | 11157.61 | 7.98 | 1.42 | 48278.48 | 5.77 | 1.15 | 45775.68 |
| 653.91 | 4.60 | 0.00 | 0.00 | 5.53 | 0.00 | 0.00 | 6.35 | 0.29 | 11065.28 | 7.92 | 1.14 | 38841.46 | 5.73 | 1.18 | 47037.41 |
| 655.51 | 4.59 | 0.00 | 0.00 | 5.51 | 0.00 | 0.00 | 6.32 | 0.29 | 10973.01 | 7.83 | 0.91 | 31196.16 | 5.70 | 1.22 | 48721.65 |
| 657.10 | 4.58 | 0.00 | 0.00 | 5.49 | 0.00 | 0.00 | 6.29 | 0.29 | 10880.81 | 7.71 | 0.73 | 25169.51 | 5.67 | 1.27 | 50863.21 |
| 658.69 | 4.58 | 0.00 | 0.00 | 5.48 | 0.00 | 0.00 | 6.26 | 0.28 | 10789.05 | 7.59 | 0.59 | 20488.96 | 5.65 | 1.34 | 53471.38 |
| 660.28 | 4.57 | 0.00 | 0.00 | 5.46 | 0.00 | 0.00 | 6.23 | 0.28 | 10697.55 | 7.47 | 0.49 | 16879.25 | 5.64 | 1.42 | 56521.04 |
| 661.88 | 4.57 | 0.00 | 0.00 | 5.45 | 0.00 | 0.00 | 6.21 | 0.28 | 10606.49 | 7.36 | 0.40 | 14101.42 | 5.64 | 1.51 | 59944.27 |
| 663.47 | 4.56 | 0.00 | 0.00 | 5.43 | 0.00 | 0.00 | 6.19 | 0.28 | 10516.06 | 7.26 | 0.34 | 11963.10 | 5.67 | 1.61 | 63573.63 |
| 665.06 | 4.55 | 0.00 | 0.00 | 5.42 | 0.00 | 0.00 | 6.16 | 0.27 | 10426.06 | 7.16 | 0.29 | 10314.01 | 5.71 | 1.72 | 67145.18 |
| 666.66 | 4.55 | 0.00 | 0.00 | 5.40 | 0.00 | 0.00 | 6.14 | 0.27 | 10336.68 | 7.07 | 0.26 | 9039.81 | 5.79 | 1.82 | 70450.63 |
| 668.25 | 4.54 | 0.00 | 0.00 | 5.39 | 0.00 | 0.00 | 6.12 | 0.27 | 10248.09 | 6.99 | 0.23 | 8053.57 | 5.89 | 1.91 | 73224.69 |
| 669.84 | 4.53 | 0.00 | 0.00 | 5.38 | 0.00 | 0.00 | 6.10 | 0.27 | 10159.94 | 6.92 | 0.20 | 7289.64 | 6.03 | 1.99 | 75161.46 |
| 671.44 | 4.53 | 0.00 | 0.00 | 5.36 | 0.00 | 0.00 | 6.08 | 0.27 | 10072.57 | 6.85 | 0.19 | 6697.96 | 6.19 | 2.05 | 75957.07 |
| 673.03 | 4.52 | 0.00 | 0.00 | 5.35 | 0.00 | 0.00 | 6.06 | 0.26 | 9986.00 | 6.79 | 0.17 | 6240.53 | 6.36 | 2.06 | 75373.46 |
| 674.62 | 4.52 | 0.00 | 0.00 | 5.34 | 0.00 | 0.00 | 6.04 | 0.26 | 9900.20 | 6.73 | 0.16 | 5888.64 | 6.55 | 2.04 | 73305.18 |
| 676.22 | 4.51 | 0.00 | 0.00 | 5.33 | 0.00 | 0.00 | 6.02 | 0.26 | 9815.00 | 6.68 | 0.16 | 5619.43 | 6.73 | 1.97 | 69822.62 |
| 677.81 | 4.51 | 0.00 | 0.00 | 5.32 | 0.00 | 0.00 | 6.00 | 0.26 | 9730.75 | 6.63 | 0.15 | 5416.01 | 6.89 | 1.86 | 65168.76 |
| 679.40 | 4.50 | 0.00 | 0.00 | 5.30 | 0.00 | 0.00 | 5.99 | 0.26 | 9647.09 | 6.58 | 0.15 | 5264.40 | 7.02 | 1.72 | 59706.61 |
| 680.99 | 4.50 | 0.00 | 0.00 | 5.29 | 0.00 | 0.00 | 5.97 | 0.25 | 9564.37 | 6.54 | 0.14 | 5154.66 | 7.12 | 1.57 | 53837.69 |
| 682.59 | 4.49 | 0.00 | 0.00 | 5.28 | 0.00 | 0.00 | 5.95 | 0.25 | 9482.22 | 6.50 | 0.14 | 5077.83 | 7.18 | 1.40 | 47925.83 |
| 684.18 | 4.49 | 0.00 | 0.00 | 5.27 | 0.00 | 0.00 | 5.94 | 0.25 | 9401.00 | 6.47 | 0.14 | 5027.62 | 7.21 | 1.24 | 42250.87 |
| 685.77 | 4.48 | 0.00 | 0.00 | 5.26 | 0.00 | 0.00 | 5.92 | 0.25 | 9320.71 | 6.43 | 0.14 | 4989.00 | 7.21 | 1.09 | 36994.96 |
| 687.36 | 4.48 | 0.00 | 0.00 | 5.25 | 0.00 | 0.00 | 5.91 | 0.25 | 9240.98 | 6.40 | 0.14 | 4950.39 | 7.19 | 0.95 | 32251.92 |
| 688.96 | 4.47 | 0.00 | 0.00 | 5.24 | 0.00 | 0.00 | 5.89 | 0.24 | 9162.16 | 6.37 | 0.14 | 4911.58 | 7.16 | 0.82 | 28050.99 |
| 690.55 | 4.47 | 0.00 | 0.00 | 5.23 | 0.00 | 0.00 | 5.88 | 0.24 | 9084.07 | 6.34 | 0.13 | 4872.96 | 7.11 | 0.72 | 24376.81 |
| 692.14 | 4.46 | 0.00 | 0.00 | 5.22 | 0.00 | 0.00 | 5.87 | 0.24 | 9006.89 | 6.32 | 0.13 | 4834.33 | 7.06 | 0.62 | 21190.12 |
| 693.74 | 4.46 | 0.00 | 0.00 | 5.21 | 0.00 | 0.00 | 5.85 | 0.24 | 8930.42 | 6.29 | 0.13 | 4796.06 | 7.00 | 0.54 | 18440.13 |
| 695.33 | 4.45 | 0.00 | 0.00 | 5.20 | 0.00 | 0.00 | 5.84 | 0.24 | 8854.67 | 6.27 | 0.13 | 4757.79 | 6.95 | 0.47 | 16073.04 |
| 696.92 | 4.45 | 0.00 | 0.00 | 5.19 | 0.00 | 0.00 | 5.83 | 0.24 | 8779.79 | 6.25 | 0.13 | 4719.87 | 6.89 | 0.41 | 14037.00 |
| 698.51 | 4.44 | 0.00 | 0.00 | 5.18 | 0.00 | 0.00 | 5.82 | 0.23 | 8705.81 | 6.22 | 0.13 | 4682.13 | 6.83 | 0.36 | 12284.78 |
| 700.10 | 4.44 | 0.00 | 0.00 | 5.17 | 0.00 | 0.00 | 5.80 | 0.23 | 8632.34 | 6.20 | 0.13 | 4644.55 | 6.78 | 0.31 | 10774.77 |
| 701.70 | 4.43 | 0.00 | 0.00 | 5.16 | 0.00 | 0.00 | 5.79 | 0.23 | 8559.74 | 6.18 | 0.13 | 4607.51 | 6.72 | 0.27 | 9470.75 |
| 703.29 | 4.43 | 0.00 | 0.00 | 5.15 | 0.00 | 0.00 | 5.78 | 0.23 | 8488.01 | 6.16 | 0.13 | 4570.63 | 6.67 | 0.24 | 8341.85 |
| 704.88 | 4.43 | 0.00 | 0.00 | 5.14 | 0.00 | 0.00 | 5.77 | 0.23 | 8416.96 | 6.14 | 0.13 | 4534.10 | 6.62 | 0.21 | 7361.92 |
| 706.47 | 4.42 | 0.00 | 0.00 | 5.14 | 0.00 | 0.00 | 5.76 | 0.23 | 8346.58 | 6.13 | 0.13 | 4497.91 | 6.58 | 0.19 | 6509.14 |
| 708.07 | 4.42 | 0.00 | 0.00 | 5.13 | 0.00 | 0.00 | 5.75 | 0.22 | 8277.06 | 6.11 | 0.12 | 4462.07 | 6.53 | 0.17 | 5764.73 |
| 709.66 | 4.41 | 0.00 | 0.00 | 5.12 | 0.00 | 0.00 | 5.73 | 0.22 | 8208.20 | 6.09 | 0.12 | 4426.56 | 6.49 | 0.15 | 5112.90 |
| 711.25 | 4.41 | 0.00 | 0.00 | 5.11 | 0.00 | 0.00 | 5.72 | 0.22 | 8140.18 | 6.07 | 0.12 | 4391.38 | 6.45 | 0.13 | 4540.85 |
| 712.84 | 4.41 | 0.00 | 0.00 | 5.10 | 0.00 | 0.00 | 5.71 | 0.22 | 8072.82 | 6.06 | 0.12 | 4356.54 | 6.41 | 0.12 | 4037.29 |
| 714.43 | 4.40 | 0.00 | 0.00 | 5.10 | 0.00 | 0.00 | 5.70 | 0.22 | 8006.11 | 6.04 | 0.12 | 4322.21 | 6.37 | 0.10 | 3592.78 |
| 716.03 | 4.40 | 0.00 | 0.00 | 5.09 | 0.00 | 0.00 | 5.69 | 0.22 | 7940.22 | 6.03 | 0.12 | 4288.03 | 6.34 | 0.09 | 3199.75 |
| 717.62 | 4.39 | 0.00 | 0.00 | 5.08 | 0.00 | 0.00 | 5.68 | 0.21 | 7874.98 | 6.01 | 0.12 | 4254.36 | 6.31 | 0.08 | 2851.18 |
| 719.21 | 4.39 | 0.00 | 0.00 | 5.07 | 0.00 | 0.00 | 5.67 | 0.21 | 7810.37 | 6.00 | 0.12 | 4220.83 | 6.27 | 0.07 | 2541.37 |
| 720.80 | 4.39 | 0.00 | 0.00 | 5.06 | 0.00 | 0.00 | 5.66 | 0.21 | 7746.40 | 5.98 | 0.12 | 4187.80 | 6.24 | 0.06 | 2265.71 |
| 722.39 | 4.38 | 0.00 | 0.00 | 5.06 | 0.00 | 0.00 | 5.66 | 0.21 | 7683.24 | 5.97 | 0.12 | 4155.09 | 6.21 | 0.06 | 2019.96 |
| 723.98 | 4.38 | 0.00 | 0.00 | 5.05 | 0.00 | 0.00 | 5.65 | 0.21 | 7620.70 | 5.96 | 0.12 | 4122.87 | 6.19 | 0.05 | 1800.47 |
| 725.58 | 4.38 | 0.00 | 0.00 | 5.04 | 0.00 | 0.00 | 5.64 | 0.21 | 7558.95 | 5.94 | 0.12 | 4090.79 | 6.16 | 0.05 | 1604.27 |
| 727.17 | 4.37 | 0.00 | 0.00 | 5.04 | 0.00 | 0.00 | 5.63 | 0.21 | 7497.65 | 5.93 | 0.11 | 4059.20 | 6.13 | 0.04 | 1428.64 |
| 728.76 | 4.37 | 0.00 | 0.00 | 5.03 | 0.00 | 0.00 | 5.62 | 0.20 | 7437.13 | 5.92 | 0.11 | 4027.92 | 6.11 | 0.04 | 1271.37 |
| 730.35 | 4.37 | 0.00 | 0.00 | 5.02 | 0.00 | 0.00 | 5.61 | 0.20 | 7377.23 | 5.91 | 0.11 | 3996.94 | 6.08 | 0.03 | 1130.43 |
| 731.94 | 4.36 | 0.00 | 0.00 | 5.02 | 0.00 | 0.00 | 5.60 | 0.20 | 7317.93 | 5.89 | 0.11 | 3966.45 | 6.06 | 0.03 | 1004.02 |
| 733.53 | 4.36 | 0.00 | 0.00 | 5.01 | 0.00 | 0.00 | 5.59 | 0.20 | 7259.22 | 5.88 | 0.11 | 3936.09 | 6.04 | 0.03 | 890.83 |
| 735.12 | 4.36 | 0.00 | 0.00 | 5.00 | 0.00 | 0.00 | 5.59 | 0.20 | 7201.12 | 5.87 | 0.11 | 3906.20 | 6.02 | 0.02 | 789.41 |
| 736.71 | 4.35 | 0.00 | 0.00 | 5.00 | 0.00 | 0.00 | 5.58 | 0.20 | 7143.77 | 5.86 | 0.11 | 3876.61 | 6.00 | 0.02 | 698.67 |
| 738.31 | 4.35 | 0.00 | 0.00 | 4.99 | 0.00 | 0.00 | 5.57 | 0.20 | 7086.85 | 5.85 | 0.11 | 3847.33 | 5.98 | 0.02 | 617.34 |
| 739.90 | 4.35 | 0.00 | 0.00 | 4.98 | 0.00 | 0.00 | 5.56 | 0.20 | 7030.68 | 5.84 | 0.11 | 3818.50 | 5.96 | 0.02 | 544.85 |
| 741.49 | 4.34 | 0.00 | 0.00 | 4.98 | 0.00 | 0.00 | 5.55 | 0.19 | 6974.92 | 5.83 | 0.11 | 3789.80 | 5.94 | 0.01 | 480.46 |
| 743.08 | 4.34 | 0.00 | 0.00 | 4.97 | 0.00 | 0.00 | 5.55 | 0.19 | 6919.91 | 5.82 | 0.11 | 3761.57 | 5.92 | 0.01 | 423.12 |
| 744.67 | 4.34 | 0.00 | 0.00 | 4.97 | 0.00 | 0.00 | 5.54 | 0.19 | 6865.30 | 5.81 | 0.11 | 3733.62 | 5.90 | 0.01 | 372.27 |
| 746.26 | 4.33 | 0.00 | 0.00 | 4.96 | 0.00 | 0.00 | 5.53 | 0.19 | 6811.43 | 5.80 | 0.11 | 3705.96 | 5.89 | 0.01 | 327.69 |
| 747.85 | 4.33 | 0.00 | 0.00 | 4.96 | 0.00 | 0.00 | 5.53 | 0.19 | 6758.13 | 5.79 | 0.11 | 3678.58 | 5.87 | 0.01 | 288.51 |
| 749.44 | 4.33 | 0.00 | 0.00 | 4.95 | 0.00 | 0.00 | 5.52 | 0.19 | 6705.22 | 5.78 | 0.10 | 3651.50 | 5.85 | 0.01 | 254.53 |
| 751.03 | 4.33 | 0.00 | 0.00 | 4.94 | 0.00 | 0.00 | 5.51 | 0.19 | 6652.87 | 5.77 | 0.10 | 3624.69 | 5.84 | 0.01 | 225.21 |
| 752.62 | 4.32 | 0.00 | 0.00 | 4.94 | 0.00 | 0.00 | 5.50 | 0.19 | 6601.25 | 5.76 | 0.10 | 3598.33 | 5.82 | 0.01 | 200.19 |
| 754.21 | 4.32 | 0.00 | 0.00 | 4.93 | 0.00 | 0.00 | 5.50 | 0.18 | 6550.00 | 5.75 | 0.10 | 3572.08 | 5.81 | 0.01 | 178.78 |
| 755.80 | 4.32 | 0.00 | 0.00 | 4.93 | 0.00 | 0.00 | 5.49 | 0.18 | 6499.32 | 5.74 | 0.10 | 3546.28 | 5.79 | 0.00 | 159.12 |
| 757.39 | 4.31 | 0.00 | 0.00 | 4.92 | 0.00 | 0.00 | 5.48 | 0.18 | 6449.17 | 5.73 | 0.10 | 3520.58 | 5.78 | 0.00 | 140.86 |
| 758.98 | 4.31 | 0.00 | 0.00 | 4.92 | 0.00 | 0.00 | 5.48 | 0.18 | 6399.56 | 5.73 | 0.10 | 3495.32 | 5.77 | 0.00 | 124.01 |
| 760.57 | 4.31 | 0.00 | 0.00 | 4.91 | 0.00 | 0.00 | 5.47 | 0.18 | 6350.33 | 5.72 | 0.10 | 3470.34 | 5.75 | 0.00 | 108.55 |
| 762.16 | 4.31 | 0.00 | 0.00 | 4.91 | 0.00 | 0.00 | 5.46 | 0.18 | 6301.64 | 5.71 | 0.10 | 3445.62 | 5.74 | 0.00 | 94.48 |
| 763.75 | 4.30 | 0.00 | 0.00 | 4.90 | 0.00 | 0.00 | 5.46 | 0.18 | 6253.47 | 5.70 | 0.10 | 3421.01 | 5.73 | 0.00 | 81.61 |
| 765.34 | 4.30 | 0.00 | 0.00 | 4.90 | 0.00 | 0.00 | 5.45 | 0.18 | 6205.84 | 5.69 | 0.10 | 3396.83 | 5.72 | 0.00 | 70.11 |
| 766.93 | 4.30 | 0.00 | 0.00 | 4.89 | 0.00 | 0.00 | 5.45 | 0.18 | 6158.73 | 5.68 | 0.10 | 3372.91 | 5.70 | 0.00 | 59.97 |
| 768.52 | 4.29 | 0.00 | 0.00 | 4.89 | 0.00 | 0.00 | 5.44 | 0.17 | 6111.98 | 5.68 | 0.10 | 3349.25 | 5.69 | 0.00 | 51.02 |
| 770.11 | 4.29 | 0.00 | 0.00 | 4.88 | 0.00 | 0.00 | 5.43 | 0.17 | 6065.75 | 5.67 | 0.10 | 3325.86 | 5.68 | 0.00 | 43.24 |
| 771.70 | 4.29 | 0.00 | 0.00 | 4.88 | 0.00 | 0.00 | 5.43 | 0.17 | 6020.04 | 5.66 | 0.10 | 3302.56 | 5.67 | 0.00 | 36.80 |
| 773.29 | 4.29 | 0.00 | 0.00 | 4.87 | 0.00 | 0.00 | 5.42 | 0.17 | 5974.68 | 5.65 | 0.10 | 3279.69 | 5.66 | 0.00 | 31.36 |
| 774.88 | 4.28 | 0.00 | 0.00 | 4.87 | 0.00 | 0.00 | 5.42 | 0.17 | 5929.83 | 5.65 | 0.10 | 3257.07 | 5.65 | 0.00 | 27.24 |
| 776.47 | 4.28 | 0.00 | 0.00 | 4.86 | 0.00 | 0.00 | 5.41 | 0.17 | 5885.32 | 5.64 | 0.09 | 3234.54 | 5.64 | 0.00 | 23.79 |
| 778.06 | 4.28 | 0.00 | 0.00 | 4.86 | 0.00 | 0.00 | 5.40 | 0.17 | 5841.33 | 5.63 | 0.09 | 3212.43 | 5.63 | 0.00 | 20.51 |
| 779.65 | 4.28 | 0.00 | 0.00 | 4.85 | 0.00 | 0.00 | 5.40 | 0.17 | 5797.83 | 5.63 | 0.09 | 3190.41 | 5.62 | 0.00 | 17.57 |
| 781.23 | 4.27 | 0.00 | 0.00 | 4.85 | 0.00 | 0.00 | 5.39 | 0.17 | 5754.67 | 5.62 | 0.09 | 3168.80 | 5.61 | 0.00 | 14.80 |
| 782.82 | 4.27 | 0.00 | 0.00 | 4.85 | 0.00 | 0.00 | 5.39 | 0.17 | 5712.01 | 5.61 | 0.09 | 3147.28 | 5.60 | 0.00 | 12.36 |

| | | | | | | | | | | | | | | |
|---|---|---|---|---|---|---|---|---|---|---|---|---|---|---|
| 784.41 | 4.27 | 0.00 | 0.00 | 4.84 | 0.00 | 0.00 | 5.38 | 0.16 | 5669.84 | 5.61 | 0.09 | 3126.01 | 5.59 | 0.00 | 10.09 |
| 786.00 | 4.27 | 0.00 | 0.00 | 4.84 | 0.00 | 0.00 | 5.38 | 0.16 | 5628.01 | 5.60 | 0.09 | 3104.98 | 5.58 | 0.00 | 8.15 |
| 787.59 | 4.27 | 0.00 | 0.00 | 4.83 | 0.00 | 0.00 | 5.37 | 0.16 | 5586.50 | 5.59 | 0.09 | 3084.20 | 5.57 | 0.00 | 6.38 |
| 789.18 | 4.26 | 0.00 | 0.00 | 4.83 | 0.00 | 0.00 | 5.37 | 0.16 | 5545.48 | 5.59 | 0.09 | 3063.50 | 5.56 | 0.00 | 4.78 |
| 790.77 | 4.26 | 0.00 | 0.00 | 4.82 | 0.00 | 0.00 | 5.36 | 0.16 | 5504.78 | 5.58 | 0.09 | 3043.20 | 5.55 | 0.00 | 3.50 |
| 792.35 | 4.26 | 0.00 | 0.00 | 4.82 | 0.00 | 0.00 | 5.36 | 0.16 | 5464.57 | 5.57 | 0.09 | 3022.99 | 5.54 | 0.00 | 2.38 |
| 793.94 | 4.26 | 0.00 | 0.00 | 4.82 | 0.00 | 0.00 | 5.35 | 0.16 | 5424.67 | 5.57 | 0.09 | 3003.02 | 5.54 | 0.00 | 1.58 |
| 795.53 | 4.25 | 0.00 | 0.00 | 4.81 | 0.00 | 0.00 | 5.35 | 0.16 | 5385.26 | 5.56 | 0.09 | 2983.28 | 5.53 | 0.00 | 0.79 |
| 797.12 | 4.25 | 0.00 | 0.00 | 4.81 | 0.00 | 0.00 | 5.34 | 0.16 | 5346.15 | 5.55 | 0.09 | 2963.78 | 5.52 | 0.00 | 0.32 |
| 798.70 | 4.25 | 0.00 | 0.00 | 4.80 | 0.00 | 0.00 | 5.34 | 0.16 | 5307.36 | 5.55 | 0.09 | 2944.51 | 5.51 | 0.00 | 0.00 |
| 800.29 | 4.25 | 0.00 | 0.00 | 4.80 | 0.00 | 0.00 | 5.33 | 0.15 | 5269.04 | 5.54 | 0.09 | 2925.32 | 5.50 | 0.00 | 0.00 |
| 801.88 | 4.25 | 0.00 | 0.00 | 4.80 | 0.00 | 0.00 | 5.33 | 0.15 | 5231.03 | 5.54 | 0.09 | 2906.37 | 5.50 | 0.00 | 0.00 |
| 803.47 | 4.24 | 0.00 | 0.00 | 4.79 | 0.00 | 0.00 | 5.32 | 0.15 | 5193.48 | 5.53 | 0.09 | 2887.65 | 5.49 | 0.00 | 0.00 |
| 805.05 | 4.24 | 0.00 | 0.00 | 4.79 | 0.00 | 0.00 | 5.32 | 0.15 | 5156.23 | 5.53 | 0.09 | 2869.15 | 5.48 | 0.00 | 0.00 |
| 806.64 | 4.24 | 0.00 | 0.00 | 4.79 | 0.00 | 0.00 | 5.31 | 0.15 | 5119.29 | 5.52 | 0.09 | 2850.73 | 5.47 | 0.00 | 0.00 |
| 808.23 | 4.24 | 0.00 | 0.00 | 4.78 | 0.00 | 0.00 | 5.31 | 0.15 | 5082.80 | 5.51 | 0.09 | 2832.54 | 5.47 | 0.00 | 0.00 |
| 809.82 | 4.24 | 0.00 | 0.00 | 4.78 | 0.00 | 0.00 | 5.30 | 0.15 | 5046.62 | 5.51 | 0.09 | 2814.57 | 5.46 | 0.00 | 0.00 |
| 811.40 | 4.23 | 0.00 | 0.00 | 4.77 | 0.00 | 0.00 | 5.30 | 0.15 | 5010.73 | 5.50 | 0.08 | 2796.68 | 5.45 | 0.00 | 0.00 |
| 812.99 | 4.23 | 0.00 | 0.00 | 4.77 | 0.00 | 0.00 | 5.29 | 0.15 | 4975.14 | 5.50 | 0.08 | 2779.17 | 5.45 | 0.00 | 0.00 |
| 814.58 | 4.23 | 0.00 | 0.00 | 4.77 | 0.00 | 0.00 | 5.29 | 0.15 | 4939.99 | 5.49 | 0.08 | 2761.57 | 5.44 | 0.00 | 0.00 |
| 816.16 | 4.23 | 0.00 | 0.00 | 4.76 | 0.00 | 0.00 | 5.29 | 0.15 | 4905.14 | 5.49 | 0.08 | 2744.34 | 5.43 | 0.00 | 0.00 |
| 817.75 | 4.23 | 0.00 | 0.00 | 4.76 | 0.00 | 0.00 | 5.28 | 0.15 | 4870.58 | 5.48 | 0.08 | 2727.18 | 5.43 | 0.00 | 0.00 |
| 819.34 | 4.22 | 0.00 | 0.00 | 4.76 | 0.00 | 0.00 | 5.28 | 0.14 | 4836.30 | 5.48 | 0.08 | 2710.25 | 5.42 | 0.00 | 0.00 |
| 820.92 | 4.22 | 0.00 | 0.00 | 4.75 | 0.00 | 0.00 | 5.27 | 0.14 | 4802.31 | 5.47 | 0.08 | 2693.53 | 5.41 | 0.00 | 0.00 |
| 822.51 | 4.22 | 0.00 | 0.00 | 4.75 | 0.00 | 0.00 | 5.27 | 0.14 | 4768.76 | 5.47 | 0.08 | 2676.88 | 5.41 | 0.00 | 0.00 |
| 824.09 | 4.22 | 0.00 | 0.00 | 4.75 | 0.00 | 0.00 | 5.26 | 0.14 | 4735.49 | 5.46 | 0.08 | 2660.44 | 5.40 | 0.00 | 0.00 |
| 825.68 | 4.22 | 0.00 | 0.00 | 4.74 | 0.00 | 0.00 | 5.26 | 0.14 | 4702.50 | 5.46 | 0.08 | 2644.22 | 5.40 | 0.00 | 0.00 |
| 827.27 | 4.21 | 0.00 | 0.00 | 4.74 | 0.00 | 0.00 | 5.26 | 0.14 | 4669.78 | 5.45 | 0.08 | 2628.06 | 5.39 | 0.00 | 0.00 |
| 828.85 | 4.21 | 0.00 | 0.00 | 4.74 | 0.00 | 0.00 | 5.25 | 0.14 | 4637.50 | 5.45 | 0.08 | 2612.12 | 5.38 | 0.00 | 0.00 |
| 830.44 | 4.21 | 0.00 | 0.00 | 4.73 | 0.00 | 0.00 | 5.25 | 0.14 | 4605.35 | 5.44 | 0.08 | 2596.24 | 5.38 | 0.00 | 0.00 |
| 832.02 | 4.21 | 0.00 | 0.00 | 4.73 | 0.00 | 0.00 | 5.24 | 0.14 | 4573.61 | 5.44 | 0.08 | 2580.57 | 5.37 | 0.00 | 0.00 |
| 833.61 | 4.21 | 0.00 | 0.00 | 4.73 | 0.00 | 0.00 | 5.24 | 0.14 | 4542.15 | 5.43 | 0.08 | 2565.11 | 5.37 | 0.00 | 0.00 |
| 835.19 | 4.21 | 0.00 | 0.00 | 4.73 | 0.00 | 0.00 | 5.24 | 0.14 | 4510.81 | 5.43 | 0.08 | 2549.71 | 5.36 | 0.00 | 0.00 |
| 836.78 | 4.20 | 0.00 | 0.00 | 4.72 | 0.00 | 0.00 | 5.23 | 0.14 | 4479.89 | 5.42 | 0.08 | 2534.52 | 5.36 | 0.00 | 0.00 |
| 838.36 | 4.20 | 0.00 | 0.00 | 4.72 | 0.00 | 0.00 | 5.23 | 0.14 | 4449.24 | 5.42 | 0.08 | 2519.38 | 5.35 | 0.00 | 0.00 |
| 839.95 | 4.20 | 0.00 | 0.00 | 4.72 | 0.00 | 0.00 | 5.23 | 0.14 | 4418.85 | 5.42 | 0.08 | 2504.45 | 5.34 | 0.00 | 0.00 |
| 841.53 | 4.20 | 0.00 | 0.00 | 4.71 | 0.00 | 0.00 | 5.22 | 0.13 | 4388.73 | 5.41 | 0.08 | 2489.73 | 5.34 | 0.00 | 0.00 |
| 843.12 | 4.20 | 0.00 | 0.00 | 4.71 | 0.00 | 0.00 | 5.22 | 0.13 | 4358.87 | 5.41 | 0.08 | 2475.07 | 5.33 | 0.00 | 0.00 |
| 844.70 | 4.20 | 0.00 | 0.00 | 4.71 | 0.00 | 0.00 | 5.21 | 0.13 | 4329.42 | 5.40 | 0.08 | 2460.61 | 5.33 | 0.00 | 0.00 |
| 846.29 | 4.19 | 0.00 | 0.00 | 4.70 | 0.00 | 0.00 | 5.21 | 0.13 | 4300.08 | 5.40 | 0.08 | 2446.20 | 5.32 | 0.00 | 0.00 |
| 847.87 | 4.19 | 0.00 | 0.00 | 4.70 | 0.00 | 0.00 | 5.21 | 0.13 | 4271.00 | 5.39 | 0.08 | 2432.00 | 5.32 | 0.00 | 0.00 |
| 849.45 | 4.19 | 0.00 | 0.00 | 4.70 | 0.00 | 0.00 | 5.20 | 0.13 | 4242.18 | 5.39 | 0.08 | 2417.85 | 5.31 | 0.00 | 0.00 |
| 851.04 | 4.19 | 0.00 | 0.00 | 4.70 | 0.00 | 0.00 | 5.20 | 0.13 | 4213.61 | 5.39 | 0.08 | 2403.90 | 5.31 | 0.00 | 0.00 |
| 852.62 | 4.19 | 0.00 | 0.00 | 4.69 | 0.00 | 0.00 | 5.20 | 0.13 | 4185.30 | 5.38 | 0.08 | 2390.00 | 5.30 | 0.00 | 0.00 |
| 854.20 | 4.19 | 0.00 | 0.00 | 4.69 | 0.00 | 0.00 | 5.19 | 0.13 | 4157.39 | 5.38 | 0.07 | 2376.30 | 5.30 | 0.00 | 0.00 |
| 855.79 | 4.18 | 0.00 | 0.00 | 4.69 | 0.00 | 0.00 | 5.19 | 0.13 | 4129.57 | 5.37 | 0.07 | 2362.65 | 5.29 | 0.00 | 0.00 |
| 857.37 | 4.18 | 0.00 | 0.00 | 4.69 | 0.00 | 0.00 | 5.19 | 0.13 | 4102.02 | 5.37 | 0.07 | 2349.20 | 5.29 | 0.00 | 0.00 |
| 858.95 | 4.18 | 0.00 | 0.00 | 4.68 | 0.00 | 0.00 | 5.18 | 0.13 | 4074.71 | 5.37 | 0.07 | 2335.95 | 5.28 | 0.00 | 0.00 |
| 860.54 | 4.18 | 0.00 | 0.00 | 4.68 | 0.00 | 0.00 | 5.18 | 0.13 | 4047.64 | 5.36 | 0.07 | 2322.74 | 5.28 | 0.00 | 0.00 |
| 862.12 | 4.18 | 0.00 | 0.00 | 4.68 | 0.00 | 0.00 | 5.18 | 0.13 | 4020.82 | 5.36 | 0.07 | 2309.59 | 5.28 | 0.00 | 0.00 |
| 863.70 | 4.18 | 0.00 | 0.00 | 4.67 | 0.00 | 0.00 | 5.17 | 0.12 | 3994.11 | 5.35 | 0.07 | 2296.63 | 5.27 | 0.00 | 0.00 |
| 865.29 | 4.17 | 0.00 | 0.00 | 4.67 | 0.00 | 0.00 | 5.17 | 0.12 | 3967.77 | 5.35 | 0.07 | 2283.71 | 5.27 | 0.00 | 0.00 |
| 866.87 | 4.17 | 0.00 | 0.00 | 4.67 | 0.00 | 0.00 | 5.17 | 0.12 | 3941.69 | 5.35 | 0.07 | 2270.99 | 5.26 | 0.00 | 0.00 |
| 868.45 | 4.17 | 0.00 | 0.00 | 4.67 | 0.00 | 0.00 | 5.16 | 0.12 | 3915.70 | 5.34 | 0.07 | 2258.31 | 5.26 | 0.00 | 0.00 |
| 870.03 | 4.17 | 0.00 | 0.00 | 4.66 | 0.00 | 0.00 | 5.16 | 0.12 | 3889.94 | 5.34 | 0.07 | 2245.83 | 5.25 | 0.00 | 0.00 |
| 871.61 | 4.17 | 0.00 | 0.00 | 4.66 | 0.00 | 0.00 | 5.16 | 0.12 | 3864.57 | 5.34 | 0.07 | 2233.39 | 5.25 | 0.00 | 0.00 |
| 873.20 | 4.17 | 0.00 | 0.00 | 4.66 | 0.00 | 0.00 | 5.15 | 0.12 | 3839.29 | 5.33 | 0.07 | 2221.14 | 5.25 | 0.00 | 0.00 |
| 874.78 | 4.17 | 0.00 | 0.00 | 4.66 | 0.00 | 0.00 | 5.15 | 0.12 | 3814.25 | 5.33 | 0.07 | 2208.94 | 5.24 | 0.00 | 0.00 |
| 876.36 | 4.16 | 0.00 | 0.00 | 4.66 | 0.00 | 0.00 | 5.15 | 0.12 | 3789.30 | 5.32 | 0.07 | 2196.92 | 5.24 | 0.00 | 0.00 |
| 877.94 | 4.16 | 0.00 | 0.00 | 4.65 | 0.00 | 0.00 | 5.14 | 0.12 | 3764.72 | 5.32 | 0.07 | 2184.81 | 5.23 | 0.00 | 0.00 |
| 879.52 | 4.16 | 0.00 | 0.00 | 4.65 | 0.00 | 0.00 | 5.14 | 0.12 | 3740.38 | 5.32 | 0.07 | 2173.02 | 5.23 | 0.00 | 0.00 |
| 881.10 | 4.16 | 0.00 | 0.00 | 4.65 | 0.00 | 0.00 | 5.14 | 0.12 | 3716.13 | 5.31 | 0.07 | 2161.28 | 5.23 | 0.00 | 0.00 |
| 882.68 | 4.16 | 0.00 | 0.00 | 4.65 | 0.00 | 0.00 | 5.13 | 0.12 | 3692.10 | 5.31 | 0.07 | 2149.57 | 5.22 | 0.00 | 0.00 |
| 884.27 | 4.16 | 0.00 | 0.00 | 4.64 | 0.00 | 0.00 | 5.13 | 0.12 | 3668.31 | 5.31 | 0.07 | 2138.06 | 5.22 | 0.00 | 0.00 |
| 885.85 | 4.16 | 0.00 | 0.00 | 4.64 | 0.00 | 0.00 | 5.13 | 0.12 | 3644.74 | 5.30 | 0.07 | 2126.58 | 5.21 | 0.00 | 0.00 |
| 887.43 | 4.15 | 0.00 | 0.00 | 4.64 | 0.00 | 0.00 | 5.12 | 0.12 | 3621.26 | 5.30 | 0.07 | 2115.15 | 5.21 | 0.00 | 0.00 |
| 889.01 | 4.15 | 0.00 | 0.00 | 4.64 | 0.00 | 0.00 | 5.12 | 0.12 | 3598.00 | 5.30 | 0.07 | 2103.90 | 5.21 | 0.00 | 0.00 |
| 890.59 | 4.15 | 0.00 | 0.00 | 4.63 | 0.00 | 0.00 | 5.12 | 0.11 | 3574.96 | 5.29 | 0.07 | 2092.69 | 5.20 | 0.00 | 0.00 |
| 892.17 | 4.15 | 0.00 | 0.00 | 4.63 | 0.00 | 0.00 | 5.12 | 0.11 | 3552.15 | 5.29 | 0.07 | 2081.66 | 5.20 | 0.00 | 0.00 |
| 893.75 | 4.15 | 0.00 | 0.00 | 4.63 | 0.00 | 0.00 | 5.11 | 0.11 | 3529.56 | 5.29 | 0.07 | 2070.66 | 5.20 | 0.00 | 0.00 |
| 895.33 | 4.15 | 0.00 | 0.00 | 4.63 | 0.00 | 0.00 | 5.11 | 0.11 | 3507.05 | 5.28 | 0.07 | 2059.85 | 5.19 | 0.00 | 0.00 |
| 896.91 | 4.15 | 0.00 | 0.00 | 4.63 | 0.00 | 0.00 | 5.11 | 0.11 | 3484.77 | 5.28 | 0.07 | 2048.94 | 5.19 | 0.00 | 0.00 |
| 898.49 | 4.15 | 0.00 | 0.00 | 4.62 | 0.00 | 0.00 | 5.10 | 0.11 | 3462.70 | 5.28 | 0.07 | 2038.34 | 5.18 | 0.00 | 0.00 |
| 900.06 | 4.14 | 0.00 | 0.00 | 4.62 | 0.00 | 0.00 | 5.10 | 0.11 | 3440.84 | 5.27 | 0.07 | 2027.65 | 5.18 | 0.00 | 0.00 |
| 901.64 | 4.14 | 0.00 | 0.00 | 4.62 | 0.00 | 0.00 | 5.10 | 0.11 | 3419.07 | 5.27 | 0.07 | 2017.13 | 5.18 | 0.00 | 0.00 |
| 903.22 | 4.14 | 0.00 | 0.00 | 4.62 | 0.00 | 0.00 | 5.10 | 0.11 | 3397.51 | 5.27 | 0.07 | 2006.78 | 5.17 | 0.00 | 0.00 |
| 904.80 | 4.14 | 0.00 | 0.00 | 4.62 | 0.00 | 0.00 | 5.09 | 0.11 | 3376.17 | 5.27 | 0.07 | 1996.34 | 5.17 | 0.00 | 0.00 |
| 906.38 | 4.14 | 0.00 | 0.00 | 4.61 | 0.00 | 0.00 | 5.09 | 0.11 | 3355.03 | 5.26 | 0.07 | 1986.07 | 5.17 | 0.00 | 0.00 |
| 907.96 | 4.14 | 0.00 | 0.00 | 4.61 | 0.00 | 0.00 | 5.09 | 0.11 | 3333.98 | 5.26 | 0.07 | 1975.97 | 5.16 | 0.00 | 0.00 |
| 909.54 | 4.14 | 0.00 | 0.00 | 4.61 | 0.00 | 0.00 | 5.09 | 0.11 | 3313.13 | 5.26 | 0.07 | 1965.77 | 5.16 | 0.00 | 0.00 |
| 911.11 | 4.14 | 0.00 | 0.00 | 4.61 | 0.00 | 0.00 | 5.08 | 0.11 | 3292.36 | 5.25 | 0.07 | 1955.89 | 5.16 | 0.00 | 0.00 |
| 912.69 | 4.13 | 0.00 | 0.00 | 4.61 | 0.00 | 0.00 | 5.08 | 0.11 | 3271.80 | 5.25 | 0.06 | 1945.90 | 5.15 | 0.00 | 0.00 |
| 914.27 | 4.13 | 0.00 | 0.00 | 4.60 | 0.00 | 0.00 | 5.08 | 0.11 | 3251.45 | 5.25 | 0.06 | 1936.08 | 5.15 | 0.00 | 0.00 |
| 915.85 | 4.13 | 0.00 | 0.00 | 4.60 | 0.00 | 0.00 | 5.08 | 0.11 | 3231.30 | 5.24 | 0.06 | 1926.29 | 5.15 | 0.00 | 0.00 |
| 917.42 | 4.13 | 0.00 | 0.00 | 4.60 | 0.00 | 0.00 | 5.07 | 0.11 | 3211.23 | 5.24 | 0.06 | 1916.55 | 5.14 | 0.00 | 0.00 |
| 919.00 | 4.13 | 0.00 | 0.00 | 4.60 | 0.00 | 0.00 | 5.07 | 0.11 | 3191.36 | 5.24 | 0.06 | 1906.97 | 5.14 | 0.00 | 0.00 |
| 920.58 | 4.13 | 0.00 | 0.00 | 4.60 | 0.00 | 0.00 | 5.07 | 0.10 | 3171.69 | 5.24 | 0.06 | 1897.42 | 5.14 | 0.00 | 0.00 |
| 922.16 | 4.13 | 0.00 | 0.00 | 4.59 | 0.00 | 0.00 | 5.06 | 0.10 | 3152.10 | 5.23 | 0.06 | 1888.04 | 5.14 | 0.00 | 0.00 |
| 923.73 | 4.13 | 0.00 | 0.00 | 4.59 | 0.00 | 0.00 | 5.06 | 0.10 | 3132.71 | 5.23 | 0.06 | 1878.56 | 5.13 | 0.00 | 0.00 |
| 925.31 | 4.12 | 0.00 | 0.00 | 4.59 | 0.00 | 0.00 | 5.06 | 0.10 | 3113.52 | 5.23 | 0.06 | 1869.39 | 5.13 | 0.00 | 0.00 |
| 926.89 | 4.12 | 0.00 | 0.00 | 4.59 | 0.00 | 0.00 | 5.06 | 0.10 | 3094.39 | 5.23 | 0.06 | 1860.11 | 5.13 | 0.00 | 0.00 |
| 928.46 | 4.12 | 0.00 | 0.00 | 4.59 | 0.00 | 0.00 | 5.05 | 0.10 | 3075.47 | 5.22 | 0.06 | 1850.99 | 5.12 | 0.00 | 0.00 |
| 930.04 | 4.12 | 0.00 | 0.00 | 4.58 | 0.00 | 0.00 | 5.05 | 0.10 | 3056.61 | 5.22 | 0.06 | 1841.91 | 5.12 | 0.00 | 0.00 |
| 931.61 | 4.12 | 0.00 | 0.00 | 4.58 | 0.00 | 0.00 | 5.05 | 0.10 | 3037.95 | 5.22 | 0.06 | 1832.86 | 5.12 | 0.00 | 0.00 |
| 933.19 | 4.12 | 0.00 | 0.00 | 4.58 | 0.00 | 0.00 | 5.05 | 0.10 | 3019.49 | 5.21 | 0.06 | 1823.98 | 5.12 | 0.00 | 0.00 |
| 934.76 | 4.12 | 0.00 | 0.00 | 4.58 | 0.00 | 0.00 | 5.05 | 0.10 | 3001.09 | 5.21 | 0.06 | 1814.99 | 5.11 | 0.00 | 0.00 |
| 936.34 | 4.12 | 0.00 | 0.00 | 4.58 | 0.00 | 0.00 | 5.04 | 0.10 | 2982.89 | 5.21 | 0.06 | 1806.30 | 5.11 | 0.00 | 0.00 |
| 937.91 | 4.12 | 0.00 | 0.00 | 4.58 | 0.00 | 0.00 | 5.04 | 0.10 | 2964.75 | 5.21 | 0.06 | 1797.50 | 5.11 | 0.00 | 0.00 |
| 939.49 | 4.11 | 0.00 | 0.00 | 4.57 | 0.00 | 0.00 | 5.04 | 0.10 | 2946.81 | 5.20 | 0.06 | 1788.87 | 5.10 | 0.00 | 0.00 |
| 941.06 | 4.11 | 0.00 | 0.00 | 4.57 | 0.00 | 0.00 | 5.04 | 0.10 | 2929.06 | 5.20 | 0.06 | 1780.27 | 5.10 | 0.00 | 0.00 |
| 942.64 | 4.11 | 0.00 | 0.00 | 4.57 | 0.00 | 0.00 | 5.03 | 0.10 | 2911.37 | 5.20 | 0.06 | 1771.70 | 5.10 | 0.00 | 0.00 |
| 944.21 | 4.11 | 0.00 | 0.00 | 4.57 | 0.00 | 0.00 | 5.03 | 0.10 | 2893.87 | 5.20 | 0.06 | 1763.29 | 5.10 | 0.00 | 0.00 |
| 945.79 | 4.11 | 0.00 | 0.00 | 4.57 | 0.00 | 0.00 | 5.03 | 0.10 | 2876.43 | 5.19 | 0.06 | 1754.90 | 5.09 | 0.00 | 0.00 |
| 947.36 | 4.11 | 0.00 | 0.00 | 4.57 | 0.00 | 0.00 | 5.03 | 0.10 | 2859.19 | 5.19 | 0.06 | 1746.55 | 5.09 | 0.00 | 0.00 |
| 948.93 | 4.11 | 0.00 | 0.00 | 4.56 | 0.00 | 0.00 | 5.02 | 0.10 | 2842.13 | 5.19 | 0.06 | 1738.36 | 5.09 | 0.00 | 0.00 |
| 950.51 | 4.11 | 0.00 | 0.00 | 4.56 | 0.00 | 0.00 | 5.02 | 0.10 | 2825.13 | 5.19 | 0.06 | 1730.06 | 5.09 | 0.00 | 0.00 |
| 952.08 | 4.11 | 0.00 | 0.00 | 4.56 | 0.00 | 0.00 | 5.02 | 0.10 | 2808.19 | 5.18 | 0.06 | 1721.92 | 5.08 | 0.00 | 0.00 |
| 953.65 | 4.11 | 0.00 | 0.00 | 4.56 | 0.00 | 0.00 | 5.02 | 0.09 | 2791.56 | 5.18 | 0.06 | 1713.94 | 5.08 | 0.00 | 0.00 |
| 955.23 | 4.10 | 0.00 | 0.00 | 4.56 | 0.00 | 0.00 | 5.02 | 0.09 | 2774.86 | 5.18 | 0.06 | 1705.86 | 5.08 | 0.00 | 0.00 |
| 956.80 | 4.10 | 0.00 | 0.00 | 4.56 | 0.00 | 0.00 | 5.01 | 0.09 | 2758.48 | 5.18 | 0.06 | 1697.93 | 5.08 | 0.00 | 0.00 |
| 958.37 | 4.10 | 0.00 | 0.00 | 4.55 | 0.00 | 0.00 | 5.01 | 0.09 | 2742.02 | 5.17 | 0.06 | 1690.03 | 5.07 | 0.00 | 0.00 |
| 959.94 | 4.10 | 0.00 | 0.00 | 4.55 | 0.00 | 0.00 | 5.01 | 0.09 | 2725.88 | 5.17 | 0.06 | 1682.16 | 5.07 | 0.00 | 0.00 |
| 961.52 | 4.10 | 0.00 | 0.00 | 4.55 | 0.00 | 0.00 | 5.01 | 0.09 | 2709.66 | 5.17 | 0.06 | 1674.44 | 5.07 | 0.00 | 0.00 |
| 963.09 | 4.10 | 0.00 | 0.00 | 4.55 | 0.00 | 0.00 | 5.00 | 0.09 | 2693.76 | 5.17 | 0.06 | 1666.62 | 5.07 | 0.00 | 0.00 |
| 964.66 | 4.10 | 0.00 | 0.00 | 4.55 | 0.00 | 0.00 | 5.00 | 0.09 | 2677.90 | 5.17 | 0.06 | 1658.95 | 5.06 | 0.00 | 0.00 |
| 966.23 | 4.10 | 0.00 | 0.00 | 4.55 | 0.00 | 0.00 | 5.00 | 0.09 | 2662.10 | 5.16 | 0.06 | 1651.44 | 5.06 | 0.00 | 0.00 |
| 967.80 | 4.10 | 0.00 | 0.00 | 4.55 | 0.00 | 0.00 | 5.00 | 0.09 | 2646.48 | 5.16 | 0.06 | 1643.83 | 5.06 | 0.00 | 0.00 |
| 969.37 | 4.10 | 0.00 | 0.00 | 4.54 | 0.00 | 0.00 | 5.00 | 0.09 | 2630.92 | 5.16 | 0.06 | 1636.37 | 5.06 | 0.00 | 0.00 |
| 970.95 | 4.09 | 0.00 | 0.00 | 4.54 | 0.00 | 0.00 | 4.99 | 0.09 | 2615.53 | 5.16 | 0.06 | 1628.93 | 5.05 | 0.00 | 0.00 |
| 972.52 | 4.09 | 0.00 | 0.00 | 4.54 | 0.00 | 0.00 | 4.99 | 0.09 | 2600.32 | 5.15 | 0.06 | 1621.52 | 5.05 | 0.00 | 0.00 |
| 974.09 | 4.09 | 0.00 | 0.00 | 4.54 | 0.00 | 0.00 | 4.99 | 0.09 | 2585.04 | 5.15 | 0.06 | 1614.26 | 5.05 | 0.00 | 0.00 |
| 975.66 | 4.09 | 0.00 | 0.00 | 4.54 | 0.00 | 0.00 | 4.99 | 0.09 | 2570.06 | 5.15 | 0.06 | 1606.90 | 5.05 | 0.00 | 0.00 |
| 977.23 | 4.09 | 0.00 | 0.00 | 4.54 | 0.00 | 0.00 | 4.99 | 0.09 | 2555.12 | 5.15 | 0.06 | 1599.69 | 5.04 | 0.00 | 0.00 |
| 978.80 | 4.09 | 0.00 | 0.00 | 4.54 | 0.00 | 0.00 | 4.98 | 0.09 | 2540.24 | 5.15 | 0.06 | 1592.63 | 5.04 | 0.00 | 0.00 |
| 980.37 | 4.09 | 0.00 | 0.00 | 4.53 | 0.00 | 0.00 | 4.98 | 0.09 | 2525.54 | 5.14 | 0.06 | 1585.46 | 5.04 | 0.00 | 0.00 |
| 981.94 | 4.09 | 0.00 | 0.00 | 4.53 | 0.00 | 0.00 | 4.98 | 0.09 | 2510.88 | 5.14 | 0.06 | 1578.32 | 5.04 | 0.00 | 0.00 |
| 983.51 | 4.09 | 0.00 | 0.00 | 4.53 | 0.00 | 0.00 | 4.98 | 0.09 | 2496.39 | 5.14 | 0.06 | 1571.33 | 5.04 | 0.00 | 0.00 |
| 985.07 | 4.09 | 0.00 | 0.00 | 4.53 | 0.00 | 0.00 | 4.98 | 0.09 | 2481.96 | 5.14 | 0.06 | 1564.36 | 5.03 | 0.00 | 0.00 |
| 986.64 | 4.09 | 0.00 | 0.00 | 4.53 | 0.00 | 0.00 | 4.97 | 0.09 | 2467.69 | 5.14 | 0.06 | 1557.54 | 5.03 | 0.00 | 0.00 |
| 988.21 | 4.08 | 0.00 | 0.00 | 4.53 | 0.00 | 0.00 | 4.97 | 0.09 | 2453.48 | 5.13 | 0.06 | 1550.62 | 5.03 | 0.00 | 0.00 |
| 989.78 | 4.08 | 0.00 | 0.00 | 4.53 | 0.00 | 0.00 | 4.97 | 0.09 | 2439.43 | 5.13 | 0.06 | 1543.85 | 5.03 | 0.00 | 0.00 |
| 991.35 | 4.08 | 0.00 | 0.00 | 4.52 | 0.00 | 0.00 | 4.97 | 0.09 | 2425.43 | 5.13 | 0.05 | 1537.10 | 5.03 | 0.00 | 0.00 |
| 992.92 | 4.08 | 0.00 | 0.00 | 4.52 | 0.00 | 0.00 | 4.97 | 0.08 | 2411.48 | 5.13 | 0.05 | 1530.37 | 5.02 | 0.00 | 0.00 |
| 994.48 | 4.08 | 0.00 | 0.00 | 4.52 | 0.00 | 0.00 | 4.96 | 0.08 | 2397.69 | 5.13 | 0.05 | 1523.66 | 5.02 | 0.00 | 0.00 |
| 996.05 | 4.08 | 0.00 | 0.00 | 4.52 | 0.00 | 0.00 | 4.96 | 0.08 | 2383.95 | 5.12 | 0.05 | 1517.10 | 5.02 | 0.00 | 0.00 |
| 997.62 | 4.08 | 0.00 | 0.00 | 4.52 | 0.00 | 0.00 | 4.96 | 0.08 | 2370.38 | 5.12 | 0.05 | 1510.43 | 5.02 | 0.00 | 0.00 |
| 999.19 | 4.08 | 0.00 | 0.00 | 4.52 | 0.00 | 0.00 | 4.96 | 0.08 | 2356.98 | 5.12 | 0.05 | 1503.91 | 5.02 | 0.00 | 0.00 |
| 1013.00 | 4.07 | 0.00 | 0.00 | 4.51 | 0.00 | 0.00 | 4.94 | 0.08 | 2241.97 | 5.10 | 0.05 | 1448.30 | 5.00 | 0.00 | 0.00 |
| 1016.40 | 4.07 | 0.00 | 0.00 | 4.50 | 0.00 | 0.00 | 4.94 | 0.08 | 2214.68 | 5.10 | 0.05 | 1435.04 | 4.99 | 0.00 | 0.00 |
| 1019.81 | 4.07 | 0.00 | 0.00 | 4.50 | 0.00 | 0.00 | 4.93 | 0.08 | 2187.82 | 5.09 | 0.05 | 1421.99 | 4.99 | 0.00 | 0.00 |
| 1023.22 | 4.07 | 0.00 | 0.00 | 4.50 | 0.00 | 0.00 | 4.93 | 0.08 | 2161.38 | 5.09 | 0.05 | 1409.15 | 4.99 | 0.00 | 0.00 |
| 1026.62 | 4.06 | 0.00 | 0.00 | 4.50 | 0.00 | 0.00 | 4.93 | 0.08 | 2135.36 | 5.09 | 0.05 | 1396.52 | 4.98 | 0.00 | 0.00 |
| 1030.03 | 4.06 | 0.00 | 0.00 | 4.49 | 0.00 | 0.00 | 4.92 | 0.08 | 2109.63 | 5.08 | 0.05 | 1384.09 | 4.98 | 0.00 | 0.00 |
| 1033.43 | 4.06 | 0.00 | 0.00 | 4.49 | 0.00 | 0.00 | 4.92 | 0.08 | 2084.31 | 5.08 | 0.05 | 1371.75 | 4.98 | 0.00 | 0.00 |
| 1036.84 | 4.06 | 0.00 | 0.00 | 4.49 | 0.00 | 0.00 | 4.92 | 0.08 | 2059.41 | 5.08 | 0.05 | 1359.61 | 4.97 | 0.00 | 0.00 |
| 1040.25 | 4.06 | 0.00 | 0.00 | 4.49 | 0.00 | 0.00 | 4.91 | 0.07 | 2034.78 | 5.07 | 0.05 | 1347.66 | 4.97 | 0.00 | 0.00 |
| 1043.66 | 4.06 | 0.00 | 0.00 | 4.49 | 0.00 | 0.00 | 4.91 | 0.07 | 2010.56 | 5.07 | 0.05 | 1335.92 | 4.96 | 0.00 | 0.00 |
| 1047.06 | 4.05 | 0.00 | 0.00 | 4.48 | 0.00 | 0.00 | 4.91 | 0.07 | 1986.73 | 5.07 | 0.05 | 1324.25 | 4.96 | 0.00 | 0.00 |
| 1050.47 | 4.05 | 0.00 | 0.00 | 4.48 | 0.00 | 0.00 | 4.90 | 0.07 | 1963.18 | 5.06 | 0.05 | 1312.77 | 4.96 | 0.00 | 0.00 |
| 1053.88 | 4.05 | 0.00 | 0.00 | 4.48 | 0.00 | 0.00 | 4.90 | 0.07 | 1940.02 | 5.06 | 0.05 | 1301.49 | 4.95 | 0.00 | 0.00 |
| 1057.29 | 4.05 | 0.00 | 0.00 | 4.48 | 0.00 | 0.00 | 4.90 | 0.07 | 1917.12 | 5.05 | 0.05 | 1290.28 | 4.95 | 0.00 | 0.00 |
| 1060.70 | 4.05 | 0.00 | 0.00 | 4.47 | 0.00 | 0.00 | 4.89 | 0.07 | 1894.61 | 5.05 | 0.05 | 1279.27 | 4.95 | 0.00 | 0.00 |
| 1064.11 | 4.05 | 0.00 | 0.00 | 4.47 | 0.00 | 0.00 | 4.89 | 0.07 | 1872.36 | 5.05 | 0.05 | 1268.43 | 4.94 | 0.00 | 0.00 |
| 1067.52 | 4.04 | 0.00 | 0.00 | 4.47 | 0.00 | 0.00 | 4.89 | 0.07 | 1850.49 | 5.04 | 0.05 | 1257.67 | 4.94 | 0.00 | 0.00 |
| 1070.93 | 4.04 | 0.00 | 0.00 | 4.47 | 0.00 | 0.00 | 4.88 | 0.07 | 1828.87 | 5.04 | 0.05 | 1247.10 | 4.94 | 0.00 | 0.00 |
| 1074.34 | 4.04 | 0.00 | 0.00 | 4.47 | 0.00 | 0.00 | 4.88 | 0.07 | 1807.51 | 5.04 | 0.05 | 1236.71 | 4.94 | 0.00 | 0.00 |
| 1077.75 | 4.04 | 0.00 | 0.00 | 4.46 | 0.00 | 0.00 | 4.88 | 0.07 | 1786.51 | 5.04 | 0.05 | 1226.38 | 4.93 | 0.00 | 0.00 |
| 1081.16 | 4.04 | 0.00 | 0.00 | 4.46 | 0.00 | 0.00 | 4.87 | 0.07 | 1765.88 | 5.03 | 0.05 | 1216.23 | 4.93 | 0.00 | 0.00 |
| 1084.58 | 4.04 | 0.00 | 0.00 | 4.46 | 0.00 | 0.00 | 4.87 | 0.07 | 1745.38 | 5.03 | 0.05 | 1206.15 | 4.93 | 0.00 | 0.00 |
| 1087.99 | 4.04 | 0.00 | 0.00 | 4.46 | 0.00 | 0.00 | 4.87 | 0.07 | 1725.35 | 5.03 | 0.05 | 1196.24 | 4.92 | 0.00 | 0.00 |
| 1091.40 | 4.03 | 0.00 | 0.00 | 4.46 | 0.00 | 0.00 | 4.86 | 0.07 | 1705.45 | 5.02 | 0.05 | 1186.52 | 4.92 | 0.00 | 0.00 |
| 1094.81 | 4.03 | 0.00 | 0.00 | 4.45 | 0.00 | 0.00 | 4.86 | 0.06 | 1685.90 | 5.02 | 0.05 | 1176.85 | 4.92 | 0.00 | 0.00 |
| 1098.23 | 4.03 | 0.00 | 0.00 | 4.45 | 0.00 | 0.00 | 4.86 | 0.06 | 1666.59 | 5.02 | 0.05 | 1167.36 | 4.92 | 0.00 | 0.00 |
| 1101.64 | 4.03 | 0.00 | 0.00 | 4.45 | 0.00 | 0.00 | 4.85 | 0.06 | 1647.51 | 5.01 | 0.05 | 1157.92 | 4.91 | 0.00 | 0.00 |
| 1105.05 | 4.03 | 0.00 | 0.00 | 4.45 | 0.00 | 0.00 | 4.85 | 0.06 | 1628.66 | 5.01 | 0.05 | 1148.66 | 4.91 | 0.00 | 0.00 |
| 1108.47 | 4.03 | 0.00 | 0.00 | 4.45 | 0.00 | 0.00 | 4.85 | 0.06 | 1610.15 | 5.01 | 0.04 | 1139.45 | 4.91 | 0.00 | 0.00 |
| 1111.88 | 4.03 | 0.00 | 0.00 | 4.45 | 0.00 | 0.00 | 4.84 | 0.06 | 1591.87 | 5.01 | 0.04 | 1130.41 | 4.90 | 0.00 | 0.00 |
| 1115.30 | 4.02 | 0.00 | 0.00 | 4.44 | 0.00 | 0.00 | 4.84 | 0.06 | 1573.81 | 5.00 | 0.04 | 1121.43 | 4.90 | 0.00 | 0.00 |
| 1118.71 | 4.02 | 0.00 | 0.00 | 4.44 | 0.00 | 0.00 | 4.84 | 0.06 | 1555.98 | 5.00 | 0.04 | 1112.62 | 4.90 | 0.00 | 0.00 |
| 1122.13 | 4.02 | 0.00 | 0.00 | 4.44 | 0.00 | 0.00 | 4.84 | 0.06 | 1538.48 | 5.00 | 0.04 | 1103.86 | 4.90 | 0.00 | 0.00 |
| 1125.54 | 4.02 | 0.00 | 0.00 | 4.44 | 0.00 | 0.00 | 4.83 | 0.06 | 1521.08 | 4.99 | 0.04 | 1095.26 | 4.89 | 0.00 | 0.00 |
| 1128.96 | 4.02 | 0.00 | 0.00 | 4.44 | 0.00 | 0.00 | 4.83 | 0.06 | 1504.01 | 4.99 | 0.04 | 1086.82 | 4.89 | 0.00 | 0.00 |
| 1132.38 | 4.02 | 0.00 | 0.00 | 4.44 | 0.00 | 0.00 | 4.83 | 0.06 | 1487.15 | 4.99 | 0.04 | 1078.33 | 4.89 | 0.00 | 0.00 |
| 1135.79 | 4.02 | 0.00 | 0.00 | 4.43 | 0.00 | 0.00 | 4.83 | 0.06 | 1470.51 | 4.99 | 0.04 | 1070.11 | 4.89 | 0.00 | 0.00 |
| 1139.21 | 4.02 | 0.00 | 0.00 | 4.43 | 0.00 | 0.00 | 4.82 | 0.06 | 1454.08 | 4.98 | 0.04 | 1061.82 | 4.89 | 0.00 | 0.00 |
| 1142.63 | 4.01 | 0.00 | 0.00 | 4.43 | 0.00 | 0.00 | 4.82 | 0.06 | 1437.85 | 4.98 | 0.04 | 1053.81 | 4.88 | 0.00 | 0.00 |
| 1146.05 | 4.01 | 0.00 | 0.00 | 4.43 | 0.00 | 0.00 | 4.82 | 0.06 | 1421.94 | 4.98 | 0.04 | 1045.73 | 4.88 | 0.00 | 0.00 |
| 1149.47 | 4.01 | 0.00 | 0.00 | 4.43 | 0.00 | 0.00 | 4.81 | 0.06 | 1406.12 | 4.98 | 0.04 | 1037.81 | 4.88 | 0.00 | 0.00 |
| 1152.88 | 4.01 | 0.00 | 0.00 | 4.43 | 0.00 | 0.00 | 4.81 | 0.06 | 1390.50 | 4.97 | 0.04 | 1030.04 | 4.88 | 0.00 | 0.00 |
| 1156.30 | 4.01 | 0.00 | 0.00 | 4.43 | 0.00 | 0.00 | 4.81 | 0.06 | 1375.20 | 4.97 | 0.04 | 1022.33 | 4.87 | 0.00 | 0.00 |
| 1159.72 | 4.01 | 0.00 | 0.00 | 4.42 | 0.00 | 0.00 | 4.81 | 0.06 | 1359.98 | 4.97 | 0.04 | 1014.65 | 4.87 | 0.00 | 0.00 |
| 1163.14 | 4.01 | 0.00 | 0.00 | 4.42 | 0.00 | 0.00 | 4.80 | 0.05 | 1344.97 | 4.97 | 0.04 | 1007.13 | 4.87 | 0.00 | 0.00 |
| 1166.56 | 4.01 | 0.00 | 0.00 | 4.42 | 0.00 | 0.00 | 4.80 | 0.05 | 1330.25 | 4.96 | 0.04 | 999.65 | 4.87 | 0.00 | 0.00 |
| 1169.98 | 4.00 | 0.00 | 0.00 | 4.42 | 0.00 | 0.00 | 4.80 | 0.05 | 1315.62 | 4.96 | 0.04 | 992.22 | 4.87 | 0.00 | 0.00 |
| 1173.41 | 4.00 | 0.00 | 0.00 | 4.42 | 0.00 | 0.00 | 4.80 | 0.05 | 1301.18 | 4.96 | 0.04 | 984.94 | 4.86 | 0.00 | 0.00 |
| 1176.83 | 4.00 | 0.00 | 0.00 | 4.42 | 0.00 | 0.00 | 4.79 | 0.05 | 1287.04 | 4.96 | 0.04 | 977.80 | 4.86 | 0.00 | 0.00 |
| 1180.25 | 4.00 | 0.00 | 0.00 | 4.42 | 0.00 | 0.00 | 4.79 | 0.05 | 1272.98 | 4.95 | 0.04 | 970.60 | 4.86 | 0.00 | 0.00 |
| 1183.67 | 4.00 | 0.00 | 0.00 | 4.42 | 0.00 | 0.00 | 4.79 | 0.05 | 1259.11 | 4.95 | 0.04 | 963.55 | 4.86 | 0.00 | 0.00 |
| 1187.09 | 4.00 | 0.00 | 0.00 | 4.41 | 0.00 | 0.00 | 4.79 | 0.05 | 1245.42 | 4.95 | 0.04 | 956.64 | 4.86 | 0.00 | 0.00 |
| 1190.51 | 4.00 | 0.00 | 0.00 | 4.41 | 0.00 | 0.00 | 4.78 | 0.05 | 1231.92 | 4.95 | 0.04 | 949.78 | 4.85 | 0.00 | 0.00 |
| 1193.94 | 4.00 | 0.00 | 0.00 | 4.41 | 0.00 | 0.00 | 4.78 | 0.05 | 1218.60 | 4.95 | 0.04 | 942.95 | 4.85 | 0.00 | 0.00 |
| 1197.36 | 4.00 | 0.00 | 0.00 | 4.41 | 0.00 | 0.00 | 4.78 | 0.05 | 1205.36 | 4.94 | 0.04 | 936.26 | 4.85 | 0.00 | 0.00 |
| 1200.78 | 3.99 | 0.00 | 0.00 | 4.41 | 0.00 | 0.00 | 4.78 | 0.05 | 1192.40 | 4.94 | 0.04 | 929.51 | 4.85 | 0.00 | 0.00 |
| 1204.21 | 3.99 | 0.00 | 0.00 | 4.41 | 0.00 | 0.00 | 4.77 | 0.05 | 1179.51 | 4.94 | 0.04 | 923.01 | 4.85 | 0.00 | 0.00 |
| 1207.63 | 3.99 | 0.00 | 0.00 | 4.41 | 0.00 | 0.00 | 4.77 | 0.05 | 1166.90 | 4.94 | 0.04 | 916.44 | 4.85 | 0.00 | 0.00 |
| 1211.06 | 3.99 | 0.00 | 0.00 | 4.41 | 0.00 | 0.00 | 4.77 | 0.05 | 1154.37 | 4.94 | 0.04 | 910.01 | 4.84 | 0.00 | 0.00 |
| 1214.48 | 3.99 | 0.00 | 0.00 | 4.40 | 0.00 | 0.00 | 4.77 | 0.05 | 1142.01 | 4.93 | 0.04 | 903.72 | 4.84 | 0.00 | 0.00 |
| 1217.91 | 3.99 | 0.00 | 0.00 | 4.40 | 0.00 | 0.00 | 4.76 | 0.05 | 1129.72 | 4.93 | 0.04 | 897.36 | 4.84 | 0.00 | 0.00 |
| 1221.33 | 3.99 | 0.00 | 0.00 | 4.40 | 0.00 | 0.00 | 4.76 | 0.05 | 1117.70 | 4.93 | 0.04 | 891.13 | 4.84 | 0.00 | 0.00 |
| 1224.76 | 3.99 | 0.00 | 0.00 | 4.40 | 0.00 | 0.00 | 4.76 | 0.05 | 1105.75 | 4.93 | 0.04 | 884.95 | 4.84 | 0.00 | 0.00 |
| 1228.19 | 3.99 | 0.00 | 0.00 | 4.40 | 0.00 | 0.00 | 4.76 | 0.05 | 1094.07 | 4.93 | 0.04 | 878.90 | 4.84 | 0.00 | 0.00 |
| 1231.61 | 3.98 | 0.00 | 0.00 | 4.40 | 0.00 | 0.00 | 4.76 | 0.05 | 1082.35 | 4.92 | 0.04 | 872.88 | 4.83 | 0.00 | 0.00 |
| 1235.04 | 3.98 | 0.00 | 0.00 | 4.40 | 0.00 | 0.00 | 4.75 | 0.05 | 1070.90 | 4.92 | 0.04 | 866.90 | 4.83 | 0.00 | 0.00 |
| 1238.47 | 3.98 | 0.00 | 0.00 | 4.40 | 0.00 | 0.00 | 4.75 | 0.05 | 1059.62 | 4.92 | 0.04 | 860.95 | 4.83 | 0.00 | 0.00 |
| 1241.90 | 3.98 | 0.00 | 0.00 | 4.40 | 0.00 | 0.00 | 4.75 | 0.05 | 1048.40 | 4.92 | 0.04 | 855.13 | 4.83 | 0.00 | 0.00 |
| 1245.32 | 3.98 | 0.00 | 0.00 | 4.40 | 0.00 | 0.00 | 4.75 | 0.04 | 1037.34 | 4.92 | 0.04 | 849.35 | 4.83 | 0.00 | 0.00 |
| 1248.75 | 3.98 | 0.00 | 0.00 | 4.39 | 0.00 | 0.00 | 4.74 | 0.04 | 1026.44 | 4.91 | 0.04 | 843.69 | 4.83 | 0.00 | 0.00 |
| 1252.18 | 3.98 | 0.00 | 0.00 | 4.39 | 0.00 | 0.00 | 4.74 | 0.04 | 1015.60 | 4.91 | 0.04 | 837.97 | 4.83 | 0.00 | 0.00 |
| 1255.61 | 3.98 | 0.00 | 0.00 | 4.39 | 0.00 | 0.00 | 4.74 | 0.04 | 1004.92 | 4.91 | 0.04 | 832.38 | 4.82 | 0.00 | 0.00 |
| 1259.04 | 3.98 | 0.00 | 0.00 | 4.39 | 0.00 | 0.00 | 4.74 | 0.04 | 994.40 | 4.91 | 0.04 | 826.92 | 4.82 | 0.00 | 0.00 |
| 1262.47 | 3.98 | 0.00 | 0.00 | 4.39 | 0.00 | 0.00 | 4.74 | 0.04 | 984.03 | 4.91 | 0.04 | 821.39 | 4.82 | 0.00 | 0.00 |
| 1265.90 | 3.97 | 0.00 | 0.00 | 4.39 | 0.00 | 0.00 | 4.73 | 0.04 | 973.72 | 4.90 | 0.04 | 815.98 | 4.82 | 0.00 | 0.00 |
| 1269.33 | 3.97 | 0.00 | 0.00 | 4.39 | 0.00 | 0.00 | 4.73 | 0.04 | 963.57 | 4.90 | 0.04 | 810.61 | 4.82 | 0.00 | 0.00 |
| 1272.76 | 3.97 | 0.00 | 0.00 | 4.39 | 0.00 | 0.00 | 4.73 | 0.04 | 953.56 | 4.90 | 0.04 | 805.27 | 4.82 | 0.00 | 0.00 |
| 1276.19 | 3.97 | 0.00 | 0.00 | 4.39 | 0.00 | 0.00 | 4.73 | 0.04 | 943.61 | 4.90 | 0.04 | 800.05 | 4.82 | 0.00 | 0.00 |
| 1279.63 | 3.97 | 0.00 | 0.00 | 4.39 | 0.00 | 0.00 | 4.73 | 0.04 | 933.82 | 4.90 | 0.04 | 794.86 | 4.81 | 0.00 | 0.00 |
| 1283.06 | 3.97 | 0.00 | 0.00 | 4.39 | 0.00 | 0.00 | 4.72 | 0.04 | 924.17 | 4.90 | 0.04 | 789.70 | 4.81 | 0.00 | 0.00 |
| 1286.49 | 3.97 | 0.00 | 0.00 | 4.38 | 0.00 | 0.00 | 4.72 | 0.04 | 914.57 | 4.89 | 0.04 | 784.56 | 4.81 | 0.00 | 0.00 |
| 1289.92 | 3.97 | 0.00 | 0.00 | 4.38 | 0.00 | 0.00 | 4.72 | 0.04 | 905.12 | 4.89 | 0.04 | 779.55 | 4.81 | 0.00 | 0.00 |
| 1293.36 | 3.97 | 0.00 | 0.00 | 4.38 | 0.00 | 0.00 | 4.72 | 0.04 | 895.82 | 4.89 | 0.04 | 774.47 | 4.81 | 0.00 | 0.00 |
| 1296.79 | 3.97 | 0.00 | 0.00 | 4.38 | 0.00 | 0.00 | 4.72 | 0.04 | 886.57 | 4.89 | 0.04 | 769.51 | 4.81 | 0.00 | 0.00 |
| 1300.22 | 3.97 | 0.00 | 0.00 | 4.38 | 0.00 | 0.00 | 4.71 | 0.04 | 877.47 | 4.89 | 0.03 | 764.68 | 4.81 | 0.00 | 0.00 |
| 1303.66 | 3.96 | 0.00 | 0.00 | 4.38 | 0.00 | 0.00 | 4.71 | 0.04 | 868.41 | 4.89 | 0.03 | 759.77 | 4.81 | 0.00 | 0.00 |
| 1307.09 | 3.96 | 0.00 | 0.00 | 4.38 | 0.00 | 0.00 | 4.71 | 0.04 | 859.59 | 4.88 | 0.03 | 754.99 | 4.81 | 0.00 | 0.00 |
| 1310.53 | 3.96 | 0.00 | 0.00 | 4.38 | 0.00 | 0.00 | 4.71 | 0.04 | 850.72 | 4.88 | 0.03 | 750.23 | 4.80 | 0.00 | 0.00 |
| 1313.96 | 3.96 | 0.00 | 0.00 | 4.38 | 0.00 | 0.00 | 4.71 | 0.04 | 842.09 | 4.88 | 0.03 | 745.49 | 4.80 | 0.00 | 0.00 |
| 1317.40 | 3.96 | 0.00 | 0.00 | 4.38 | 0.00 | 0.00 | 4.70 | 0.04 | 833.50 | 4.88 | 0.03 | 740.88 | 4.80 | 0.00 | 0.00 |
| 1320.83 | 3.96 | 0.00 | 0.00 | 4.38 | 0.00 | 0.00 | 4.70 | 0.04 | 824.96 | 4.88 | 0.03 | 736.19 | 4.80 | 0.00 | 0.00 |
| 1324.27 | 3.96 | 0.00 | 0.00 | 4.38 | 0.00 | 0.00 | 4.70 | 0.04 | 816.55 | 4.88 | 0.03 | 731.62 | 4.80 | 0.00 | 0.00 |
| 1327.71 | 3.96 | 0.00 | 0.00 | 4.38 | 0.00 | 0.00 | 4.70 | 0.04 | 808.29 | 4.87 | 0.03 | 727.17 | 4.80 | 0.00 | 0.00 |
| 1331.14 | 3.96 | 0.00 | 0.00 | 4.37 | 0.00 | 0.00 | 4.70 | 0.04 | 800.06 | 4.87 | 0.03 | 722.65 | 4.80 | 0.00 | 0.00 |
| 1334.58 | 3.96 | 0.00 | 0.00 | 4.37 | 0.00 | 0.00 | 4.69 | 0.04 | 791.98 | 4.87 | 0.03 | 718.16 | 4.80 | 0.00 | 0.00 |
| 1338.02 | 3.96 | 0.00 | 0.00 | 4.37 | 0.00 | 0.00 | 4.69 | 0.04 | 783.93 | 4.87 | 0.03 | 713.78 | 4.80 | 0.00 | 0.00 |
| 1341.46 | 3.95 | 0.00 | 0.00 | 4.37 | 0.00 | 0.00 | 4.69 | 0.04 | 776.02 | 4.87 | 0.03 | 709.42 | 4.79 | 0.00 | 0.00 |
| 1344.90 | 3.95 | 0.00 | 0.00 | 4.37 | 0.00 | 0.00 | 4.69 | 0.04 | 768.24 | 4.87 | 0.03 | 705.08 | 4.79 | 0.00 | 0.00 |
| 1348.33 | 3.95 | 0.00 | 0.00 | 4.37 | 0.00 | 0.00 | 4.69 | 0.04 | 760.51 | 4.87 | 0.03 | 700.86 | 4.79 | 0.00 | 0.00 |
| 1351.77 | 3.95 | 0.00 | 0.00 | 4.37 | 0.00 | 0.00 | 4.69 | 0.04 | 752.81 | 4.86 | 0.03 | 696.56 | 4.79 | 0.00 | 0.00 |
| 1355.21 | 3.95 | 0.00 | 0.00 | 4.37 | 0.00 | 0.00 | 4.68 | 0.03 | 745.24 | 4.86 | 0.03 | 692.39 | 4.79 | 0.00 | 0.00 |
| 1358.65 | 3.95 | 0.00 | 0.00 | 4.37 | 0.00 | 0.00 | 4.68 | 0.03 | 737.80 | 4.86 | 0.03 | 688.23 | 4.79 | 0.00 | 0.00 |
| 1362.09 | 3.95 | 0.00 | 0.00 | 4.37 | 0.00 | 0.00 | 4.68 | 0.03 | 730.41 | 4.86 | 0.03 | 684.09 | 4.79 | 0.00 | 0.00 |
| 1365.53 | 3.95 | 0.00 | 0.00 | 4.37 | 0.00 | 0.00 | 4.68 | 0.03 | 723.14 | 4.86 | 0.03 | 680.07 | 4.79 | 0.00 | 0.00 |
| 1368.97 | 3.95 | 0.00 | 0.00 | 4.37 | 0.00 | 0.00 | 4.68 | 0.03 | 715.90 | 4.86 | 0.03 | 675.97 | 4.79 | 0.00 | 0.00 |
| 1372.42 | 3.95 | 0.00 | 0.00 | 4.37 | 0.00 | 0.00 | 4.67 | 0.03 | 708.70 | 4.86 | 0.03 | 671.99 | 4.79 | 0.00 | 0.00 |
| 1375.86 | 3.95 | 0.00 | 0.00 | 4.37 | 0.00 | 0.00 | 4.67 | 0.03 | 701.73 | 4.85 | 0.03 | 668.02 | 4.79 | 0.00 | 0.00 |
| 1379.30 | 3.94 | 0.00 | 0.00 | 4.37 | 0.00 | 0.00 | 4.67 | 0.03 | 694.69 | 4.85 | 0.03 | 664.08 | 4.78 | 0.00 | 0.00 |
| 1382.74 | 3.94 | 0.00 | 0.00 | 4.37 | 0.00 | 0.00 | 4.67 | 0.03 | 687.78 | 4.85 | 0.03 | 660.15 | 4.78 | 0.00 | 0.00 |
| 1386.18 | 3.94 | 0.00 | 0.00 | 4.36 | 0.00 | 0.00 | 4.67 | 0.03 | 681.00 | 4.85 | 0.03 | 656.34 | 4.78 | 0.00 | 0.00 |
| 1389.63 | 3.94 | 0.00 | 0.00 | 4.36 | 0.00 | 0.00 | 4.67 | 0.03 | 674.24 | 4.85 | 0.03 | 652.45 | 4.78 | 0.00 | 0.00 |
| 1393.07 | 3.94 | 0.00 | 0.00 | 4.36 | 0.00 | 0.00 | 4.66 | 0.03 | 667.62 | 4.85 | 0.03 | 648.67 | 4.78 | 0.00 | 0.00 |
| 1396.51 | 3.94 | 0.00 | 0.00 | 4.36 | 0.00 | 0.00 | 4.66 | 0.03 | 661.02 | 4.85 | 0.03 | 644.91 | 4.78 | 0.00 | 0.00 |
| 1399.96 | 3.94 | 0.00 | 0.00 | 4.36 | 0.00 | 0.00 | 4.66 | 0.03 | 654.46 | 4.84 | 0.03 | 641.17 | 4.78 | 0.00 | 0.00 |
| 1403.40 | 3.94 | 0.00 | 0.00 | 4.36 | 0.00 | 0.00 | 4.66 | 0.03 | 648.02 | 4.84 | 0.03 | 637.54 | 4.78 | 0.00 | 0.00 |
| 1406.85 | 3.94 | 0.00 | 0.00 | 4.36 | 0.00 | 0.00 | 4.66 | 0.03 | 641.70 | 4.84 | 0.03 | 633.84 | 4.78 | 0.00 | 0.00 |
| 1410.29 | 3.94 | 0.00 | 0.00 | 4.36 | 0.00 | 0.00 | 4.66 | 0.03 | 635.41 | 4.84 | 0.03 | 630.24 | 4.78 | 0.00 | 0.00 |
| 1413.74 | 3.94 | 0.00 | 0.00 | 4.36 | 0.00 | 0.00 | 4.65 | 0.03 | 629.15 | 4.84 | 0.03 | 626.66 | 4.78 | 0.00 | 0.00 |
| 1417.18 | 3.94 | 0.00 | 0.00 | 4.36 | 0.00 | 0.00 | 4.65 | 0.03 | 623.01 | 4.84 | 0.03 | 623.09 | 4.78 | 0.00 | 0.00 |
| 1420.63 | 3.94 | 0.00 | 0.00 | 4.36 | 0.00 | 0.00 | 4.65 | 0.03 | 616.90 | 4.84 | 0.03 | 619.55 | 4.78 | 0.00 | 0.00 |
| 1424.07 | 3.93 | 0.00 | 0.00 | 4.36 | 0.00 | 0.00 | 4.65 | 0.03 | 610.90 | 4.84 | 0.03 | 616.02 | 4.78 | 0.00 | 0.00 |
| 1427.52 | 3.93 | 0.00 | 0.00 | 4.36 | 0.00 | 0.00 | 4.65 | 0.03 | 604.94 | 4.83 | 0.03 | 612.60 | 4.77 | 0.00 | 0.00 |
| 1430.97 | 3.93 | 0.00 | 0.00 | 4.36 | 0.00 | 0.00 | 4.65 | 0.03 | 599.09 | 4.83 | 0.03 | 609.10 | 4.77 | 0.00 | 0.00 |
| 1434.42 | 3.93 | 0.00 | 0.00 | 4.36 | 0.00 | 0.00 | 4.64 | 0.03 | 593.27 | 4.83 | 0.03 | 605.71 | 4.77 | 0.00 | 0.00 |
| 1437.86 | 3.93 | 0.00 | 0.00 | 4.36 | 0.00 | 0.00 | 4.64 | 0.03 | 587.48 | 4.83 | 0.03 | 602.33 | 4.77 | 0.00 | 0.00 |
| 1441.31 | 3.93 | 0.00 | 0.00 | 4.36 | 0.00 | 0.00 | 4.64 | 0.03 | 581.80 | 4.83 | 0.03 | 598.98 | 4.77 | 0.00 | 0.00 |
| 1444.76 | 3.93 | 0.00 | 0.00 | 4.36 | 0.00 | 0.00 | 4.64 | 0.03 | 576.15 | 4.83 | 0.03 | 595.63 | 4.77 | 0.00 | 0.00 |
| 1448.21 | 3.93 | 0.00 | 0.00 | 4.36 | 0.00 | 0.00 | 4.64 | 0.03 | 570.61 | 4.83 | 0.03 | 592.39 | 4.77 | 0.00 | 0.00 |
| 1451.66 | 3.93 | 0.00 | 0.00 | 4.36 | 0.00 | 0.00 | 4.64 | 0.03 | 565.10 | 4.83 | 0.03 | 589.08 | 4.77 | 0.00 | 0.00 |
| 1455.11 | 3.93 | 0.00 | 0.00 | 4.36 | 0.00 | 0.00 | 4.63 | 0.03 | 559.62 | 4.82 | 0.03 | 585.87 | 4.77 | 0.00 | 0.00 |
| 1458.56 | 3.93 | 0.00 | 0.00 | 4.36 | 0.00 | 0.00 | 4.63 | 0.03 | 554.24 | 4.82 | 0.03 | 582.67 | 4.77 | 0.00 | 0.00 |
| 1462.01 | 3.93 | 0.00 | 0.00 | 4.36 | 0.00 | 0.00 | 4.63 | 0.03 | 548.89 | 4.82 | 0.03 | 579.41 | 4.77 | 0.00 | 0.00 |
| 1465.46 | 3.93 | 0.00 | 0.00 | 4.35 | 0.00 | 0.00 | 4.63 | 0.03 | 543.66 | 4.82 | 0.03 | 576.33 | 4.77 | 0.00 | 0.00 |
| 1468.91 | 3.92 | 0.00 | 0.00 | 4.35 | 0.00 | 0.00 | 4.63 | 0.03 | 538.45 | 4.82 | 0.03 | 573.18 | 4.77 | 0.00 | 0.00 |
| 1472.36 | 3.92 | 0.00 | 0.00 | 4.35 | 0.00 | 0.00 | 4.63 | 0.03 | 533.26 | 4.82 | 0.03 | 570.04 | 4.77 | 0.00 | 0.00 |
| 1475.81 | 3.92 | 0.00 | 0.00 | 4.35 | 0.00 | 0.00 | 4.62 | 0.03 | 528.18 | 4.82 | 0.03 | 567.01 | 4.77 | 0.00 | 0.00 |
| 1479.26 | 3.92 | 0.00 | 0.00 | 4.35 | 0.00 | 0.00 | 4.62 | 0.03 | 523.21 | 4.82 | 0.03 | 563.90 | 4.77 | 0.00 | 0.00 |
| 1482.72 | 3.92 | 0.00 | 0.00 | 4.35 | 0.00 | 0.00 | 4.62 | 0.03 | 518.18 | 4.82 | 0.03 | 560.89 | 4.77 | 0.00 | 0.00 |
| 1486.17 | 3.92 | 0.00 | 0.00 | 4.35 | 0.00 | 0.00 | 4.62 | 0.03 | 513.25 | 4.81 | 0.03 | 557.90 | 4.76 | 0.00 | 0.00 |
| 1489.62 | 3.92 | 0.00 | 0.00 | 4.35 | 0.00 | 0.00 | 4.62 | 0.03 | 508.35 | 4.81 | 0.03 | 554.92 | 4.76 | 0.00 | 0.00 |
| 1493.07 | 3.92 | 0.00 | 0.00 | 4.35 | 0.00 | 0.00 | 4.62 | 0.03 | 503.56 | 4.81 | 0.03 | 551.95 | 4.76 | 0.00 | 0.00 |
| 1496.53 | 3.92 | 0.00 | 0.00 | 4.35 | 0.00 | 0.00 | 4.61 | 0.03 | 498.70 | 4.81 | 0.03 | 549.00 | 4.76 | 0.00 | 0.00 |
| 1499.98 | 3.92 | 0.00 | 0.00 | 4.35 | 0.00 | 0.00 | 4.61 | 0.03 | 494.03 | 4.81 | 0.03 | 546.14 | 4.76 | 0.00 | 0.00 |
| 1503.44 | 3.92 | 0.00 | 0.00 | 4.35 | 0.00 | 0.00 | 4.61 | 0.03 | 489.30 | 4.81 | 0.03 | 543.21 | 4.76 | 0.00 | 0.00 |
| 1506.89 | 3.92 | 0.00 | 0.00 | 4.35 | 0.00 | 0.00 | 4.61 | 0.02 | 484.68 | 4.81 | 0.03 | 540.38 | 4.76 | 0.00 | 0.00 |
| 1510.35 | 3.92 | 0.00 | 0.00 | 4.35 | 0.00 | 0.00 | 4.61 | 0.02 | 480.07 | 4.81 | 0.03 | 537.57 | 4.76 | 0.00 | 0.00 |
| 1513.80 | 3.91 | 0.00 | 0.00 | 4.35 | 0.00 | 0.00 | 4.61 | 0.02 | 475.49 | 4.81 | 0.03 | 534.76 | 4.76 | 0.00 | 0.00 |
| 1517.26 | 3.91 | 0.00 | 0.00 | 4.35 | 0.00 | 0.00 | 4.61 | 0.02 | 470.93 | 4.80 | 0.03 | 531.97 | 4.76 | 0.00 | 0.00 |
| 1520.71 | 3.91 | 0.00 | 0.00 | 4.35 | 0.00 | 0.00 | 4.60 | 0.02 | 466.47 | 4.80 | 0.03 | 529.19 | 4.76 | 0.00 | 0.00 |
| 1524.17 | 3.91 | 0.00 | 0.00 | 4.35 | 0.00 | 0.00 | 4.60 | 0.02 | 462.03 | 4.80 | 0.03 | 526.43 | 4.76 | 0.00 | 0.00 |
| 1527.63 | 3.91 | 0.00 | 0.00 | 4.35 | 0.00 | 0.00 | 4.60 | 0.02 | 457.70 | 4.80 | 0.03 | 523.67 | 4.76 | 0.00 | 0.00 |
| 1531.09 | 3.91 | 0.00 | 0.00 | 4.35 | 0.00 | 0.00 | 4.60 | 0.02 | 453.30 | 4.80 | 0.03 | 521.01 | 4.76 | 0.00 | 0.00 |
| 1534.54 | 3.91 | 0.00 | 0.00 | 4.35 | 0.00 | 0.00 | 4.60 | 0.02 | 449.00 | 4.80 | 0.03 | 518.36 | 4.76 | 0.00 | 0.00 |
| 1538.00 | 3.91 | 0.00 | 0.00 | 4.35 | 0.00 | 0.00 | 4.60 | 0.02 | 444.72 | 4.80 | 0.03 | 515.65 | 4.76 | 0.00 | 0.00 |
| 1541.46 | 3.91 | 0.00 | 0.00 | 4.35 | 0.00 | 0.00 | 4.60 | 0.02 | 440.55 | 4.80 | 0.03 | 513.02 | 4.76 | 0.00 | 0.00 |
| 1544.92 | 3.91 | 0.00 | 0.00 | 4.35 | 0.00 | 0.00 | 4.59 | 0.02 | 436.39 | 4.80 | 0.03 | 510.41 | 4.76 | 0.00 | 0.00 |
| 1548.38 | 3.91 | 0.00 | 0.00 | 4.35 | 0.00 | 0.00 | 4.59 | 0.02 | 432.25 | 4.80 | 0.03 | 507.81 | 4.76 | 0.00 | 0.00 |
| 1551.84 | 3.91 | 0.00 | 0.00 | 4.35 | 0.00 | 0.00 | 4.59 | 0.02 | 428.13 | 4.79 | 0.03 | 505.22 | 4.76 | 0.00 | 0.00 |
| 1555.30 | 3.91 | 0.00 | 0.00 | 4.35 | 0.00 | 0.00 | 4.59 | 0.02 | 424.02 | 4.79 | 0.03 | 502.64 | 4.76 | 0.00 | 0.00 |
| 1558.76 | 3.91 | 0.00 | 0.00 | 4.35 | 0.00 | 0.00 | 4.59 | 0.02 | 420.02 | 4.79 | 0.03 | 500.15 | 4.76 | 0.00 | 0.00 |
| 1562.22 | 3.90 | 0.00 | 0.00 | 4.35 | 0.00 | 0.00 | 4.59 | 0.02 | 416.03 | 4.79 | 0.03 | 497.60 | 4.76 | 0.00 | 0.00 |
| 1565.68 | 3.90 | 0.00 | 0.00 | 4.35 | 0.00 | 0.00 | 4.58 | 0.02 | 412.06 | 4.79 | 0.03 | 495.13 | 4.76 | 0.00 | 0.00 |
| 1569.14 | 3.90 | 0.00 | 0.00 | 4.35 | 0.00 | 0.00 | 4.58 | 0.02 | 408.19 | 4.79 | 0.03 | 492.68 | 4.76 | 0.00 | 0.00 |
| 1572.60 | 3.90 | 0.00 | 0.00 | 4.35 | 0.00 | 0.00 | 4.58 | 0.02 | 404.34 | 4.79 | 0.03 | 490.16 | 4.76 | 0.00 | 0.00 |
| 1576.06 | 3.90 | 0.00 | 0.00 | 4.35 | 0.00 | 0.00 | 4.58 | 0.02 | 400.50 | 4.79 | 0.03 | 487.73 | 4.76 | 0.00 | 0.00 |
| 1579.52 | 3.90 | 0.00 | 0.00 | 4.35 | 0.00 | 0.00 | 4.58 | 0.02 | 396.68 | 4.79 | 0.03 | 485.30 | 4.76 | 0.00 | 0.00 |
| 1582.99 | 3.90 | 0.00 | 0.00 | 4.35 | 0.00 | 0.00 | 4.58 | 0.02 | 392.87 | 4.79 | 0.03 | 482.89 | 4.75 | 0.00 | 0.00 |
| 1586.45 | 3.90 | 0.00 | 0.00 | 4.35 | 0.00 | 0.00 | 4.58 | 0.02 | 389.16 | 4.78 | 0.03 | 480.57 | 4.75 | 0.00 | 0.00 |
| 1589.91 | 3.90 | 0.00 | 0.00 | 4.35 | 0.00 | 0.00 | 4.57 | 0.02 | 385.47 | 4.78 | 0.03 | 478.18 | 4.75 | 0.00 | 0.00 |
| 1593.37 | 3.90 | 0.00 | 0.00 | 4.35 | 0.00 | 0.00 | 4.57 | 0.02 | 381.79 | 4.78 | 0.03 | 475.80 | 4.75 | 0.00 | 0.00 |
| 1596.84 | 3.90 | 0.00 | 0.00 | 4.35 | 0.00 | 0.00 | 4.57 | 0.02 | 378.21 | 4.78 | 0.03 | 473.51 | 4.75 | 0.00 | 0.00 |
| 1600.30 | 3.90 | 0.00 | 0.00 | 4.35 | 0.00 | 0.00 | 4.57 | 0.02 | 374.64 | 4.78 | 0.03 | 471.15 | 4.75 | 0.00 | 0.00 |
| 1603.77 | 3.90 | 0.00 | 0.00 | 4.35 | 0.00 | 0.00 | 4.57 | 0.02 | 371.01 | 4.78 | 0.03 | 468.88 | 4.75 | 0.00 | 0.00 |
| 1607.23 | 3.90 | 0.00 | 0.00 | 4.35 | 0.00 | 0.00 | 4.57 | 0.02 | 367.55 | 4.78 | 0.03 | 466.62 | 4.75 | 0.00 | 0.00 |
| 1610.70 | 3.90 | 0.00 | 0.00 | 4.35 | 0.00 | 0.00 | 4.57 | 0.02 | 364.03 | 4.78 | 0.03 | 464.36 | 4.75 | 0.00 | 0.00 |
| 1614.16 | 3.89 | 0.00 | 0.00 | 4.35 | 0.00 | 0.00 | 4.56 | 0.02 | 360.60 | 4.78 | 0.03 | 462.12 | 4.75 | 0.00 | 0.00 |
| 1617.63 | 3.89 | 0.00 | 0.00 | 4.35 | 0.00 | 0.00 | 4.56 | 0.02 | 357.11 | 4.78 | 0.03 | 459.89 | 4.75 | 0.00 | 0.00 |
| 1621.10 | 3.89 | 0.00 | 0.00 | 4.35 | 0.00 | 0.00 | 4.56 | 0.02 | 353.79 | 4.78 | 0.03 | 457.66 | 4.75 | 0.00 | 0.00 |
| 1624.56 | 3.89 | 0.00 | 0.00 | 4.35 | 0.00 | 0.00 | 4.56 | 0.02 | 350.41 | 4.77 | 0.03 | 455.45 | 4.75 | 0.00 | 0.00 |
| 1628.03 | 3.89 | 0.00 | 0.00 | 4.35 | 0.00 | 0.00 | 4.56 | 0.02 | 347.04 | 4.77 | 0.03 | 453.25 | 4.75 | 0.00 | 0.00 |
| 1631.50 | 3.89 | 0.00 | 0.00 | 4.35 | 0.00 | 0.00 | 4.56 | 0.02 | 343.76 | 4.77 | 0.03 | 451.13 | 4.75 | 0.00 | 0.00 |
| 1634.96 | 3.89 | 0.00 | 0.00 | 4.35 | 0.00 | 0.00 | 4.56 | 0.02 | 340.49 | 4.77 | 0.03 | 448.94 | 4.75 | 0.00 | 0.00 |
| 1638.43 | 3.89 | 0.00 | 0.00 | 4.35 | 0.00 | 0.00 | 4.56 | 0.02 | 337.24 | 4.77 | 0.03 | 446.84 | 4.75 | 0.00 | 0.00 |
| 1641.90 | 3.89 | 0.00 | 0.00 | 4.35 | 0.00 | 0.00 | 4.55 | 0.02 | 334.08 | 4.77 | 0.03 | 444.67 | 4.75 | 0.00 | 0.00 |
| 1645.37 | 3.89 | 0.00 | 0.00 | 4.35 | 0.00 | 0.00 | 4.55 | 0.02 | 330.85 | 4.77 | 0.03 | 442.59 | 4.75 | 0.00 | 0.00 |
| 1648.84 | 3.89 | 0.00 | 0.00 | 4.35 | 0.00 | 0.00 | 4.55 | 0.02 | 327.72 | 4.77 | 0.03 | 440.51 | 4.75 | 0.00 | 0.00 |
| 1652.31 | 3.89 | 0.00 | 0.00 | 4.35 | 0.00 | 0.00 | 4.55 | 0.02 | 324.60 | 4.77 | 0.03 | 438.45 | 4.75 | 0.00 | 0.00 |

| | | | | | | | | | | | | | |
|---|---|---|---|---|---|---|---|---|---|---|---|---|---|
| 1655.78 | 3.89 | 0.00 | 0.00 | 4.35 | 0.00 | 0.00 | 4.55 | 0.02 | 321.49 | 4.77 | 0.03 | 436.39 | 4.75 | 0.00 | 0.00 |
| 1659.25 | 3.89 | 0.00 | 0.00 | 4.35 | 0.00 | 0.00 | 4.55 | 0.02 | 318.39 | 4.77 | 0.03 | 434.34 | 4.75 | 0.00 | 0.00 |
| 1662.72 | 3.89 | 0.00 | 0.00 | 4.34 | 0.00 | 0.00 | 4.55 | 0.02 | 315.38 | 4.76 | 0.02 | 432.30 | 4.75 | 0.00 | 0.00 |
| 1666.19 | 3.88 | 0.00 | 0.00 | 4.34 | 0.00 | 0.00 | 4.54 | 0.02 | 312.39 | 4.76 | 0.02 | 430.27 | 4.75 | 0.00 | 0.00 |
| 1669.66 | 3.88 | 0.00 | 0.00 | 4.34 | 0.00 | 0.00 | 4.54 | 0.02 | 309.41 | 4.76 | 0.02 | 428.32 | 4.75 | 0.00 | 0.00 |
| 1673.13 | 3.88 | 0.00 | 0.00 | 4.34 | 0.00 | 0.00 | 4.54 | 0.02 | 306.44 | 4.76 | 0.02 | 426.31 | 4.75 | 0.00 | 0.00 |
| 1676.60 | 3.88 | 0.00 | 0.00 | 4.34 | 0.00 | 0.00 | 4.54 | 0.02 | 303.55 | 4.76 | 0.02 | 424.37 | 4.75 | 0.00 | 0.00 |
| 1680.08 | 3.88 | 0.00 | 0.00 | 4.34 | 0.00 | 0.00 | 4.54 | 0.02 | 300.61 | 4.76 | 0.02 | 422.38 | 4.75 | 0.00 | 0.00 |
| 1683.55 | 3.88 | 0.00 | 0.00 | 4.34 | 0.00 | 0.00 | 4.54 | 0.02 | 297.75 | 4.76 | 0.02 | 420.46 | 4.75 | 0.00 | 0.00 |
| 1687.02 | 3.88 | 0.00 | 0.00 | 4.34 | 0.00 | 0.00 | 4.54 | 0.02 | 294.90 | 4.76 | 0.02 | 418.55 | 4.75 | 0.00 | 0.00 |

**Table S4.** Dielectric functions and absorption coefficients of 2D DJ perovskites

| Wavelength (nm) | $m=1$ | | | $m=2$ | | | $m=3$ | | | $m=4$ | | |
|---|---|---|---|---|---|---|---|---|---|---|---|---|
| | $\varepsilon_r$ | $\varepsilon_i$ | $\alpha$ | $\varepsilon_r$ | $\varepsilon_i$ | $\alpha$ | $\varepsilon_r$ | $\varepsilon_i$ | $\alpha$ | $\varepsilon_r$ | $\varepsilon_i$ | $\alpha$ |
| 370.91 | 3.34 | 4.56 | 364346.24 | 3.97 | 3.76 | 293456.26 | 2.39 | 4.52 | 395460.50 | 3.64 | 5.18 | 393043.19 |
| 372.49 | 3.47 | 4.59 | 360258.24 | 4.03 | 3.75 | 290075.38 | 2.54 | 4.59 | 392832.73 | 3.77 | 5.15 | 385671.98 |
| 374.08 | 3.61 | 4.61 | 355739.77 | 4.08 | 3.74 | 286639.99 | 2.69 | 4.65 | 389300.10 | 3.89 | 5.11 | 378166.68 |
| 375.66 | 3.75 | 4.62 | 350829.46 | 4.14 | 3.73 | 283151.85 | 2.84 | 4.70 | 384909.02 | 4.01 | 5.07 | 370569.65 |
| 377.24 | 3.89 | 4.63 | 345561.81 | 4.19 | 3.71 | 279612.22 | 3.00 | 4.73 | 379723.20 | 4.12 | 5.02 | 362921.30 |
| 378.82 | 4.03 | 4.62 | 339962.90 | 4.24 | 3.70 | 276023.11 | 3.16 | 4.74 | 373822.85 | 4.23 | 4.97 | 355260.02 |
| 380.40 | 4.17 | 4.61 | 334050.76 | 4.30 | 3.68 | 272385.81 | 3.32 | 4.74 | 367300.86 | 4.33 | 4.91 | 347621.72 |
| 381.99 | 4.31 | 4.59 | 327834.85 | 4.35 | 3.66 | 268702.89 | 3.47 | 4.73 | 360260.29 | 4.43 | 4.85 | 340038.94 |
| 383.57 | 4.45 | 4.57 | 321315.67 | 4.40 | 3.64 | 264976.88 | 3.62 | 4.71 | 352810.58 | 4.51 | 4.78 | 332541.95 |
| 385.15 | 4.59 | 4.53 | 314488.16 | 4.45 | 3.61 | 261209.29 | 3.75 | 4.67 | 345064.07 | 4.60 | 4.71 | 325156.71 |
| 386.74 | 4.73 | 4.49 | 307340.86 | 4.51 | 3.59 | 257403.90 | 3.89 | 4.62 | 337131.56 | 4.67 | 4.64 | 317906.81 |
| 388.32 | 4.87 | 4.44 | 299862.16 | 4.56 | 3.56 | 253562.81 | 4.01 | 4.57 | 329118.27 | 4.74 | 4.57 | 310812.64 |
| 389.90 | 5.00 | 4.37 | 292040.29 | 4.60 | 3.54 | 249689.32 | 4.12 | 4.51 | 321122.99 | 4.81 | 4.50 | 303890.80 |
| 391.49 | 5.13 | 4.30 | 283869.79 | 4.65 | 3.51 | 245787.05 | 4.22 | 4.44 | 313232.63 | 4.86 | 4.43 | 297155.09 |
| 393.07 | 5.26 | 4.22 | 275354.29 | 4.70 | 3.47 | 241859.58 | 4.31 | 4.37 | 305523.02 | 4.92 | 4.36 | 290617.37 |
| 394.65 | 5.38 | 4.13 | 266509.65 | 4.74 | 3.44 | 237910.01 | 4.39 | 4.30 | 298056.41 | 4.97 | 4.29 | 284286.18 |
| 396.24 | 5.49 | 4.03 | 257367.81 | 4.79 | 3.41 | 233943.12 | 4.46 | 4.22 | 290882.48 | 5.01 | 4.22 | 278167.56 |
| 397.82 | 5.60 | 3.91 | 247977.48 | 4.83 | 3.37 | 229962.90 | 4.53 | 4.15 | 284036.61 | 5.05 | 4.15 | 272265.94 |
| 399.40 | 5.69 | 3.79 | 238402.93 | 4.87 | 3.33 | 225973.58 | 4.58 | 4.08 | 277541.97 | 5.08 | 4.08 | 266583.79 |
| 400.99 | 5.78 | 3.67 | 228724.11 | 4.91 | 3.30 | 221980.30 | 4.63 | 4.02 | 271409.46 | 5.11 | 4.01 | 261121.65 |
| 402.57 | 5.85 | 3.53 | 219030.86 | 4.95 | 3.26 | 217987.47 | 4.67 | 3.95 | 265638.40 | 5.14 | 3.95 | 255878.97 |
| 404.16 | 5.90 | 3.40 | 209419.45 | 4.98 | 3.22 | 214001.00 | 4.71 | 3.89 | 260219.52 | 5.16 | 3.89 | 250853.74 |
| 405.74 | 5.95 | 3.26 | 199987.11 | 5.02 | 3.17 | 210025.50 | 4.74 | 3.84 | 255134.83 | 5.18 | 3.83 | 246042.87 |
| 407.33 | 5.98 | 3.12 | 190825.18 | 5.05 | 3.13 | 206066.95 | 4.77 | 3.78 | 250359.63 | 5.20 | 3.77 | 241442.49 |
| 408.91 | 5.99 | 2.98 | 182015.67 | 5.08 | 3.09 | 202131.03 | 4.80 | 3.73 | 245863.51 | 5.21 | 3.72 | 237048.54 |
| 410.50 | 6.00 | 2.85 | 173627.98 | 5.11 | 3.05 | 198223.93 | 4.83 | 3.68 | 241612.64 | 5.22 | 3.66 | 232855.58 |
| 412.08 | 5.99 | 2.72 | 165715.77 | 5.13 | 3.00 | 194351.68 | 4.85 | 3.64 | 237572.24 | 5.23 | 3.61 | 228858.32 |
| 413.67 | 5.97 | 2.60 | 158315.81 | 5.16 | 2.96 | 190520.91 | 4.87 | 3.60 | 233705.74 | 5.24 | 3.57 | 225050.72 |
| 415.25 | 5.94 | 2.49 | 151450.69 | 5.18 | 2.91 | 186737.78 | 4.90 | 3.56 | 229979.81 | 5.25 | 3.52 | 221426.83 |
| 416.84 | 5.91 | 2.39 | 145128.48 | 5.20 | 2.86 | 183009.62 | 4.92 | 3.52 | 226362.63 | 5.25 | 3.48 | 217980.56 |
| 418.42 | 5.87 | 2.29 | 139344.19 | 5.21 | 2.82 | 179342.96 | 4.95 | 3.48 | 222827.67 | 5.26 | 3.43 | 214705.02 |
| 420.01 | 5.83 | 2.20 | 134084.60 | 5.23 | 2.77 | 175744.94 | 4.97 | 3.44 | 219353.28 | 5.26 | 3.39 | 211594.29 |
| 421.59 | 5.78 | 2.12 | 129327.36 | 5.24 | 2.73 | 172222.79 | 4.99 | 3.40 | 215923.64 | 5.26 | 3.36 | 208642.13 |
| 423.18 | 5.73 | 2.05 | 125046.96 | 5.25 | 2.68 | 168783.72 | 5.01 | 3.36 | 212530.09 | 5.26 | 3.32 | 205842.45 |
| 424.77 | 5.68 | 1.98 | 121212.32 | 5.26 | 2.64 | 165435.07 | 5.03 | 3.33 | 209168.59 | 5.26 | 3.29 | 203188.72 |
| 426.35 | 5.62 | 1.92 | 117792.23 | 5.26 | 2.60 | 162184.35 | 5.05 | 3.29 | 205842.09 | 5.26 | 3.26 | 200675.26 |
| 427.94 | 5.57 | 1.87 | 114753.91 | 5.27 | 2.55 | 159039.32 | 5.07 | 3.25 | 202557.54 | 5.26 | 3.23 | 198295.80 |
| 429.52 | 5.52 | 1.82 | 112065.93 | 5.27 | 2.51 | 156007.58 | 5.09 | 3.21 | 199325.81 | 5.26 | 3.20 | 196045.28 |
| 431.11 | 5.47 | 1.78 | 109697.32 | 5.26 | 2.47 | 153096.34 | 5.10 | 3.17 | 196160.72 | 5.26 | 3.18 | 193917.72 |
| 432.70 | 5.41 | 1.75 | 107618.99 | 5.26 | 2.43 | 150312.96 | 5.11 | 3.13 | 193076.59 | 5.26 | 3.15 | 191907.94 |
| 434.28 | 5.36 | 1.71 | 105803.49 | 5.25 | 2.40 | 147665.33 | 5.12 | 3.10 | 190089.28 | 5.25 | 3.13 | 190010.57 |
| 435.87 | 5.31 | 1.69 | 104225.73 | 5.25 | 2.36 | 145160.02 | 5.13 | 3.06 | 187212.72 | 5.25 | 3.11 | 188220.93 |
| 437.46 | 5.27 | 1.66 | 102862.47 | 5.24 | 2.33 | 142804.68 | 5.14 | 3.03 | 184461.00 | 5.25 | 3.09 | 186534.15 |
| 439.05 | 5.22 | 1.64 | 101693.34 | 5.23 | 2.30 | 140605.72 | 5.14 | 2.99 | 181845.05 | 5.25 | 3.08 | 184945.39 |
| 440.63 | 5.18 | 1.63 | 100699.14 | 5.22 | 2.27 | 138568.85 | 5.14 | 2.96 | 179373.95 | 5.25 | 3.06 | 183450.45 |
| 442.22 | 5.13 | 1.61 | 99863.20 | 5.20 | 2.24 | 136700.28 | 5.14 | 2.93 | 177055.00 | 5.25 | 3.05 | 182044.95 |
| 443.81 | 5.09 | 1.60 | 99170.53 | 5.19 | 2.22 | 135004.71 | 5.14 | 2.90 | 174891.86 | 5.25 | 3.04 | 180724.46 |
| 445.39 | 5.05 | 1.59 | 98608.58 | 5.17 | 2.20 | 133485.65 | 5.14 | 2.88 | 172886.31 | 5.25 | 3.03 | 179485.58 |
| 446.98 | 5.01 | 1.58 | 98165.58 | 5.16 | 2.18 | 132146.03 | 5.14 | 2.85 | 171038.00 | 5.25 | 3.02 | 178323.42 |
| 448.57 | 4.97 | 1.58 | 97829.33 | 5.14 | 2.17 | 130987.28 | 5.13 | 2.83 | 169344.31 | 5.25 | 3.01 | 177235.11 |
| 450.16 | 4.94 | 1.57 | 97551.54 | 5.13 | 2.15 | 130008.34 | 5.13 | 2.82 | 167800.95 | 5.25 | 3.00 | 176216.08 |
| 451.75 | 4.90 | 1.57 | 97318.22 | 5.11 | 2.14 | 129206.18 | 5.12 | 2.80 | 166401.60 | 5.26 | 3.00 | 175262.82 |
| 453.33 | 4.87 | 1.57 | 97135.59 | 5.10 | 2.14 | 128575.65 | 5.12 | 2.79 | 165139.01 | 5.26 | 2.99 | 174371.69 |
| 454.92 | 4.83 | 1.56 | 97010.26 | 5.08 | 2.14 | 128107.45 | 5.12 | 2.78 | 164004.56 | 5.27 | 2.99 | 173538.97 |
| 456.51 | 4.80 | 1.56 | 96949.66 | 5.07 | 2.14 | 127789.76 | 5.11 | 2.77 | 162988.30 | 5.27 | 2.99 | 172760.14 |
| 458.10 | 4.77 | 1.56 | 96962.12 | 5.06 | 2.14 | 127605.60 | 5.11 | 2.76 | 162080.33 | 5.28 | 2.99 | 172031.71 |
| 459.69 | 4.73 | 1.57 | 97057.86 | 5.06 | 2.14 | 127533.96 | 5.11 | 2.76 | 161269.06 | 5.29 | 2.99 | 171349.52 |
| 461.27 | 4.70 | 1.57 | 97248.25 | 5.05 | 2.15 | 127549.12 | 5.11 | 2.75 | 160542.83 | 5.29 | 2.99 | 170708.17 |
| 462.86 | 4.66 | 1.57 | 97545.89 | 5.05 | 2.16 | 127620.07 | 5.11 | 2.75 | 159889.81 | 5.31 | 2.99 | 170103.88 |
| 464.45 | 4.62 | 1.58 | 97965.65 | 5.06 | 2.17 | 127711.06 | 5.12 | 2.75 | 159297.14 | 5.32 | 2.99 | 169530.38 |
| 466.04 | 4.59 | 1.59 | 98522.19 | 5.07 | 2.18 | 127782.01 | 5.12 | 2.75 | 158753.03 | 5.33 | 3.00 | 168982.40 |
| 467.63 | 4.55 | 1.60 | 99231.84 | 5.08 | 2.19 | 127790.38 | 5.13 | 2.76 | 158243.74 | 5.35 | 3.00 | 168452.86 |
| 469.22 | 4.51 | 1.61 | 100111.79 | 5.09 | 2.20 | 127691.55 | 5.14 | 2.76 | 157757.11 | 5.36 | 3.01 | 167934.61 |
| 470.81 | 4.47 | 1.63 | 101178.19 | 5.11 | 2.21 | 127440.97 | 5.15 | 2.76 | 157280.19 | 5.38 | 3.01 | 167419.07 |
| 472.40 | 4.43 | 1.65 | 102447.27 | 5.14 | 2.21 | 126998.17 | 5.17 | 2.77 | 156800.04 | 5.40 | 3.02 | 166897.09 |
| 473.99 | 4.39 | 1.67 | 103934.99 | 5.17 | 2.21 | 126327.68 | 5.18 | 2.77 | 156304.26 | 5.43 | 3.03 | 166357.65 |
| 475.58 | 4.35 | 1.70 | 105655.29 | 5.20 | 2.21 | 125403.25 | 5.20 | 2.78 | 155780.83 | 5.45 | 3.03 | 165788.47 |
| 477.16 | 4.31 | 1.73 | 107621.11 | 5.23 | 2.20 | 124209.80 | 5.22 | 2.78 | 155217.68 | 5.48 | 3.04 | 165175.92 |
| 478.75 | 4.27 | 1.76 | 109844.97 | 5.26 | 2.19 | 122744.75 | 5.24 | 2.78 | 154603.69 | 5.51 | 3.05 | 164504.46 |
| 480.34 | 4.22 | 1.80 | 112337.37 | 5.30 | 2.17 | 121019.93 | 5.26 | 2.79 | 153928.04 | 5.55 | 3.05 | 163757.33 |
| 481.93 | 4.18 | 1.85 | 115109.75 | 5.33 | 2.15 | 119060.11 | 5.29 | 2.79 | 153180.42 | 5.58 | 3.05 | 162915.74 |
| 483.52 | 4.14 | 1.90 | 118173.88 | 5.36 | 2.12 | 116902.13 | 5.32 | 2.79 | 152351.63 | 5.62 | 3.06 | 161959.82 |
| 485.11 | 4.10 | 1.95 | 121542.71 | 5.38 | 2.09 | 114592.01 | 5.35 | 2.79 | 151434.01 | 5.67 | 3.06 | 160868.76 |
| 486.70 | 4.06 | 2.01 | 125232.10 | 5.40 | 2.06 | 112181.57 | 5.38 | 2.79 | 150419.57 | 5.71 | 3.05 | 159621.60 |
| 488.29 | 4.03 | 2.08 | 129261.78 | 5.42 | 2.02 | 109723.74 | 5.41 | 2.78 | 149302.86 | 5.76 | 3.05 | 158197.75 |
| 489.88 | 3.99 | 2.15 | 133655.16 | 5.44 | 1.98 | 107231.35 | 5.45 | 2.78 | 148078.74 | 5.81 | 3.04 | 156579.74 |
| 491.47 | 3.95 | 2.23 | 138442.43 | 5.45 | 1.94 | 104706.99 | 5.48 | 2.77 | 146743.83 | 5.86 | 3.02 | 154752.48 |
| 493.06 | 3.92 | 2.32 | 143659.27 | 5.45 | 1.90 | 102204.76 | 5.52 | 2.76 | 145296.48 | 5.92 | 3.00 | 152706.62 |
| 494.65 | 3.88 | 2.42 | 149349.29 | 5.45 | 1.86 | 99769.66 | 5.55 | 2.74 | 143736.16 | 5.97 | 2.98 | 150439.86 |
| 496.24 | 3.85 | 2.53 | 155562.40 | 5.45 | 1.82 | 97437.76 | 5.59 | 2.73 | 142063.80 | 6.02 | 2.95 | 147957.71 |
| 497.83 | 3.82 | 2.65 | 162356.65 | 5.44 | 1.78 | 95237.60 | 5.63 | 2.71 | 140282.45 | 6.07 | 2.91 | 145275.58 |
| 499.43 | 3.80 | 2.78 | 169796.19 | 5.43 | 1.75 | 93189.30 | 5.66 | 2.69 | 138395.97 | 6.12 | 2.87 | 142417.81 |
| 501.02 | 3.78 | 2.94 | 177949.35 | 5.42 | 1.72 | 91307.73 | 5.70 | 2.66 | 136410.58 | 6.16 | 2.83 | 139417.63 |
| 502.61 | 3.76 | 3.11 | 186883.36 | 5.40 | 1.69 | 89600.81 | 5.73 | 2.64 | 134333.39 | 6.20 | 2.78 | 136315.33 |
| 504.20 | 3.76 | 3.30 | 196657.35 | 5.38 | 1.66 | 88072.85 | 5.76 | 2.61 | 132172.94 | 6.24 | 2.73 | 133155.68 |
| 505.79 | 3.77 | 3.53 | 207306.10 | 5.36 | 1.63 | 86724.40 | 5.80 | 2.58 | 129939.52 | 6.27 | 2.68 | 129984.24 |
| 507.38 | 3.81 | 3.79 | 218814.25 | 5.34 | 1.61 | 85553.79 | 5.83 | 2.54 | 127644.34 | 6.29 | 2.62 | 126845.84 |
| 508.97 | 3.88 | 4.08 | 231078.41 | 5.31 | 1.60 | 84556.63 | 5.85 | 2.51 | 125299.55 | 6.31 | 2.57 | 123779.65 |
| 510.56 | 3.99 | 4.42 | 243835.31 | 5.29 | 1.58 | 83727.84 | 5.88 | 2.47 | 122918.41 | 6.33 | 2.52 | 120818.68 |
| 512.15 | 4.18 | 4.80 | 256559.55 | 5.26 | 1.57 | 83062.19 | 5.90 | 2.43 | 120515.24 | 6.34 | 2.46 | 117987.75 |
| 513.74 | 4.46 | 5.22 | 268311.02 | 5.24 | 1.56 | 82553.02 | 5.92 | 2.40 | 118104.18 | 6.35 | 2.42 | 115302.97 |
| 515.33 | 4.89 | 5.66 | 277532.45 | 5.22 | 1.56 | 82194.74 | 5.94 | 2.36 | 115701.16 | 6.35 | 2.37 | 112773.02 |
| 516.93 | 5.51 | 6.07 | 281874.87 | 5.19 | 1.55 | 81981.82 | 5.95 | 2.32 | 113321.21 | 6.35 | 2.33 | 110398.44 |
| 518.52 | 6.35 | 6.36 | 278244.90 | 5.17 | 1.55 | 81908.81 | 5.96 | 2.28 | 110981.04 | 6.35 | 2.28 | 108175.09 |
| 520.11 | 7.38 | 6.39 | 263513.35 | 5.14 | 1.56 | 81971.03 | 5.97 | 2.24 | 108696.52 | 6.35 | 2.25 | 106093.64 |
| 521.70 | 8.48 | 6.03 | 236219.48 | 5.12 | 1.56 | 82165.33 | 5.98 | 2.20 | 106484.67 | 6.35 | 2.21 | 104141.69 |
| 523.29 | 9.41 | 5.25 | 198552.68 | 5.10 | 1.57 | 82488.09 | 5.98 | 2.16 | 104361.63 | 6.35 | 2.18 | 102304.82 |
| 524.88 | 9.97 | 4.21 | 156389.85 | 5.07 | 1.58 | 82937.68 | 5.98 | 2.12 | 102344.88 | 6.35 | 2.15 | 100568.20 |
| 526.47 | 10.12 | 3.14 | 116430.82 | 5.05 | 1.59 | 83512.70 | 5.98 | 2.09 | 100450.82 | 6.35 | 2.12 | 98916.29 |
| 528.07 | 9.95 | 2.22 | 83047.78 | 5.03 | 1.61 | 84212.73 | 5.97 | 2.06 | 98696.92 | 6.35 | 2.09 | 97334.31 |
| 529.66 | 9.60 | 1.51 | 57518.69 | 5.01 | 1.63 | 85038.54 | 5.97 | 2.03 | 97099.17 | 6.35 | 2.06 | 95808.03 |
| 531.25 | 9.19 | 1.00 | 39120.89 | 5.00 | 1.65 | 85992.07 | 5.96 | 2.00 | 95673.76 | 6.35 | 2.03 | 94324.52 |
| 532.84 | 8.77 | 0.66 | 26399.23 | 4.98 | 1.67 | 87075.73 | 5.95 | 1.98 | 94435.40 | 6.35 | 2.01 | 92872.04 |
| 534.43 | 8.39 | 0.44 | 17895.88 | 4.96 | 1.70 | 88293.28 | 5.94 | 1.96 | 93395.80 | 6.35 | 1.98 | 91440.19 |
| 536.03 | 8.05 | 0.30 | 12418.34 | 4.95 | 1.73 | 89649.62 | 5.93 | 1.95 | 92562.81 | 6.36 | 1.96 | 90019.65 |
| 537.62 | 7.76 | 0.22 | 9071.75 | 4.93 | 1.76 | 91150.54 | 5.92 | 1.94 | 91937.18 | 6.36 | 1.93 | 88602.62 |
| 539.21 | 7.51 | 0.17 | 7211.78 | 4.92 | 1.80 | 92802.67 | 5.91 | 1.93 | 91508.38 | 6.37 | 1.91 | 87182.48 |
| 540.80 | 7.30 | 0.15 | 6382.49 | 4.91 | 1.84 | 94614.42 | 5.91 | 1.93 | 91250.61 | 6.37 | 1.88 | 85753.77 |
| 542.39 | 7.12 | 0.14 | 6249.78 | 4.91 | 1.88 | 96593.93 | 5.92 | 1.94 | 91115.96 | 6.38 | 1.86 | 84310.72 |
| 543.99 | 6.98 | 0.14 | 6279.30 | 4.90 | 1.93 | 98750.99 | 5.93 | 1.94 | 91028.74 | 6.38 | 1.83 | 82831.76 |
| 545.58 | 6.85 | 0.14 | 6298.97 | 4.90 | 1.98 | 101095.13 | 5.95 | 1.95 | 90882.84 | 6.39 | 1.80 | 81312.59 |
| 547.17 | 6.74 | 0.14 | 6310.72 | 4.90 | 2.04 | 103636.18 | 5.98 | 1.95 | 90544.65 | 6.39 | 1.77 | 79757.72 |
| 548.76 | 6.64 | 0.14 | 6315.76 | 4.91 | 2.10 | 106382.29 | 6.02 | 1.95 | 89866.29 | 6.40 | 1.74 | 78170.64 |
| 550.36 | 6.56 | 0.14 | 6315.30 | 4.92 | 2.17 | 109339.66 | 6.06 | 1.94 | 88711.67 | 6.40 | 1.71 | 76556.66 |
| 551.95 | 6.48 | 0.14 | 6310.05 | 4.94 | 2.25 | 112510.11 | 6.11 | 1.91 | 86988.82 | 6.41 | 1.68 | 74920.33 |
| 553.54 | 6.40 | 0.14 | 6300.30 | 4.97 | 2.34 | 115887.48 | 6.16 | 1.87 | 84680.66 | 6.41 | 1.65 | 73267.78 |
| 555.13 | 6.34 | 0.14 | 6287.20 | 5.01 | 2.43 | 119453.69 | 6.20 | 1.82 | 81855.55 | 6.41 | 1.61 | 71605.66 |
| 556.73 | 6.27 | 0.14 | 6270.57 | 5.06 | 2.53 | 123171.70 | 6.23 | 1.76 | 78654.80 | 6.41 | 1.58 | 69941.58 |
| 558.32 | 6.22 | 0.14 | 6251.33 | 5.13 | 2.63 | 126976.50 | 6.25 | 1.69 | 75258.43 | 6.41 | 1.55 | 68283.57 |
| 559.91 | 6.16 | 0.14 | 6229.29 | 5.22 | 2.75 | 130761.21 | 6.26 | 1.61 | 71846.82 | 6.40 | 1.51 | 66640.60 |
| 561.50 | 6.11 | 0.14 | 6205.13 | 5.34 | 2.87 | 134359.98 | 6.25 | 1.54 | 68571.62 | 6.40 | 1.48 | 65022.85 |
| 563.10 | 6.06 | 0.14 | 6178.88 | 5.48 | 2.98 | 137526.03 | 6.24 | 1.48 | 65541.02 | 6.39 | 1.45 | 63440.59 |
| 564.69 | 6.02 | 0.14 | 6150.99 | 5.67 | 3.10 | 139909.67 | 6.21 | 1.42 | 62821.01 | 6.38 | 1.41 | 61905.50 |
| 566.28 | 5.98 | 0.13 | 6121.49 | 5.90 | 3.19 | 141039.80 | 6.18 | 1.36 | 60442.29 | 6.36 | 1.38 | 60429.20 |
| 567.88 | 5.94 | 0.13 | 6090.38 | 6.18 | 3.25 | 140331.12 | 6.14 | 1.32 | 58411.51 | 6.35 | 1.35 | 59024.04 |
| 569.47 | 5.90 | 0.13 | 6058.12 | 6.50 | 3.26 | 137134.70 | 6.10 | 1.28 | 56720.46 | 6.33 | 1.32 | 57702.66 |
| 571.06 | 5.86 | 0.13 | 6024.72 | 6.85 | 3.19 | 130867.41 | 6.06 | 1.24 | 55352.81 | 6.32 | 1.30 | 56477.50 |
| 572.65 | 5.83 | 0.13 | 5990.40 | 7.21 | 3.03 | 121222.21 | 6.01 | 1.22 | 54290.58 | 6.30 | 1.27 | 55360.57 |
| 574.25 | 5.80 | 0.13 | 5955.19 | 7.53 | 2.76 | 108401.64 | 5.97 | 1.20 | 53515.43 | 6.28 | 1.25 | 54364.06 |
| 575.84 | 5.77 | 0.13 | 5919.30 | 7.78 | 2.41 | 93235.89 | 5.92 | 1.19 | 53012.11 | 6.25 | 1.23 | 53498.53 |
| 577.43 | 5.74 | 0.13 | 5882.73 | 7.93 | 2.01 | 77046.23 | 5.88 | 1.18 | 52768.14 | 6.23 | 1.22 | 52773.58 |
| 579.03 | 5.71 | 0.13 | 5845.71 | 7.96 | 1.60 | 61280.27 | 5.83 | 1.18 | 52774.68 | 6.21 | 1.20 | 52197.17 |
| 580.62 | 5.68 | 0.13 | 5808.03 | 7.90 | 1.23 | 47102.87 | 5.79 | 1.19 | 53026.59 | 6.19 | 1.20 | 51774.98 |
| 582.21 | 5.65 | 0.13 | 5769.91 | 7.76 | 0.91 | 35182.52 | 5.75 | 1.20 | 53522.75 | 6.17 | 1.19 | 51510.49 |
| 583.80 | 5.63 | 0.13 | 5731.57 | 7.59 | 0.66 | 25699.83 | 5.71 | 1.21 | 54264.82 | 6.15 | 1.19 | 51404.59 |
| 585.40 | 5.60 | 0.13 | 5693.00 | 7.40 | 0.47 | 18501.62 | 5.67 | 1.23 | 55257.64 | 6.13 | 1.19 | 51453.80 |
| 586.99 | 5.58 | 0.12 | 5654.01 | 7.21 | 0.33 | 13263.18 | 5.64 | 1.26 | 56508.83 | 6.12 | 1.20 | 51650.91 |
| 588.58 | 5.56 | 0.12 | 5615.01 | 7.03 | 0.24 | 9614.02 | 5.61 | 1.30 | 58027.36 | 6.11 | 1.21 | 51948.55 |
| 590.18 | 5.54 | 0.12 | 5575.79 | 6.87 | 0.18 | 7205.95 | 5.58 | 1.34 | 59823.68 | 6.11 | 1.22 | 52206.92 |
| 591.77 | 5.52 | 0.12 | 5536.57 | 6.73 | 0.14 | 5742.10 | 5.56 | 1.39 | 61906.96 | 6.11 | 1.23 | 52407.82 |
| 593.36 | 5.49 | 0.12 | 5497.36 | 6.60 | 0.12 | 4983.00 | 5.54 | 1.44 | 64283.16 | 6.11 | 1.23 | 52541.99 |
| 594.96 | 5.48 | 0.12 | 5457.93 | 6.49 | 0.11 | 4742.21 | 5.53 | 1.50 | 66951.45 | 6.12 | 1.24 | 52597.51 |
| 596.55 | 5.46 | 0.12 | 5418.71 | 6.40 | 0.11 | 4740.92 | 5.53 | 1.58 | 69897.42 | 6.14 | 1.24 | 52560.67 |
| 598.14 | 5.44 | 0.12 | 5379.48 | 6.32 | 0.11 | 4738.37 | 5.55 | 1.66 | 73086.11 | 6.15 | 1.24 | 52416.04 |
| 599.74 | 5.42 | 0.12 | 5340.26 | 6.25 | 0.11 | 4734.59 | 5.58 | 1.74 | 76449.78 | 6.17 | 1.24 | 52145.83 |
| 601.33 | 5.40 | 0.12 | 5301.04 | 6.19 | 0.11 | 4729.98 | 5.62 | 1.84 | 79873.77 | 6.20 | 1.24 | 51732.01 |
| 602.92 | 5.38 | 0.12 | 5262.23 | 6.13 | 0.11 | 4724.98 | 5.70 | 1.93 | 83180.49 | 6.22 | 1.23 | 51155.13 |
| 604.52 | 5.37 | 0.12 | 5223.42 | 6.08 | 0.11 | 4718.97 | 5.79 | 2.02 | 86113.84 | 6.25 | 1.22 | 50396.24 |
| 606.11 | 5.35 | 0.12 | 5184.60 | 6.03 | 0.11 | 4711.75 | 5.93 | 2.11 | 88336.03 | 6.27 | 1.20 | 49437.98 |
| 607.70 | 5.34 | 0.11 | 5146.20 | 5.99 | 0.11 | 4703.32 | 6.09 | 2.17 | 89442.74 | 6.30 | 1.18 | 48265.97 |
| 609.30 | 5.32 | 0.11 | 5107.99 | 5.95 | 0.11 | 4693.91 | 6.28 | 2.19 | 89016.59 | 6.33 | 1.15 | 46870.71 |
| 610.89 | 5.31 | 0.11 | 5069.78 | 5.91 | 0.11 | 4683.72 | 6.49 | 2.18 | 86717.71 | 6.36 | 1.11 | 45250.11 |
| 612.48 | 5.29 | 0.11 | 5031.98 | 5.87 | 0.11 | 4672.77 | 6.71 | 2.11 | 82398.34 | 6.38 | 1.07 | 43411.22 |
| 614.08 | 5.28 | 0.11 | 4994.37 | 5.84 | 0.11 | 4661.06 | 6.91 | 1.98 | 76192.10 | 6.40 | 1.03 | 41372.31 |
| 615.67 | 5.26 | 0.11 | 4956.95 | 5.81 | 0.11 | 4648.79 | 7.08 | 1.80 | 68523.53 | 6.41 | 0.97 | 39164.55 |
| 617.26 | 5.25 | 0.11 | 4919.73 | 5.78 | 0.11 | 4636.18 | 7.19 | 1.59 | 60014.43 | 6.42 | 0.92 | 36831.37 |
| 618.86 | 5.24 | 0.11 | 4882.90 | 5.75 | 0.11 | 4623.02 | 7.26 | 1.37 | 51327.63 | 6.42 | 0.86 | 34427.77 |
| 620.45 | 5.22 | 0.11 | 4846.26 | 5.72 | 0.11 | 4609.53 | 7.28 | 1.15 | 43022.08 | 6.41 | 0.80 | 32017.89 |
| 622.04 | 5.21 | 0.11 | 4810.01 | 5.69 | 0.11 | 4595.70 | 7.25 | 0.95 | 35474.02 | 6.39 | 0.74 | 29670.87 |
| 623.64 | 5.20 | 0.11 | 4773.95 | 5.67 | 0.11 | 4581.34 | 7.19 | 0.77 | 28872.88 | 6.36 | 0.69 | 27456.12 |
| 625.23 | 5.19 | 0.11 | 4738.07 | 5.65 | 0.11 | 4566.85 | 7.12 | 0.62 | 23261.28 | 6.33 | 0.64 | 25439.18 |
| 626.83 | 5.17 | 0.11 | 4702.57 | 5.62 | 0.11 | 4552.23 | 7.03 | 0.49 | 18586.78 | 6.28 | 0.59 | 23677.48 |
| 628.42 | 5.16 | 0.11 | 4667.46 | 5.60 | 0.11 | 4537.29 | 6.94 | 0.39 | 14749.26 | 6.24 | 0.56 | 22217.67 |
| 630.01 | 5.15 | 0.11 | 4632.52 | 5.58 | 0.11 | 4522.03 | 6.85 | 0.31 | 11630.66 | 6.18 | 0.53 | 21094.75 |
| 631.61 | 5.14 | 0.10 | 4597.75 | 5.56 | 0.11 | 4506.64 | 6.76 | 0.24 | 9115.52 | 6.13 | 0.51 | 20331.84 |
| 633.20 | 5.13 | 0.10 | 4563.56 | 5.54 | 0.11 | 4491.13 | 6.67 | 0.18 | 7098.46 | 6.07 | 0.50 | 19941.71 |
| 634.79 | 5.12 | 0.10 | 4529.54 | 5.52 | 0.11 | 4475.51 | 6.59 | 0.14 | 5489.24 | 6.02 | 0.49 | 19929.46 |
| 636.39 | 5.11 | 0.10 | 4495.69 | 5.50 | 0.11 | 4459.76 | 6.51 | 0.11 | 4211.73 | 5.96 | 0.50 | 20292.86 |
| 637.98 | 5.10 | 0.10 | 4462.41 | 5.48 | 0.11 | 4443.89 | 6.44 | 0.08 | 3203.74 | 5.91 | 0.52 | 21026.52 |
| 639.57 | 5.09 | 0.10 | 4429.09 | 5.47 | 0.11 | 4428.11 | 6.37 | 0.06 | 2414.36 | 5.86 | 0.55 | 22120.80 |
| 641.17 | 5.08 | 0.10 | 4396.33 | 5.45 | 0.11 | 4412.01 | 6.31 | 0.05 | 1802.74 | 5.82 | 0.58 | 23561.64 |
| 642.76 | 5.07 | 0.10 | 4363.93 | 5.43 | 0.10 | 4395.79 | 6.25 | 0.03 | 1335.51 | 5.79 | 0.62 | 25329.08 |
| 644.35 | 5.06 | 0.10 | 4331.69 | 5.42 | 0.10 | 4379.66 | 6.20 | 0.03 | 985.84 | 5.77 | 0.68 | 27353.38 |
| 645.95 | 5.05 | 0.10 | 4299.61 | 5.40 | 0.10 | 4363.60 | 6.15 | 0.02 | 732.06 | 5.76 | 0.73 | 29520.61 |
| 647.54 | 5.04 | 0.10 | 4268.07 | 5.39 | 0.10 | 4347.24 | 6.10 | 0.01 | 556.96 | 5.77 | 0.79 | 31773.06 |
| 649.13 | 5.03 | 0.10 | 4236.69 | 5.37 | 0.10 | 4330.96 | 6.06 | 0.01 | 446.41 | 5.79 | 0.85 | 34011.92 |
| 650.73 | 5.02 | 0.10 | 4205.66 | 5.36 | 0.10 | 4314.75 | 6.02 | 0.01 | 387.58 | 5.82 | 0.90 | 36084.74 |
| 652.32 | 5.01 | 0.10 | 4174.97 | 5.35 | 0.10 | 4298.24 | 5.98 | 0.01 | 367.37 | 5.88 | 0.95 | 37785.10 |
| 653.91 | 5.01 | 0.10 | 4144.62 | 5.33 | 0.10 | 4282.00 | 5.95 | 0.01 | 356.86 | 5.95 | 0.99 | 38872.02 |
| 655.51 | 5.00 | 0.10 | 4114.42 | 5.32 | 0.10 | 4265.65 | 5.92 | 0.01 | 346.99 | 6.04 | 1.01 | 39115.88 |
| 657.10 | 4.99 | 0.10 | 4084.56 | 5.31 | 0.10 | 4249.38 | 5.89 | 0.01 | 337.73 | 6.13 | 1.00 | 38366.04 |
| 658.69 | 4.98 | 0.09 | 4055.03 | 5.30 | 0.10 | 4233.00 | 5.86 | 0.01 | 328.52 | 6.22 | 0.96 | 36611.21 |
| 660.28 | 4.97 | 0.09 | 4025.84 | 5.29 | 0.10 | 4216.70 | 5.83 | 0.01 | 319.54 | 6.30 | 0.90 | 33998.68 |
| 661.88 | 4.97 | 0.09 | 3996.78 | 5.27 | 0.10 | 4200.28 | 5.80 | 0.01 | 310.61 | 6.37 | 0.82 | 30795.94 |
| 663.47 | 4.96 | 0.09 | 3968.25 | 5.26 | 0.10 | 4183.95 | 5.78 | 0.01 | 302.10 | 6.40 | 0.73 | 27313.11 |
| 665.06 | 4.95 | 0.09 | 3939.85 | 5.25 | 0.10 | 4167.69 | 5.76 | 0.01 | 293.63 | 6.42 | 0.64 | 23826.75 |
| 666.66 | 4.94 | 0.09 | 3911.59 | 5.24 | 0.10 | 4151.51 | 5.73 | 0.01 | 285.39 | 6.42 | 0.55 | 20537.07 |
| 668.25 | 4.94 | 0.09 | 3883.83 | 5.23 | 0.10 | 4135.22 | 5.71 | 0.01 | 277.18 | 6.41 | 0.47 | 17561.12 |
| 669.84 | 4.93 | 0.09 | 3856.22 | 5.22 | 0.10 | 4119.01 | 5.69 | 0.01 | 269.40 | 6.38 | 0.40 | 14946.41 |
| 671.44 | 4.92 | 0.09 | 3828.91 | 5.21 | 0.10 | 4102.88 | 5.67 | 0.01 | 261.64 | 6.35 | 0.34 | 12693.89 |
| 673.03 | 4.91 | 0.09 | 3801.93 | 5.20 | 0.10 | 4086.82 | 5.65 | 0.01 | 254.12 | 6.31 | 0.29 | 10777.85 |
| 674.62 | 4.91 | 0.09 | 3775.26 | 5.19 | 0.10 | 4070.65 | 5.64 | 0.01 | 246.62 | 6.27 | 0.25 | 9160.33 |
| 676.22 | 4.90 | 0.09 | 3748.71 | 5.18 | 0.10 | 4054.56 | 5.62 | 0.01 | 239.54 | 6.23 | 0.21 | 7800.19 |
| 677.81 | 4.89 | 0.09 | 3722.48 | 5.17 | 0.10 | 4038.73 | 5.60 | 0.01 | 232.49 | 6.20 | 0.18 | 6657.98 |
| 679.40 | 4.89 | 0.09 | 3696.55 | 5.16 | 0.10 | 4022.60 | 5.58 | 0.01 | 225.47 | 6.16 | 0.15 | 5698.69 |
| 680.99 | 4.88 | 0.09 | 3670.93 | 5.15 | 0.10 | 4006.73 | 5.57 | 0.01 | 218.85 | 6.12 | 0.13 | 4891.70 |
| 682.59 | 4.87 | 0.09 | 3645.43 | 5.15 | 0.10 | 3990.94 | 5.55 | 0.01 | 212.27 | 6.09 | 0.11 | 4211.27 |
| 684.18 | 4.87 | 0.09 | 3620.23 | 5.14 | 0.10 | 3975.04 | 5.54 | 0.01 | 205.89 | 6.06 | 0.10 | 3636.13 |
| 685.77 | 4.86 | 0.09 | 3595.33 | 5.13 | 0.10 | 3959.39 | 5.52 | 0.01 | 199.55 | 6.02 | 0.08 | 3148.32 |
| 687.36 | 4.85 | 0.09 | 3570.74 | 5.12 | 0.10 | 3943.64 | 5.51 | 0.00 | 193.42 | 5.99 | 0.07 | 2733.33 |
| 688.96 | 4.85 | 0.09 | 3546.25 | 5.11 | 0.10 | 3927.96 | 5.49 | 0.00 | 187.50 | 5.97 | 0.06 | 2379.37 |
| 690.55 | 4.84 | 0.09 | 3522.06 | 5.10 | 0.10 | 3912.53 | 5.48 | 0.00 | 181.61 | 5.94 | 0.06 | 2076.35 |
| 692.14 | 4.84 | 0.08 | 3498.16 | 5.10 | 0.10 | 3896.99 | 5.46 | 0.00 | 175.93 | 5.91 | 0.05 | 1816.30 |
| 693.74 | 4.83 | 0.08 | 3474.37 | 5.09 | 0.10 | 3881.53 | 5.45 | 0.00 | 170.45 | 5.89 | 0.04 | 1592.41 |
| 695.33 | 4.83 | 0.08 | 3450.88 | 5.08 | 0.10 | 3866.13 | 5.44 | 0.00 | 165.00 | 5.86 | 0.04 | 1399.18 |
| 696.92 | 4.82 | 0.08 | 3427.67 | 5.07 | 0.10 | 3850.63 | 5.43 | 0.00 | 159.76 | 5.84 | 0.03 | 1232.26 |
| 698.51 | 4.81 | 0.08 | 3404.74 | 5.07 | 0.10 | 3835.37 | 5.41 | 0.00 | 154.54 | 5.82 | 0.03 | 1087.33 |
| 700.10 | 4.81 | 0.08 | 3381.92 | 5.06 | 0.10 | 3820.19 | 5.40 | 0.00 | 149.52 | 5.80 | 0.03 | 961.54 |
| 701.70 | 4.80 | 0.08 | 3359.39 | 5.05 | 0.10 | 3805.07 | 5.39 | 0.00 | 144.70 | 5.78 | 0.02 | 852.27 |
| 703.29 | 4.80 | 0.08 | 3336.95 | 5.05 | 0.10 | 3789.85 | 5.38 | 0.00 | 139.91 | 5.76 | 0.02 | 757.07 |
| 704.88 | 4.79 | 0.08 | 3314.98 | 5.04 | 0.10 | 3774.87 | 5.37 | 0.00 | 135.31 | 5.74 | 0.02 | 674.06 |
| 706.47 | 4.79 | 0.08 | 3293.10 | 5.03 | 0.09 | 3759.96 | 5.36 | 0.00 | 130.74 | 5.73 | 0.02 | 601.57 |
| 708.07 | 4.78 | 0.08 | 3271.32 | 5.03 | 0.09 | 3744.94 | 5.35 | 0.00 | 126.18 | 5.71 | 0.01 | 538.46 |
| 709.66 | 4.78 | 0.08 | 3249.82 | 5.02 | 0.09 | 3730.16 | 5.34 | 0.00 | 122.01 | 5.69 | 0.01 | 483.24 |
| 711.25 | 4.77 | 0.08 | 3228.59 | 5.01 | 0.09 | 3715.46 | 5.33 | 0.00 | 117.67 | 5.68 | 0.01 | 435.16 |
| 712.84 | 4.77 | 0.08 | 3207.63 | 5.01 | 0.09 | 3700.64 | 5.32 | 0.00 | 113.70 | 5.66 | 0.01 | 393.29 |
| 714.43 | 4.76 | 0.08 | 3186.76 | 5.00 | 0.09 | 3686.06 | 5.31 | 0.00 | 109.58 | 5.65 | 0.01 | 356.71 |
| 716.03 | 4.76 | 0.08 | 3166.17 | 4.99 | 0.09 | 3671.55 | 5.30 | 0.00 | 105.83 | 5.63 | 0.01 | 325.20 |
| 717.62 | 4.75 | 0.08 | 3145.66 | 4.99 | 0.09 | 3656.92 | 5.29 | 0.00 | 101.92 | 5.62 | 0.01 | 297.69 |
| 719.21 | 4.75 | 0.08 | 3125.42 | 4.98 | 0.09 | 3642.54 | 5.28 | 0.00 | 98.20 | 5.61 | 0.01 | 274.14 |
| 720.80 | 4.74 | 0.08 | 3105.45 | 4.98 | 0.09 | 3628.05 | 5.27 | 0.00 | 94.67 | 5.59 | 0.01 | 253.84 |
| 722.39 | 4.74 | 0.08 | 3085.56 | 4.97 | 0.09 | 3613.79 | 5.26 | 0.00 | 91.15 | 5.58 | 0.01 | 236.58 |
| 723.98 | 4.73 | 0.08 | 3065.94 | 4.96 | 0.09 | 3599.60 | 5.25 | 0.00 | 87.83 | 5.57 | 0.01 | 222.00 |
| 725.58 | 4.73 | 0.08 | 3046.57 | 4.96 | 0.09 | 3585.47 | 5.24 | 0.00 | 84.52 | 5.56 | 0.01 | 209.74 |
| 727.17 | 4.72 | 0.08 | 3027.29 | 4.95 | 0.09 | 3571.23 | 5.23 | 0.00 | 81.22 | 5.55 | 0.01 | 198.56 |
| 728.76 | 4.72 | 0.08 | 3008.26 | 4.95 | 0.09 | 3557.23 | 5.22 | 0.00 | 78.11 | 5.53 | 0.01 | 187.61 |
| 730.35 | 4.71 | 0.08 | 2989.32 | 4.94 | 0.09 | 3543.28 | 5.22 | 0.00 | 75.02 | 5.52 | 0.00 | 177.22 |
| 731.94 | 4.71 | 0.08 | 2970.63 | 4.94 | 0.09 | 3529.40 | 5.21 | 0.00 | 72.11 | 5.51 | 0.00 | 167.05 |
| 733.53 | 4.71 | 0.07 | 2952.03 | 4.93 | 0.09 | 3515.58 | 5.20 | 0.00 | 69.21 | 5.50 | 0.00 | 157.27 |
| 735.12 | 4.70 | 0.07 | 2933.68 | 4.93 | 0.09 | 3501.81 | 5.19 | 0.00 | 66.33 | 5.49 | 0.00 | 147.87 |
| 736.71 | 4.70 | 0.07 | 2915.40 | 4.92 | 0.09 | 3488.11 | 5.18 | 0.00 | 63.62 | 5.48 | 0.00 | 138.68 |
| 738.31 | 4.69 | 0.07 | 2897.55 | 4.92 | 0.09 | 3474.47 | 5.18 | 0.00 | 61.10 | 5.47 | 0.00 | 130.04 |
| 739.90 | 4.69 | 0.07 | 2879.60 | 4.91 | 0.09 | 3460.88 | 5.17 | 0.00 | 58.42 | 5.46 | 0.00 | 121.61 |
| 741.49 | 4.68 | 0.07 | 2861.90 | 4.91 | 0.09 | 3447.36 | 5.16 | 0.00 | 55.93 | 5.45 | 0.00 | 113.55 |
| 743.08 | 4.68 | 0.07 | 2844.44 | 4.90 | 0.09 | 3433.89 | 5.15 | 0.00 | 53.61 | 5.45 | 0.00 | 105.70 |
| 744.67 | 4.68 | 0.07 | 2827.06 | 4.90 | 0.09 | 3420.48 | 5.15 | 0.00 | 51.13 | 5.44 | 0.00 | 98.21 |
| 746.26 | 4.67 | 0.07 | 2809.92 | 4.89 | 0.09 | 3407.30 | 5.14 | 0.00 | 48.83 | 5.43 | 0.00 | 91.10 |
| 747.85 | 4.67 | 0.07 | 2792.86 | 4.89 | 0.09 | 3394.00 | 5.13 | 0.00 | 46.71 | 5.42 | 0.00 | 84.35 |
| 749.44 | 4.66 | 0.07 | 2776.03 | 4.88 | 0.09 | 3380.93 | 5.13 | 0.00 | 44.60 | 5.41 | 0.00 | 77.80 |
| 751.03 | 4.66 | 0.07 | 2759.28 | 4.88 | 0.09 | 3367.75 | 5.12 | 0.00 | 42.50 | 5.40 | 0.00 | 71.61 |
| 752.62 | 4.66 | 0.07 | 2742.76 | 4.88 | 0.09 | 3354.79 | 5.11 | 0.00 | 40.41 | 5.40 | 0.00 | 65.62 |
| 754.21 | 4.65 | 0.07 | 2726.48 | 4.87 | 0.09 | 3341.72 | 5.11 | 0.00 | 38.49 | 5.39 | 0.00 | 59.98 |
| 755.80 | 4.65 | 0.07 | 2710.11 | 4.87 | 0.09 | 3328.87 | 5.10 | 0.00 | 36.58 | 5.38 | 0.00 | 54.70 |
| 757.39 | 4.65 | 0.07 | 2694.13 | 4.86 | 0.09 | 3316.07 | 5.09 | 0.00 | 34.84 | 5.37 | 0.00 | 49.61 |
| 758.98 | 4.64 | 0.07 | 2678.06 | 4.86 | 0.09 | 3303.33 | 5.09 | 0.00 | 32.95 | 5.36 | 0.00 | 44.87 |
| 760.57 | 4.64 | 0.07 | 2662.38 | 4.85 | 0.09 | 3290.48 | 5.08 | 0.00 | 31.23 | 5.36 | 0.00 | 40.31 |
| 762.16 | 4.63 | 0.07 | 2646.61 | 4.85 | 0.09 | 3277.85 | 5.07 | 0.00 | 29.68 | 5.35 | 0.00 | 36.11 |
| 763.75 | 4.63 | 0.07 | 2631.23 | 4.85 | 0.09 | 3265.26 | 5.07 | 0.00 | 27.97 | 5.34 | 0.00 | 32.08 |
| 765.34 | 4.63 | 0.07 | 2615.75 | 4.84 | 0.09 | 3252.90 | 5.06 | 0.00 | 26.44 | 5.34 | 0.00 | 28.24 |
| 766.93 | 4.62 | 0.07 | 2600.50 | 4.84 | 0.09 | 3240.42 | 5.06 | 0.00 | 25.07 | 5.33 | 0.00 | 24.74 |
| 768.52 | 4.62 | 0.07 | 2585.47 | 4.83 | 0.09 | 3228.00 | 5.05 | 0.00 | 23.55 | 5.32 | 0.00 | 21.58 |
| 770.11 | 4.62 | 0.07 | 2570.51 | 4.83 | 0.09 | 3215.63 | 5.04 | 0.00 | 22.19 | 5.31 | 0.00 | 18.60 |
| 771.70 | 4.61 | 0.07 | 2555.61 | 4.83 | 0.09 | 3203.47 | 5.04 | 0.00 | 20.84 | 5.31 | 0.00 | 15.80 |
| 773.29 | 4.61 | 0.07 | 2540.93 | 4.82 | 0.09 | 3191.20 | 5.03 | 0.00 | 19.66 | 5.30 | 0.00 | 13.33 |
| 774.88 | 4.61 | 0.07 | 2526.48 | 4.82 | 0.09 | 3179.14 | 5.03 | 0.00 | 18.33 | 5.30 | 0.00 | 11.03 |
| 776.47 | 4.60 | 0.07 | 2512.09 | 4.81 | 0.09 | 3166.97 | 5.02 | 0.00 | 17.16 | 5.29 | 0.00 | 8.90 |
| 778.06 | 4.60 | 0.07 | 2497.75 | 4.81 | 0.09 | 3155.01 | 5.02 | 0.00 | 15.99 | 5.28 | 0.00 | 7.11 |
| 779.65 | 4.60 | 0.07 | 2483.47 | 4.81 | 0.09 | 3143.10 | 5.01 | 0.00 | 14.99 | 5.28 | 0.00 | 5.48 |
| 781.23 | 4.59 | 0.07 | 2469.58 | 4.80 | 0.09 | 3131.24 | 5.01 | 0.00 | 13.99 | 5.27 | 0.00 | 4.02 |
| 782.82 | 4.59 | 0.07 | 2455.58 | 4.80 | 0.09 | 3119.43 | 5.00 | 0.00 | 12.84 | 5.26 | 0.00 | 2.89 |
| 784.41 | 4.59 | 0.07 | 2441.79 | 4.80 | 0.08 | 3107.66 | 4.99 | 0.00 | 12.02 | 5.26 | 0.00 | 1.92 |
| 786.00 | 4.58 | 0.07 | 2428.07 | 4.79 | 0.08 | 3095.95 | 4.99 | 0.00 | 11.03 | 5.25 | 0.00 | 1.12 |
| 787.59 | 4.58 | 0.06 | 2414.55 | 4.79 | 0.08 | 3084.28 | 4.98 | 0.00 | 10.21 | 5.25 | 0.00 | 0.48 |
| 789.18 | 4.58 | 0.06 | 2401.10 | 4.79 | 0.08 | 3072.66 | 4.98 | 0.00 | 9.39 | 5.24 | 0.00 | 0.16 |
| 790.77 | 4.58 | 0.06 | 2387.85 | 4.78 | 0.08 | 3061.24 | 4.97 | 0.00 | 8.58 | 5.24 | 0.00 | 0.00 |
| 792.35 | 4.57 | 0.06 | 2374.66 | 4.78 | 0.08 | 3049.72 | 4.97 | 0.00 | 7.77 | 5.23 | 0.00 | 0.00 |
| 793.94 | 4.57 | 0.06 | 2361.52 | 4.78 | 0.08 | 3038.39 | 4.96 | 0.00 | 7.12 | 5.22 | 0.00 | 0.00 |
| 795.53 | 4.57 | 0.06 | 2348.44 | 4.77 | 0.08 | 3027.12 | 4.96 | 0.00 | 6.48 | 5.22 | 0.00 | 0.00 |
| 797.12 | 4.56 | 0.06 | 2335.72 | 4.77 | 0.08 | 3015.73 | 4.96 | 0.00 | 5.83 | 5.21 | 0.00 | 0.00 |
| 798.70 | 4.56 | 0.06 | 2322.90 | 4.77 | 0.08 | 3004.54 | 4.95 | 0.00 | 5.19 | 5.21 | 0.00 | 0.00 |
| 800.29 | 4.56 | 0.06 | 2310.28 | 4.76 | 0.08 | 2993.40 | 4.95 | 0.00 | 4.55 | 5.20 | 0.00 | 0.00 |
| 801.88 | 4.55 | 0.06 | 2297.72 | 4.76 | 0.08 | 2982.30 | 4.94 | 0.00 | 4.07 | 5.20 | 0.00 | 0.00 |
| 803.47 | 4.55 | 0.06 | 2285.36 | 4.76 | 0.08 | 2971.25 | 4.94 | 0.00 | 3.60 | 5.19 | 0.00 | 0.00 |
| 805.05 | 4.55 | 0.06 | 2272.89 | 4.75 | 0.08 | 2960.24 | 4.93 | 0.00 | 3.12 | 5.19 | 0.00 | 0.00 |
| 806.64 | 4.55 | 0.06 | 2260.79 | 4.75 | 0.08 | 2949.28 | 4.93 | 0.00 | 2.80 | 5.18 | 0.00 | 0.00 |
| 808.23 | 4.54 | 0.06 | 2248.58 | 4.75 | 0.08 | 2938.51 | 4.92 | 0.00 | 2.33 | 5.18 | 0.00 | 0.00 |
| 809.82 | 4.54 | 0.06 | 2236.57 | 4.74 | 0.08 | 2927.63 | 4.92 | 0.00 | 2.02 | 5.17 | 0.00 | 0.00 |
| 811.40 | 4.54 | 0.06 | 2224.76 | 4.74 | 0.08 | 2916.95 | 4.91 | 0.00 | 1.70 | 5.17 | 0.00 | 0.00 |
| 812.99 | 4.54 | 0.06 | 2212.85 | 4.74 | 0.08 | 2906.16 | 4.91 | 0.00 | 1.39 | 5.16 | 0.00 | 0.00 |
| 814.58 | 4.53 | 0.06 | 2201.13 | 4.74 | 0.08 | 2895.56 | 4.91 | 0.00 | 1.08 | 5.16 | 0.00 | 0.00 |
| 816.16 | 4.53 | 0.06 | 2189.62 | 4.73 | 0.08 | 2885.00 | 4.90 | 0.00 | 0.92 | 5.16 | 0.00 | 0.00 |
| 817.75 | 4.53 | 0.06 | 2177.99 | 4.73 | 0.08 | 2874.49 | 4.90 | 0.00 | 0.61 | 5.15 | 0.00 | 0.00 |
| 819.34 | 4.52 | 0.06 | 2166.57 | 4.73 | 0.08 | 2864.02 | 4.89 | 0.00 | 0.46 | 5.15 | 0.00 | 0.00 |
| 820.92 | 4.52 | 0.06 | 2155.34 | 4.72 | 0.08 | 2853.59 | 4.89 | 0.00 | 0.31 | 5.14 | 0.00 | 0.00 |
| 822.51 | 4.52 | 0.06 | 2144.01 | 4.72 | 0.08 | 2843.19 | 4.89 | 0.00 | 0.15 | 5.14 | 0.00 | 0.00 |
| 824.09 | 4.52 | 0.06 | 2132.87 | 4.72 | 0.08 | 2832.84 | 4.88 | 0.00 | 0.15 | 5.13 | 0.00 | 0.00 |
| 825.68 | 4.51 | 0.06 | 2121.77 | 4.72 | 0.08 | 2822.53 | 4.88 | 0.00 | 0.00 | 5.13 | 0.00 | 0.00 |
| 827.27 | 4.51 | 0.06 | 2110.87 | 4.71 | 0.08 | 2812.41 | 4.87 | 0.00 | 0.00 | 5.12 | 0.00 | 0.00 |
| 828.85 | 4.51 | 0.06 | 2100.01 | 4.71 | 0.08 | 2802.18 | 4.87 | 0.00 | 0.00 | 5.12 | 0.00 | 0.00 |
| 830.44 | 4.51 | 0.06 | 2089.19 | 4.71 | 0.08 | 2792.14 | 4.87 | 0.00 | 0.00 | 5.12 | 0.00 | 0.00 |
| 832.02 | 4.50 | 0.06 | 2078.56 | 4.70 | 0.08 | 2781.99 | 4.86 | 0.00 | 0.00 | 5.11 | 0.00 | 0.00 |
| 833.61 | 4.50 | 0.06 | 2067.98 | 4.70 | 0.08 | 2772.03 | 4.86 | 0.00 | 0.00 | 5.11 | 0.00 | 0.00 |
| 835.19 | 4.50 | 0.06 | 2057.44 | 4.70 | 0.08 | 2762.10 | 4.86 | 0.00 | 0.00 | 5.10 | 0.00 | 0.00 |
| 836.78 | 4.50 | 0.06 | 2046.93 | 4.70 | 0.08 | 2752.22 | 4.85 | 0.00 | 0.00 | 5.10 | 0.00 | 0.00 |
| 838.36 | 4.49 | 0.06 | 2036.62 | 4.69 | 0.08 | 2742.37 | 4.85 | 0.00 | 0.00 | 5.10 | 0.00 | 0.00 |
| 839.95 | 4.49 | 0.06 | 2026.34 | 4.69 | 0.08 | 2732.56 | 4.84 | 0.00 | 0.00 | 5.09 | 0.00 | 0.00 |
| 841.53 | 4.49 | 0.06 | 2016.11 | 4.69 | 0.08 | 2722.78 | 4.84 | 0.00 | 0.00 | 5.09 | 0.00 | 0.00 |
| 843.12 | 4.49 | 0.06 | 2006.06 | 4.69 | 0.08 | 2713.04 | 4.84 | 0.00 | 0.00 | 5.08 | 0.00 | 0.00 |
| 844.70 | 4.49 | 0.06 | 1996.05 | 4.68 | 0.08 | 2703.49 | 4.83 | 0.00 | 0.00 | 5.08 | 0.00 | 0.00 |
| 846.29 | 4.48 | 0.06 | 1986.08 | 4.68 | 0.08 | 2693.83 | 4.83 | 0.00 | 0.00 | 5.08 | 0.00 | 0.00 |
| 847.87 | 4.48 | 0.06 | 1976.14 | 4.68 | 0.08 | 2684.35 | 4.83 | 0.00 | 0.00 | 5.07 | 0.00 | 0.00 |
| 849.45 | 4.48 | 0.06 | 1966.39 | 4.68 | 0.08 | 2674.91 | 4.82 | 0.00 | 0.00 | 5.07 | 0.00 | 0.00 |
| 851.04 | 4.48 | 0.06 | 1956.68 | 4.67 | 0.08 | 2665.35 | 4.82 | 0.00 | 0.00 | 5.07 | 0.00 | 0.00 |
| 852.62 | 4.47 | 0.06 | 1947.00 | 4.67 | 0.08 | 2655.98 | 4.82 | 0.00 | 0.00 | 5.06 | 0.00 | 0.00 |
| 854.20 | 4.47 | 0.06 | 1937.51 | 4.67 | 0.08 | 2646.64 | 4.81 | 0.00 | 0.00 | 5.06 | 0.00 | 0.00 |
| 855.79 | 4.47 | 0.06 | 1927.90 | 4.67 | 0.08 | 2637.34 | 4.81 | 0.00 | 0.00 | 5.06 | 0.00 | 0.00 |
| 857.37 | 4.47 | 0.06 | 1918.63 | 4.67 | 0.08 | 2628.08 | 4.81 | 0.00 | 0.00 | 5.05 | 0.00 | 0.00 |
| 858.95 | 4.47 | 0.06 | 1909.24 | 4.66 | 0.08 | 2618.84 | 4.80 | 0.00 | 0.00 | 5.05 | 0.00 | 0.00 |
| 860.54 | 4.46 | 0.05 | 1899.89 | 4.66 | 0.08 | 2609.65 | 4.80 | 0.00 | 0.00 | 5.05 | 0.00 | 0.00 |
| 862.12 | 4.46 | 0.05 | 1890.72 | 4.66 | 0.08 | 2600.63 | 4.80 | 0.00 | 0.00 | 5.04 | 0.00 | 0.00 |
| 863.70 | 4.46 | 0.05 | 1881.58 | 4.66 | 0.08 | 2591.50 | 4.79 | 0.00 | 0.00 | 5.04 | 0.00 | 0.00 |
| 865.29 | 4.46 | 0.05 | 1872.62 | 4.65 | 0.08 | 2582.55 | 4.79 | 0.00 | 0.00 | 5.03 | 0.00 | 0.00 |
| 866.87 | 4.45 | 0.05 | 1863.55 | 4.65 | 0.08 | 2573.48 | 4.79 | 0.00 | 0.00 | 5.03 | 0.00 | 0.00 |
| 868.45 | 4.45 | 0.05 | 1854.65 | 4.65 | 0.08 | 2564.60 | 4.78 | 0.00 | 0.00 | 5.03 | 0.00 | 0.00 |
| 870.03 | 4.45 | 0.05 | 1845.79 | 4.65 | 0.08 | 2555.75 | 4.78 | 0.00 | 0.00 | 5.03 | 0.00 | 0.00 |
| 871.61 | 4.45 | 0.05 | 1836.97 | 4.64 | 0.08 | 2546.93 | 4.78 | 0.00 | 0.00 | 5.02 | 0.00 | 0.00 |
| 873.20 | 4.45 | 0.05 | 1828.31 | 4.64 | 0.08 | 2538.14 | 4.78 | 0.00 | 0.00 | 5.02 | 0.00 | 0.00 |
| 874.78 | 4.44 | 0.05 | 1819.69 | 4.64 | 0.08 | 2529.39 | 4.77 | 0.00 | 0.00 | 5.02 | 0.00 | 0.00 |
| 876.36 | 4.44 | 0.05 | 1811.10 | 4.64 | 0.08 | 2520.66 | 4.77 | 0.00 | 0.00 | 5.01 | 0.00 | 0.00 |
| 877.94 | 4.44 | 0.05 | 1802.55 | 4.64 | 0.08 | 2511.97 | 4.77 | 0.00 | 0.00 | 5.01 | 0.00 | 0.00 |
| 879.52 | 4.44 | 0.05 | 1794.02 | 4.63 | 0.08 | 2503.31 | 4.76 | 0.00 | 0.00 | 5.01 | 0.00 | 0.00 |
| 881.10 | 4.44 | 0.05 | 1785.67 | 4.63 | 0.08 | 2494.83 | 4.76 | 0.00 | 0.00 | 5.00 | 0.00 | 0.00 |
| 882.68 | 4.43 | 0.05 | 1777.35 | 4.63 | 0.08 | 2486.23 | 4.76 | 0.00 | 0.00 | 5.00 | 0.00 | 0.00 |
| 884.27 | 4.43 | 0.05 | 1769.05 | 4.63 | 0.08 | 2477.81 | 4.75 | 0.00 | 0.00 | 5.00 | 0.00 | 0.00 |
| 885.85 | 4.43 | 0.05 | 1760.93 | 4.63 | 0.07 | 2469.28 | 4.75 | 0.00 | 0.00 | 4.99 | 0.00 | 0.00 |
| 887.43 | 4.43 | 0.05 | 1752.70 | 4.62 | 0.07 | 2460.91 | 4.75 | 0.00 | 0.00 | 4.99 | 0.00 | 0.00 |
| 889.01 | 4.43 | 0.05 | 1744.64 | 4.62 | 0.07 | 2452.58 | 4.75 | 0.00 | 0.00 | 4.99 | 0.00 | 0.00 |
| 890.59 | 4.42 | 0.05 | 1736.60 | 4.62 | 0.07 | 2444.28 | 4.74 | 0.00 | 0.00 | 4.99 | 0.00 | 0.00 |
| 892.17 | 4.42 | 0.05 | 1728.60 | 4.62 | 0.07 | 2436.01 | 4.74 | 0.00 | 0.00 | 4.98 | 0.00 | 0.00 |
| 893.75 | 4.42 | 0.05 | 1720.76 | 4.62 | 0.07 | 2427.76 | 4.74 | 0.00 | 0.00 | 4.98 | 0.00 | 0.00 |
| 895.33 | 4.42 | 0.05 | 1712.81 | 4.61 | 0.07 | 2419.55 | 4.74 | 0.00 | 0.00 | 4.98 | 0.00 | 0.00 |
| 896.91 | 4.42 | 0.05 | 1705.03 | 4.61 | 0.07 | 2411.37 | 4.73 | 0.00 | 0.00 | 4.97 | 0.00 | 0.00 |
| 898.49 | 4.42 | 0.05 | 1697.28 | 4.61 | 0.07 | 2403.35 | 4.73 | 0.00 | 0.00 | 4.97 | 0.00 | 0.00 |
| 900.06 | 4.41 | 0.05 | 1689.70 | 4.61 | 0.07 | 2395.23 | 4.73 | 0.00 | 0.00 | 4.97 | 0.00 | 0.00 |
| 901.64 | 4.41 | 0.05 | 1682.00 | 4.61 | 0.07 | 2387.27 | 4.72 | 0.00 | 0.00 | 4.97 | 0.00 | 0.00 |
| 903.22 | 4.41 | 0.05 | 1674.47 | 4.60 | 0.07 | 2379.20 | 4.72 | 0.00 | 0.00 | 4.96 | 0.00 | 0.00 |
| 904.80 | 4.41 | 0.05 | 1666.97 | 4.60 | 0.07 | 2371.30 | 4.72 | 0.00 | 0.00 | 4.96 | 0.00 | 0.00 |
| 906.38 | 4.41 | 0.05 | 1659.49 | 4.60 | 0.07 | 2363.43 | 4.72 | 0.00 | 0.00 | 4.96 | 0.00 | 0.00 |
| 907.96 | 4.40 | 0.05 | 1652.04 | 4.60 | 0.07 | 2355.45 | 4.71 | 0.00 | 0.00 | 4.95 | 0.00 | 0.00 |
| 909.54 | 4.40 | 0.05 | 1644.61 | 4.60 | 0.07 | 2347.63 | 4.71 | 0.00 | 0.00 | 4.95 | 0.00 | 0.00 |
| 911.11 | 4.40 | 0.05 | 1637.35 | 4.60 | 0.07 | 2339.84 | 4.71 | 0.00 | 0.00 | 4.95 | 0.00 | 0.00 |
| 912.69 | 4.40 | 0.05 | 1630.11 | 4.59 | 0.07 | 2332.08 | 4.71 | 0.00 | 0.00 | 4.95 | 0.00 | 0.00 |
| 914.27 | 4.40 | 0.05 | 1622.90 | 4.59 | 0.07 | 2324.34 | 4.70 | 0.00 | 0.00 | 4.94 | 0.00 | 0.00 |
| 915.85 | 4.40 | 0.05 | 1615.72 | 4.59 | 0.07 | 2316.77 | 4.70 | 0.00 | 0.00 | 4.94 | 0.00 | 0.00 |
| 917.42 | 4.39 | 0.05 | 1608.69 | 4.59 | 0.07 | 2309.09 | 4.70 | 0.00 | 0.00 | 4.94 | 0.00 | 0.00 |
| 919.00 | 4.39 | 0.05 | 1601.56 | 4.59 | 0.07 | 2301.44 | 4.70 | 0.00 | 0.00 | 4.94 | 0.00 | 0.00 |
| 920.58 | 4.39 | 0.05 | 1594.58 | 4.58 | 0.07 | 2293.94 | 4.69 | 0.00 | 0.00 | 4.93 | 0.00 | 0.00 |
| 922.16 | 4.39 | 0.05 | 1587.63 | 4.58 | 0.07 | 2286.34 | 4.69 | 0.00 | 0.00 | 4.93 | 0.00 | 0.00 |
| 923.73 | 4.39 | 0.05 | 1580.71 | 4.58 | 0.07 | 2278.90 | 4.69 | 0.00 | 0.00 | 4.93 | 0.00 | 0.00 |
| 925.31 | 4.39 | 0.05 | 1573.94 | 4.58 | 0.07 | 2271.49 | 4.69 | 0.00 | 0.00 | 4.93 | 0.00 | 0.00 |
| 926.89 | 4.38 | 0.05 | 1567.06 | 4.58 | 0.07 | 2264.10 | 4.69 | 0.00 | 0.00 | 4.92 | 0.00 | 0.00 |
| 928.46 | 4.38 | 0.05 | 1560.34 | 4.58 | 0.07 | 2256.61 | 4.68 | 0.00 | 0.00 | 4.92 | 0.00 | 0.00 |
| 930.04 | 4.38 | 0.05 | 1553.64 | 4.57 | 0.07 | 2249.27 | 4.68 | 0.00 | 0.00 | 4.92 | 0.00 | 0.00 |
| 931.61 | 4.38 | 0.05 | 1546.97 | 4.57 | 0.07 | 2241.96 | 4.68 | 0.00 | 0.00 | 4.92 | 0.00 | 0.00 |
| 933.19 | 4.38 | 0.05 | 1540.32 | 4.57 | 0.07 | 2234.67 | 4.68 | 0.00 | 0.00 | 4.91 | 0.00 | 0.00 |
| 934.76 | 4.38 | 0.05 | 1533.69 | 4.57 | 0.07 | 2227.55 | 4.67 | 0.00 | 0.00 | 4.91 | 0.00 | 0.00 |
| 936.34 | 4.37 | 0.05 | 1527.22 | 4.57 | 0.07 | 2220.31 | 4.67 | 0.00 | 0.00 | 4.91 | 0.00 | 0.00 |
| 937.91 | 4.37 | 0.05 | 1520.63 | 4.57 | 0.07 | 2213.10 | 4.67 | 0.00 | 0.00 | 4.91 | 0.00 | 0.00 |
| 939.49 | 4.37 | 0.05 | 1514.21 | 4.56 | 0.07 | 2206.04 | 4.67 | 0.00 | 0.00 | 4.90 | 0.00 | 0.00 |
| 941.06 | 4.37 | 0.05 | 1507.80 | 4.56 | 0.07 | 2198.88 | 4.66 | 0.00 | 0.00 | 4.90 | 0.00 | 0.00 |
| 942.64 | 4.37 | 0.05 | 1501.55 | 4.56 | 0.07 | 2191.88 | 4.66 | 0.00 | 0.00 | 4.90 | 0.00 | 0.00 |
| 944.21 | 4.37 | 0.05 | 1495.19 | 4.56 | 0.07 | 2184.76 | 4.66 | 0.00 | 0.00 | 4.90 | 0.00 | 0.00 |
| 945.79 | 4.37 | 0.05 | 1488.85 | 4.56 | 0.07 | 2177.80 | 4.66 | 0.00 | 0.00 | 4.90 | 0.00 | 0.00 |
| 947.36 | 4.36 | 0.05 | 1482.66 | 4.56 | 0.07 | 2170.87 | 4.66 | 0.00 | 0.00 | 4.89 | 0.00 | 0.00 |
| 948.93 | 4.36 | 0.05 | 1476.49 | 4.55 | 0.07 | 2163.96 | 4.65 | 0.00 | 0.00 | 4.89 | 0.00 | 0.00 |
| 950.51 | 4.36 | 0.05 | 1470.35 | 4.55 | 0.07 | 2157.07 | 4.65 | 0.00 | 0.00 | 4.89 | 0.00 | 0.00 |
| 952.08 | 4.36 | 0.05 | 1464.22 | 4.55 | 0.07 | 2150.21 | 4.65 | 0.00 | 0.00 | 4.89 | 0.00 | 0.00 |
| 953.65 | 4.36 | 0.05 | 1458.12 | 4.55 | 0.07 | 2143.37 | 4.65 | 0.00 | 0.00 | 4.88 | 0.00 | 0.00 |
| 955.23 | 4.36 | 0.05 | 1452.17 | 4.55 | 0.07 | 2136.55 | 4.65 | 0.00 | 0.00 | 4.88 | 0.00 | 0.00 |
| 956.80 | 4.35 | 0.05 | 1446.23 | 4.55 | 0.07 | 2129.76 | 4.64 | 0.00 | 0.00 | 4.88 | 0.00 | 0.00 |
| 958.37 | 4.35 | 0.05 | 1440.19 | 4.55 | 0.07 | 2123.11 | 4.64 | 0.00 | 0.00 | 4.88 | 0.00 | 0.00 |
| 959.94 | 4.35 | 0.05 | 1434.30 | 4.54 | 0.07 | 2116.37 | 4.64 | 0.00 | 0.00 | 4.88 | 0.00 | 0.00 |
| 961.52 | 4.35 | 0.05 | 1428.42 | 4.54 | 0.07 | 2109.64 | 4.64 | 0.00 | 0.00 | 4.87 | 0.00 | 0.00 |
| 963.09 | 4.35 | 0.05 | 1422.57 | 4.54 | 0.07 | 2103.06 | 4.63 | 0.00 | 0.00 | 4.87 | 0.00 | 0.00 |
| 964.66 | 4.35 | 0.05 | 1416.86 | 4.54 | 0.07 | 2096.51 | 4.63 | 0.00 | 0.00 | 4.87 | 0.00 | 0.00 |
| 966.23 | 4.35 | 0.05 | 1411.05 | 4.54 | 0.07 | 2089.85 | 4.63 | 0.00 | 0.00 | 4.87 | 0.00 | 0.00 |
| 967.80 | 4.34 | 0.05 | 1405.38 | 4.54 | 0.07 | 2083.34 | 4.63 | 0.00 | 0.00 | 4.86 | 0.00 | 0.00 |
| 969.37 | 4.34 | 0.05 | 1399.73 | 4.53 | 0.07 | 2076.85 | 4.63 | 0.00 | 0.00 | 4.86 | 0.00 | 0.00 |
| 970.95 | 4.34 | 0.04 | 1394.11 | 4.53 | 0.07 | 2070.39 | 4.62 | 0.00 | 0.00 | 4.86 | 0.00 | 0.00 |
| 972.52 | 4.34 | 0.04 | 1388.49 | 4.53 | 0.07 | 2063.94 | 4.62 | 0.00 | 0.00 | 4.86 | 0.00 | 0.00 |
| 974.09 | 4.34 | 0.04 | 1382.90 | 4.53 | 0.07 | 2057.52 | 4.62 | 0.00 | 0.00 | 4.86 | 0.00 | 0.00 |
| 975.66 | 4.34 | 0.04 | 1377.33 | 4.53 | 0.07 | 2051.12 | 4.62 | 0.00 | 0.00 | 4.85 | 0.00 | 0.00 |
| 977.23 | 4.34 | 0.04 | 1371.90 | 4.53 | 0.07 | 2044.74 | 4.62 | 0.00 | 0.00 | 4.85 | 0.00 | 0.00 |
| 978.80 | 4.33 | 0.04 | 1366.36 | 4.53 | 0.07 | 2038.38 | 4.62 | 0.00 | 0.00 | 4.85 | 0.00 | 0.00 |
| 980.37 | 4.33 | 0.04 | 1360.97 | 4.52 | 0.07 | 2032.16 | 4.61 | 0.00 | 0.00 | 4.85 | 0.00 | 0.00 |
| 981.94 | 4.33 | 0.04 | 1355.60 | 4.52 | 0.07 | 2025.85 | 4.61 | 0.00 | 0.00 | 4.85 | 0.00 | 0.00 |
| 983.51 | 4.33 | 0.04 | 1350.24 | 4.52 | 0.07 | 2019.67 | 4.61 | 0.00 | 0.00 | 4.84 | 0.00 | 0.00 |
| 985.07 | 4.33 | 0.04 | 1344.90 | 4.52 | 0.07 | 2013.40 | 4.61 | 0.00 | 0.00 | 4.84 | 0.00 | 0.00 |
| 986.64 | 4.33 | 0.04 | 1339.58 | 4.52 | 0.07 | 2007.27 | 4.61 | 0.00 | 0.00 | 4.84 | 0.00 | 0.00 |
| 988.21 | 4.33 | 0.04 | 1334.40 | 4.52 | 0.07 | 2001.16 | 4.60 | 0.00 | 0.00 | 4.84 | 0.00 | 0.00 |
| 989.78 | 4.33 | 0.04 | 1329.11 | 4.52 | 0.07 | 1994.94 | 4.60 | 0.00 | 0.00 | 4.84 | 0.00 | 0.00 |
| 991.35 | 4.32 | 0.04 | 1323.97 | 4.51 | 0.07 | 1988.87 | 4.60 | 0.00 | 0.00 | 4.83 | 0.00 | 0.00 |
| 992.92 | 4.32 | 0.04 | 1318.71 | 4.51 | 0.07 | 1982.82 | 4.60 | 0.00 | 0.00 | 4.83 | 0.00 | 0.00 |
| 994.48 | 4.32 | 0.04 | 1313.60 | 4.51 | 0.07 | 1976.78 | 4.60 | 0.00 | 0.00 | 4.83 | 0.00 | 0.00 |
| 996.05 | 4.32 | 0.04 | 1308.51 | 4.51 | 0.07 | 1970.77 | 4.59 | 0.00 | 0.00 | 4.83 | 0.00 | 0.00 |
| 997.62 | 4.32 | 0.04 | 1303.56 | 4.51 | 0.07 | 1964.78 | 4.59 | 0.00 | 0.00 | 4.83 | 0.00 | 0.00 |
| 999.19 | 4.32 | 0.04 | 1298.49 | 4.51 | 0.07 | 1958.81 | 4.59 | 0.00 | 0.00 | 4.83 | 0.00 | 0.00 |
| 1013.00 | 4.31 | 0.04 | 1255.48 | 4.50 | 0.07 | 1907.54 | 4.58 | 0.00 | 0.00 | 4.81 | 0.00 | 0.00 |
| 1016.40 | 4.30 | 0.04 | 1245.22 | 4.49 | 0.06 | 1895.21 | 4.57 | 0.00 | 0.00 | 4.81 | 0.00 | 0.00 |
| 1019.81 | 4.30 | 0.04 | 1235.02 | 4.49 | 0.06 | 1882.97 | 4.57 | 0.00 | 0.00 | 4.80 | 0.00 | 0.00 |
| 1023.22 | 4.30 | 0.04 | 1225.02 | 4.49 | 0.06 | 1870.81 | 4.56 | 0.00 | 0.00 | 4.80 | 0.00 | 0.00 |
| 1026.62 | 4.30 | 0.04 | 1215.20 | 4.49 | 0.06 | 1858.85 | 4.56 | 0.00 | 0.00 | 4.79 | 0.00 | 0.00 |
| 1030.03 | 4.29 | 0.04 | 1205.45 | 4.48 | 0.06 | 1846.97 | 4.56 | 0.00 | 0.00 | 4.79 | 0.00 | 0.00 |
| 1033.43 | 4.29 | 0.04 | 1195.76 | 4.48 | 0.06 | 1835.17 | 4.55 | 0.00 | 0.00 | 4.79 | 0.00 | 0.00 |
| 1036.84 | 4.29 | 0.04 | 1186.26 | 4.48 | 0.06 | 1823.56 | 4.55 | 0.00 | 0.00 | 4.78 | 0.00 | 0.00 |
| 1040.25 | 4.29 | 0.04 | 1176.82 | 4.48 | 0.06 | 1812.03 | 4.55 | 0.00 | 0.00 | 4.78 | 0.00 | 0.00 |
| 1043.66 | 4.28 | 0.04 | 1167.56 | 4.47 | 0.06 | 1800.58 | 4.54 | 0.00 | 0.00 | 4.78 | 0.00 | 0.00 |
| 1047.06 | 4.28 | 0.04 | 1158.48 | 4.47 | 0.06 | 1789.20 | 4.54 | 0.00 | 0.00 | 4.77 | 0.00 | 0.00 |
| 1050.47 | 4.28 | 0.04 | 1149.34 | 4.47 | 0.06 | 1778.01 | 4.54 | 0.00 | 0.00 | 4.77 | 0.00 | 0.00 |
| 1053.88 | 4.28 | 0.04 | 1140.37 | 4.46 | 0.06 | 1766.89 | 4.53 | 0.00 | 0.00 | 4.77 | 0.00 | 0.00 |
| 1057.29 | 4.27 | 0.04 | 1131.59 | 4.46 | 0.06 | 1755.85 | 4.53 | 0.00 | 0.00 | 4.76 | 0.00 | 0.00 |
| 1060.70 | 4.27 | 0.04 | 1122.86 | 4.46 | 0.06 | 1744.87 | 4.53 | 0.00 | 0.00 | 4.76 | 0.00 | 0.00 |
| 1064.11 | 4.27 | 0.04 | 1114.30 | 4.46 | 0.06 | 1734.09 | 4.52 | 0.00 | 0.00 | 4.76 | 0.00 | 0.00 |
| 1067.52 | 4.27 | 0.04 | 1105.68 | 4.46 | 0.06 | 1723.37 | 4.52 | 0.00 | 0.00 | 4.75 | 0.00 | 0.00 |
| 1070.93 | 4.26 | 0.04 | 1097.35 | 4.45 | 0.06 | 1712.72 | 4.52 | 0.00 | 0.00 | 4.75 | 0.00 | 0.00 |
| 1074.34 | 4.26 | 0.04 | 1088.95 | 4.45 | 0.06 | 1702.25 | 4.51 | 0.00 | 0.00 | 4.75 | 0.00 | 0.00 |
| 1077.75 | 4.26 | 0.04 | 1080.72 | 4.45 | 0.06 | 1691.74 | 4.51 | 0.00 | 0.00 | 4.74 | 0.00 | 0.00 |
| 1081.16 | 4.26 | 0.04 | 1072.67 | 4.45 | 0.06 | 1681.40 | 4.51 | 0.00 | 0.00 | 4.74 | 0.00 | 0.00 |
| 1084.58 | 4.26 | 0.04 | 1064.66 | 4.44 | 0.06 | 1671.13 | 4.51 | 0.00 | 0.00 | 4.74 | 0.00 | 0.00 |
| 1087.99 | 4.25 | 0.04 | 1056.70 | 4.44 | 0.06 | 1661.04 | 4.50 | 0.00 | 0.00 | 4.73 | 0.00 | 0.00 |
| 1091.40 | 4.25 | 0.04 | 1048.79 | 4.44 | 0.06 | 1651.01 | 4.50 | 0.00 | 0.00 | 4.73 | 0.00 | 0.00 |
| 1094.81 | 4.25 | 0.04 | 1041.04 | 4.44 | 0.06 | 1641.04 | 4.50 | 0.00 | 0.00 | 4.73 | 0.00 | 0.00 |
| 1098.23 | 4.25 | 0.04 | 1033.46 | 4.43 | 0.06 | 1631.14 | 4.49 | 0.00 | 0.00 | 4.72 | 0.00 | 0.00 |
| 1101.64 | 4.24 | 0.04 | 1025.81 | 4.43 | 0.06 | 1621.29 | 4.49 | 0.00 | 0.00 | 4.72 | 0.00 | 0.00 |
| 1105.05 | 4.24 | 0.04 | 1018.32 | 4.43 | 0.06 | 1611.62 | 4.49 | 0.00 | 0.00 | 4.72 | 0.00 | 0.00 |
| 1108.47 | 4.24 | 0.04 | 1010.99 | 4.43 | 0.06 | 1602.01 | 4.49 | 0.00 | 0.00 | 4.72 | 0.00 | 0.00 |
| 1111.88 | 4.24 | 0.04 | 1003.59 | 4.43 | 0.06 | 1592.46 | 4.48 | 0.00 | 0.00 | 4.71 | 0.00 | 0.00 |
| 1115.30 | 4.24 | 0.04 | 996.35 | 4.42 | 0.06 | 1582.96 | 4.48 | 0.00 | 0.00 | 4.71 | 0.00 | 0.00 |
| 1118.71 | 4.23 | 0.04 | 989.27 | 4.42 | 0.06 | 1573.64 | 4.48 | 0.00 | 0.00 | 4.71 | 0.00 | 0.00 |
| 1122.13 | 4.23 | 0.04 | 982.11 | 4.42 | 0.06 | 1564.37 | 4.47 | 0.00 | 0.00 | 4.70 | 0.00 | 0.00 |
| 1125.54 | 4.23 | 0.04 | 975.11 | 4.42 | 0.06 | 1555.15 | 4.47 | 0.00 | 0.00 | 4.70 | 0.00 | 0.00 |
| 1128.96 | 4.23 | 0.04 | 968.27 | 4.41 | 0.06 | 1546.00 | 4.47 | 0.00 | 0.00 | 4.70 | 0.00 | 0.00 |
| 1132.38 | 4.23 | 0.04 | 961.35 | 4.41 | 0.06 | 1537.00 | 4.47 | 0.00 | 0.00 | 4.70 | 0.00 | 0.00 |
| 1135.79 | 4.22 | 0.04 | 954.59 | 4.41 | 0.06 | 1527.96 | 4.46 | 0.00 | 0.00 | 4.69 | 0.00 | 0.00 |
| 1139.21 | 4.22 | 0.04 | 947.97 | 4.41 | 0.06 | 1519.07 | 4.46 | 0.00 | 0.00 | 4.69 | 0.00 | 0.00 |
| 1142.63 | 4.22 | 0.04 | 941.29 | 4.41 | 0.06 | 1510.24 | 4.46 | 0.00 | 0.00 | 4.69 | 0.00 | 0.00 |
| 1146.05 | 4.22 | 0.04 | 934.75 | 4.40 | 0.06 | 1501.57 | 4.46 | 0.00 | 0.00 | 4.68 | 0.00 | 0.00 |
| 1149.47 | 4.22 | 0.03 | 928.26 | 4.40 | 0.06 | 1492.84 | 4.45 | 0.00 | 0.00 | 4.68 | 0.00 | 0.00 |
| 1152.88 | 4.21 | 0.03 | 921.91 | 4.40 | 0.06 | 1484.27 | 4.45 | 0.00 | 0.00 | 4.68 | 0.00 | 0.00 |
| 1156.30 | 4.21 | 0.03 | 915.49 | 4.40 | 0.06 | 1475.75 | 4.45 | 0.00 | 0.00 | 4.68 | 0.00 | 0.00 |
| 1159.72 | 4.21 | 0.03 | 909.21 | 4.40 | 0.06 | 1467.28 | 4.44 | 0.00 | 0.00 | 4.67 | 0.00 | 0.00 |
| 1163.14 | 4.21 | 0.03 | 903.08 | 4.39 | 0.06 | 1458.97 | 4.44 | 0.00 | 0.00 | 4.67 | 0.00 | 0.00 |
| 1166.56 | 4.21 | 0.03 | 896.88 | 4.39 | 0.06 | 1450.71 | 4.44 | 0.00 | 0.00 | 4.67 | 0.00 | 0.00 |
| 1169.98 | 4.20 | 0.03 | 890.82 | 4.39 | 0.06 | 1442.39 | 4.44 | 0.00 | 0.00 | 4.67 | 0.00 | 0.00 |
| 1173.41 | 4.20 | 0.03 | 884.80 | 4.39 | 0.06 | 1434.22 | 4.43 | 0.00 | 0.00 | 4.66 | 0.00 | 0.00 |
| 1176.83 | 4.20 | 0.03 | 878.91 | 4.39 | 0.06 | 1426.21 | 4.43 | 0.00 | 0.00 | 4.66 | 0.00 | 0.00 |
| 1180.25 | 4.20 | 0.03 | 872.96 | 4.38 | 0.06 | 1418.13 | 4.43 | 0.00 | 0.00 | 4.66 | 0.00 | 0.00 |
| 1183.67 | 4.20 | 0.03 | 867.15 | 4.38 | 0.06 | 1410.21 | 4.43 | 0.00 | 0.00 | 4.66 | 0.00 | 0.00 |
| 1187.09 | 4.20 | 0.03 | 861.36 | 4.38 | 0.06 | 1402.34 | 4.42 | 0.00 | 0.00 | 4.65 | 0.00 | 0.00 |
| 1190.51 | 4.19 | 0.03 | 855.72 | 4.38 | 0.06 | 1394.51 | 4.42 | 0.00 | 0.00 | 4.65 | 0.00 | 0.00 |
| 1193.94 | 4.19 | 0.03 | 850.01 | 4.38 | 0.06 | 1386.72 | 4.42 | 0.00 | 0.00 | 4.65 | 0.00 | 0.00 |
| 1197.36 | 4.19 | 0.03 | 844.43 | 4.37 | 0.05 | 1378.98 | 4.42 | 0.00 | 0.00 | 4.65 | 0.00 | 0.00 |
| 1200.78 | 4.19 | 0.03 | 838.88 | 4.37 | 0.05 | 1371.38 | 4.42 | 0.00 | 0.00 | 4.64 | 0.00 | 0.00 |
| 1204.21 | 4.19 | 0.03 | 833.47 | 4.37 | 0.05 | 1363.83 | 4.41 | 0.00 | 0.00 | 4.64 | 0.00 | 0.00 |
| 1207.63 | 4.18 | 0.03 | 827.99 | 4.37 | 0.05 | 1356.32 | 4.41 | 0.00 | 0.00 | 4.64 | 0.00 | 0.00 |
| 1211.06 | 4.18 | 0.03 | 822.64 | 4.37 | 0.05 | 1348.86 | 4.41 | 0.00 | 0.00 | 4.64 | 0.00 | 0.00 |
| 1214.48 | 4.18 | 0.03 | 817.31 | 4.37 | 0.05 | 1341.53 | 4.41 | 0.00 | 0.00 | 4.63 | 0.00 | 0.00 |
| 1217.91 | 4.18 | 0.03 | 812.13 | 4.36 | 0.05 | 1334.15 | 4.40 | 0.00 | 0.00 | 4.63 | 0.00 | 0.00 |
| 1221.33 | 4.18 | 0.03 | 806.87 | 4.36 | 0.05 | 1326.91 | 4.40 | 0.00 | 0.00 | 4.63 | 0.00 | 0.00 |
| 1224.76 | 4.18 | 0.03 | 801.74 | 4.36 | 0.05 | 1319.71 | 4.40 | 0.00 | 0.00 | 4.63 | 0.00 | 0.00 |
| 1228.19 | 4.17 | 0.03 | 796.64 | 4.36 | 0.05 | 1312.55 | 4.40 | 0.00 | 0.00 | 4.62 | 0.00 | 0.00 |
| 1231.61 | 4.17 | 0.03 | 791.56 | 4.36 | 0.05 | 1305.43 | 4.39 | 0.00 | 0.00 | 4.62 | 0.00 | 0.00 |
| 1235.04 | 4.17 | 0.03 | 786.62 | 4.35 | 0.05 | 1298.45 | 4.39 | 0.00 | 0.00 | 4.62 | 0.00 | 0.00 |
| 1238.47 | 4.17 | 0.03 | 781.60 | 4.35 | 0.05 | 1291.41 | 4.39 | 0.00 | 0.00 | 4.62 | 0.00 | 0.00 |
| 1241.90 | 4.17 | 0.03 | 776.71 | 4.35 | 0.05 | 1284.50 | 4.39 | 0.00 | 0.00 | 4.61 | 0.00 | 0.00 |
| 1245.32 | 4.17 | 0.03 | 771.85 | 4.35 | 0.05 | 1277.64 | 4.39 | 0.00 | 0.00 | 4.61 | 0.00 | 0.00 |
| 1248.75 | 4.16 | 0.03 | 767.01 | 4.35 | 0.05 | 1270.81 | 4.38 | 0.00 | 0.00 | 4.61 | 0.00 | 0.00 |
| 1252.18 | 4.16 | 0.03 | 762.30 | 4.35 | 0.05 | 1264.12 | 4.38 | 0.00 | 0.00 | 4.61 | 0.00 | 0.00 |
| 1255.61 | 4.16 | 0.03 | 757.52 | 4.34 | 0.05 | 1257.36 | 4.38 | 0.00 | 0.00 | 4.61 | 0.00 | 0.00 |
| 1259.04 | 4.16 | 0.03 | 752.86 | 4.34 | 0.05 | 1250.75 | 4.38 | 0.00 | 0.00 | 4.60 | 0.00 | 0.00 |
| 1262.47 | 4.16 | 0.03 | 748.23 | 4.34 | 0.05 | 1244.16 | 4.37 | 0.00 | 0.00 | 4.60 | 0.00 | 0.00 |
| 1265.90 | 4.16 | 0.03 | 743.62 | 4.34 | 0.05 | 1237.62 | 4.37 | 0.00 | 0.00 | 4.60 | 0.00 | 0.00 |
| 1269.33 | 4.15 | 0.03 | 739.14 | 4.34 | 0.05 | 1231.10 | 4.37 | 0.00 | 0.00 | 4.60 | 0.00 | 0.00 |
| 1272.76 | 4.15 | 0.03 | 734.58 | 4.34 | 0.05 | 1224.72 | 4.37 | 0.00 | 0.00 | 4.59 | 0.00 | 0.00 |
| 1276.19 | 4.15 | 0.03 | 730.14 | 4.33 | 0.05 | 1218.28 | 4.37 | 0.00 | 0.00 | 4.59 | 0.00 | 0.00 |
| 1279.63 | 4.15 | 0.03 | 725.73 | 4.33 | 0.05 | 1211.97 | 4.36 | 0.00 | 0.00 | 4.59 | 0.00 | 0.00 |
| 1283.06 | 4.15 | 0.03 | 721.34 | 4.33 | 0.05 | 1205.69 | 4.36 | 0.00 | 0.00 | 4.59 | 0.00 | 0.00 |
| 1286.49 | 4.15 | 0.03 | 717.07 | 4.33 | 0.05 | 1199.45 | 4.36 | 0.00 | 0.00 | 4.59 | 0.00 | 0.00 |
| 1289.92 | 4.15 | 0.03 | 712.73 | 4.33 | 0.05 | 1193.23 | 4.36 | 0.00 | 0.00 | 4.58 | 0.00 | 0.00 |
| 1293.36 | 4.14 | 0.03 | 708.50 | 4.32 | 0.05 | 1187.06 | 4.36 | 0.00 | 0.00 | 4.58 | 0.00 | 0.00 |
| 1296.79 | 4.14 | 0.03 | 704.30 | 4.32 | 0.05 | 1181.01 | 4.35 | 0.00 | 0.00 | 4.58 | 0.00 | 0.00 |
| 1300.22 | 4.14 | 0.03 | 700.12 | 4.32 | 0.05 | 1174.89 | 4.35 | 0.00 | 0.00 | 4.58 | 0.00 | 0.00 |
| 1303.66 | 4.14 | 0.03 | 695.97 | 4.32 | 0.05 | 1168.91 | 4.35 | 0.00 | 0.00 | 4.57 | 0.00 | 0.00 |
| 1307.09 | 4.14 | 0.03 | 691.93 | 4.32 | 0.05 | 1162.95 | 4.35 | 0.00 | 0.00 | 4.57 | 0.00 | 0.00 |
| 1310.53 | 4.14 | 0.03 | 687.81 | 4.32 | 0.05 | 1157.03 | 4.35 | 0.00 | 0.00 | 4.57 | 0.00 | 0.00 |
| 1313.96 | 4.13 | 0.03 | 683.81 | 4.32 | 0.05 | 1151.13 | 4.34 | 0.00 | 0.00 | 4.57 | 0.00 | 0.00 |
| 1317.40 | 4.13 | 0.03 | 679.84 | 4.31 | 0.05 | 1145.36 | 4.34 | 0.00 | 0.00 | 4.57 | 0.00 | 0.00 |
| 1320.83 | 4.13 | 0.03 | 675.88 | 4.31 | 0.05 | 1139.53 | 4.34 | 0.00 | 0.00 | 4.56 | 0.00 | 0.00 |
| 1324.27 | 4.13 | 0.03 | 671.94 | 4.31 | 0.05 | 1133.82 | 4.34 | 0.00 | 0.00 | 4.56 | 0.00 | 0.00 |
| 1327.71 | 4.13 | 0.03 | 668.03 | 4.31 | 0.05 | 1128.14 | 4.34 | 0.00 | 0.00 | 4.56 | 0.00 | 0.00 |
| 1331.14 | 4.13 | 0.03 | 664.23 | 4.31 | 0.05 | 1122.49 | 4.33 | 0.00 | 0.00 | 4.56 | 0.00 | 0.00 |
| 1334.58 | 4.13 | 0.03 | 660.44 | 4.31 | 0.05 | 1116.87 | 4.33 | 0.00 | 0.00 | 4.56 | 0.00 | 0.00 |
| 1338.02 | 4.12 | 0.03 | 656.59 | 4.30 | 0.05 | 1111.28 | 4.33 | 0.00 | 0.00 | 4.55 | 0.00 | 0.00 |
| 1341.46 | 4.12 | 0.03 | 652.84 | 4.30 | 0.05 | 1105.81 | 4.33 | 0.00 | 0.00 | 4.55 | 0.00 | 0.00 |
| 1344.90 | 4.12 | 0.03 | 649.21 | 4.30 | 0.05 | 1100.27 | 4.33 | 0.00 | 0.00 | 4.55 | 0.00 | 0.00 |
| 1348.33 | 4.12 | 0.03 | 645.51 | 4.30 | 0.05 | 1094.86 | 4.32 | 0.00 | 0.00 | 4.55 | 0.00 | 0.00 |
| 1351.77 | 4.12 | 0.03 | 641.82 | 4.30 | 0.05 | 1089.47 | 4.32 | 0.00 | 0.00 | 4.55 | 0.00 | 0.00 |
| 1355.21 | 4.12 | 0.03 | 638.24 | 4.30 | 0.05 | 1084.11 | 4.32 | 0.00 | 0.00 | 4.54 | 0.00 | 0.00 |
| 1358.65 | 4.12 | 0.03 | 634.69 | 4.29 | 0.05 | 1078.77 | 4.32 | 0.00 | 0.00 | 4.54 | 0.00 | 0.00 |
| 1362.09 | 4.11 | 0.03 | 631.05 | 4.29 | 0.05 | 1073.47 | 4.32 | 0.00 | 0.00 | 4.54 | 0.00 | 0.00 |
| 1365.53 | 4.11 | 0.03 | 627.53 | 4.29 | 0.05 | 1068.18 | 4.31 | 0.00 | 0.00 | 4.54 | 0.00 | 0.00 |
| 1368.97 | 4.11 | 0.03 | 624.12 | 4.29 | 0.05 | 1063.02 | 4.31 | 0.00 | 0.00 | 4.54 | 0.00 | 0.00 |
| 1372.42 | 4.11 | 0.03 | 620.63 | 4.29 | 0.05 | 1057.79 | 4.31 | 0.00 | 0.00 | 4.53 | 0.00 | 0.00 |
| 1375.86 | 4.11 | 0.03 | 617.16 | 4.29 | 0.05 | 1052.68 | 4.31 | 0.00 | 0.00 | 4.53 | 0.00 | 0.00 |
| 1379.30 | 4.11 | 0.03 | 613.80 | 4.28 | 0.05 | 1047.60 | 4.31 | 0.00 | 0.00 | 4.53 | 0.00 | 0.00 |
| 1382.74 | 4.11 | 0.03 | 610.45 | 4.28 | 0.05 | 1042.53 | 4.31 | 0.00 | 0.00 | 4.53 | 0.00 | 0.00 |
| 1386.18 | 4.10 | 0.03 | 607.03 | 4.28 | 0.05 | 1037.50 | 4.30 | 0.00 | 0.00 | 4.53 | 0.00 | 0.00 |
| 1389.63 | 4.10 | 0.03 | 603.72 | 4.28 | 0.05 | 1032.58 | 4.30 | 0.00 | 0.00 | 4.52 | 0.00 | 0.00 |
| 1393.07 | 4.10 | 0.03 | 600.43 | 4.28 | 0.05 | 1027.59 | 4.30 | 0.00 | 0.00 | 4.52 | 0.00 | 0.00 |
| 1396.51 | 4.10 | 0.03 | 597.24 | 4.28 | 0.05 | 1022.72 | 4.30 | 0.00 | 0.00 | 4.52 | 0.00 | 0.00 |
| 1399.96 | 4.10 | 0.03 | 593.97 | 4.28 | 0.05 | 1017.78 | 4.30 | 0.00 | 0.00 | 4.52 | 0.00 | 0.00 |
| 1403.40 | 4.10 | 0.03 | 590.81 | 4.27 | 0.05 | 1012.95 | 4.29 | 0.00 | 0.00 | 4.52 | 0.00 | 0.00 |
| 1406.85 | 4.10 | 0.03 | 587.58 | 4.27 | 0.05 | 1008.15 | 4.29 | 0.00 | 0.00 | 4.51 | 0.00 | 0.00 |
| 1410.29 | 4.10 | 0.03 | 584.45 | 4.27 | 0.05 | 1003.37 | 4.29 | 0.00 | 0.00 | 4.51 | 0.00 | 0.00 |
| 1413.74 | 4.09 | 0.03 | 581.34 | 4.27 | 0.05 | 998.61 | 4.29 | 0.00 | 0.00 | 4.51 | 0.00 | 0.00 |
| 1417.18 | 4.09 | 0.03 | 578.24 | 4.27 | 0.05 | 993.97 | 4.29 | 0.00 | 0.00 | 4.51 | 0.00 | 0.00 |
| 1420.63 | 4.09 | 0.03 | 575.16 | 4.27 | 0.05 | 989.26 | 4.29 | 0.00 | 0.00 | 4.51 | 0.00 | 0.00 |
| 1424.07 | 4.09 | 0.03 | 572.09 | 4.26 | 0.05 | 984.66 | 4.28 | 0.00 | 0.00 | 4.51 | 0.00 | 0.00 |
| 1427.52 | 4.09 | 0.03 | 569.04 | 4.26 | 0.05 | 979.99 | 4.28 | 0.00 | 0.00 | 4.50 | 0.00 | 0.00 |
| 1430.97 | 4.09 | 0.03 | 566.08 | 4.26 | 0.05 | 975.44 | 4.28 | 0.00 | 0.00 | 4.50 | 0.00 | 0.00 |
| 1434.42 | 4.09 | 0.03 | 563.06 | 4.26 | 0.05 | 970.90 | 4.28 | 0.00 | 0.00 | 4.50 | 0.00 | 0.00 |
| 1437.86 | 4.08 | 0.03 | 560.14 | 4.26 | 0.05 | 966.39 | 4.28 | 0.00 | 0.00 | 4.50 | 0.00 | 0.00 |
| 1441.31 | 4.08 | 0.03 | 557.23 | 4.26 | 0.05 | 961.90 | 4.27 | 0.00 | 0.00 | 4.50 | 0.00 | 0.00 |
| 1444.76 | 4.08 | 0.03 | 554.33 | 4.26 | 0.05 | 957.43 | 4.27 | 0.00 | 0.00 | 4.49 | 0.00 | 0.00 |
| 1448.21 | 4.08 | 0.03 | 551.45 | 4.25 | 0.05 | 953.07 | 4.27 | 0.00 | 0.00 | 4.49 | 0.00 | 0.00 |
| 1451.66 | 4.08 | 0.03 | 548.58 | 4.25 | 0.05 | 948.64 | 4.27 | 0.00 | 0.00 | 4.49 | 0.00 | 0.00 |
| 1455.11 | 4.08 | 0.03 | 545.73 | 4.25 | 0.05 | 944.32 | 4.27 | 0.00 | 0.00 | 4.49 | 0.00 | 0.00 |
| 1458.56 | 4.08 | 0.03 | 542.89 | 4.25 | 0.04 | 940.01 | 4.27 | 0.00 | 0.00 | 4.49 | 0.00 | 0.00 |
| 1462.01 | 4.08 | 0.03 | 540.14 | 4.25 | 0.04 | 935.65 | 4.26 | 0.00 | 0.00 | 4.48 | 0.00 | 0.00 |
| 1465.46 | 4.07 | 0.03 | 537.33 | 4.25 | 0.04 | 931.39 | 4.26 | 0.00 | 0.00 | 4.48 | 0.00 | 0.00 |
| 1468.91 | 4.07 | 0.03 | 534.61 | 4.25 | 0.04 | 927.15 | 4.26 | 0.00 | 0.00 | 4.48 | 0.00 | 0.00 |
| 1472.36 | 4.07 | 0.03 | 531.91 | 4.24 | 0.04 | 923.01 | 4.26 | 0.00 | 0.00 | 4.48 | 0.00 | 0.00 |
| 1475.81 | 4.07 | 0.03 | 529.22 | 4.24 | 0.04 | 918.81 | 4.26 | 0.00 | 0.00 | 4.48 | 0.00 | 0.00 |
| 1479.26 | 4.07 | 0.03 | 526.54 | 4.24 | 0.04 | 914.63 | 4.26 | 0.00 | 0.00 | 4.48 | 0.00 | 0.00 |
| 1482.72 | 4.07 | 0.02 | 523.87 | 4.24 | 0.04 | 910.55 | 4.25 | 0.00 | 0.00 | 4.47 | 0.00 | 0.00 |
| 1486.17 | 4.07 | 0.02 | 521.22 | 4.24 | 0.04 | 906.40 | 4.25 | 0.00 | 0.00 | 4.47 | 0.00 | 0.00 |
| 1489.62 | 4.07 | 0.02 | 518.58 | 4.24 | 0.04 | 902.36 | 4.25 | 0.00 | 0.00 | 4.47 | 0.00 | 0.00 |
| 1493.07 | 4.06 | 0.02 | 515.95 | 4.23 | 0.04 | 898.34 | 4.25 | 0.00 | 0.00 | 4.47 | 0.00 | 0.00 |
| 1496.53 | 4.06 | 0.02 | 513.41 | 4.23 | 0.04 | 894.33 | 4.25 | 0.00 | 0.00 | 4.47 | 0.00 | 0.00 |
| 1499.98 | 4.06 | 0.02 | 510.89 | 4.23 | 0.04 | 890.35 | 4.25 | 0.00 | 0.00 | 4.46 | 0.00 | 0.00 |
| 1503.44 | 4.06 | 0.02 | 508.29 | 4.23 | 0.04 | 886.38 | 4.24 | 0.00 | 0.00 | 4.46 | 0.00 | 0.00 |
| 1506.89 | 4.06 | 0.02 | 505.79 | 4.23 | 0.04 | 882.43 | 4.24 | 0.00 | 0.00 | 4.46 | 0.00 | 0.00 |
| 1510.35 | 4.06 | 0.02 | 503.31 | 4.23 | 0.04 | 878.50 | 4.24 | 0.00 | 0.00 | 4.46 | 0.00 | 0.00 |
| 1513.80 | 4.06 | 0.02 | 500.83 | 4.23 | 0.04 | 874.67 | 4.24 | 0.00 | 0.00 | 4.46 | 0.00 | 0.00 |
| 1517.26 | 4.06 | 0.02 | 498.36 | 4.22 | 0.04 | 870.77 | 4.24 | 0.00 | 0.00 | 4.46 | 0.00 | 0.00 |
| 1520.71 | 4.05 | 0.02 | 495.91 | 4.22 | 0.04 | 866.97 | 4.23 | 0.00 | 0.00 | 4.45 | 0.00 | 0.00 |
| 1524.17 | 4.05 | 0.02 | 493.47 | 4.22 | 0.04 | 863.19 | 4.23 | 0.00 | 0.00 | 4.45 | 0.00 | 0.00 |
| 1527.63 | 4.05 | 0.02 | 491.03 | 4.22 | 0.04 | 859.43 | 4.23 | 0.00 | 0.00 | 4.45 | 0.00 | 0.00 |
| 1531.09 | 4.05 | 0.02 | 488.69 | 4.22 | 0.04 | 855.68 | 4.23 | 0.00 | 0.00 | 4.45 | 0.00 | 0.00 |
| 1534.54 | 4.05 | 0.02 | 486.28 | 4.22 | 0.04 | 851.96 | 4.23 | 0.00 | 0.00 | 4.45 | 0.00 | 0.00 |
| 1538.00 | 4.05 | 0.02 | 483.96 | 4.22 | 0.04 | 848.24 | 4.23 | 0.00 | 0.00 | 4.45 | 0.00 | 0.00 |
| 1541.46 | 4.05 | 0.02 | 481.65 | 4.21 | 0.04 | 844.55 | 4.22 | 0.00 | 0.00 | 4.44 | 0.00 | 0.00 |
| 1544.92 | 4.05 | 0.02 | 479.28 | 4.21 | 0.04 | 840.87 | 4.22 | 0.00 | 0.00 | 4.44 | 0.00 | 0.00 |
| 1548.38 | 4.04 | 0.02 | 476.99 | 4.21 | 0.04 | 837.28 | 4.22 | 0.00 | 0.00 | 4.44 | 0.00 | 0.00 |
| 1551.84 | 4.04 | 0.02 | 474.71 | 4.21 | 0.04 | 833.64 | 4.22 | 0.00 | 0.00 | 4.44 | 0.00 | 0.00 |
| 1555.30 | 4.04 | 0.02 | 472.44 | 4.21 | 0.04 | 830.09 | 4.22 | 0.00 | 0.00 | 4.44 | 0.00 | 0.00 |
| 1558.76 | 4.04 | 0.02 | 470.18 | 4.21 | 0.04 | 826.47 | 4.22 | 0.00 | 0.00 | 4.43 | 0.00 | 0.00 |
| 1562.22 | 4.04 | 0.02 | 467.94 | 4.21 | 0.04 | 822.95 | 4.21 | 0.00 | 0.00 | 4.43 | 0.00 | 0.00 |
| 1565.68 | 4.04 | 0.02 | 465.78 | 4.20 | 0.04 | 819.45 | 4.21 | 0.00 | 0.00 | 4.43 | 0.00 | 0.00 |
| 1569.14 | 4.04 | 0.02 | 463.55 | 4.20 | 0.04 | 815.96 | 4.21 | 0.00 | 0.00 | 4.43 | 0.00 | 0.00 |
| 1572.60 | 4.04 | 0.02 | 461.33 | 4.20 | 0.04 | 812.48 | 4.21 | 0.00 | 0.00 | 4.43 | 0.00 | 0.00 |
| 1576.06 | 4.03 | 0.02 | 459.20 | 4.20 | 0.04 | 809.02 | 4.21 | 0.00 | 0.00 | 4.43 | 0.00 | 0.00 |
| 1579.52 | 4.03 | 0.02 | 457.00 | 4.20 | 0.04 | 805.58 | 4.21 | 0.00 | 0.00 | 4.42 | 0.00 | 0.00 |
| 1582.99 | 4.03 | 0.02 | 454.89 | 4.20 | 0.04 | 802.23 | 4.21 | 0.00 | 0.00 | 4.42 | 0.00 | 0.00 |
| 1586.45 | 4.03 | 0.02 | 452.79 | 4.20 | 0.04 | 798.82 | 4.20 | 0.00 | 0.00 | 4.42 | 0.00 | 0.00 |
| 1589.91 | 4.03 | 0.02 | 450.70 | 4.19 | 0.04 | 795.50 | 4.20 | 0.00 | 0.00 | 4.42 | 0.00 | 0.00 |
| 1593.37 | 4.03 | 0.02 | 448.61 | 4.19 | 0.04 | 792.11 | 4.20 | 0.00 | 0.00 | 4.42 | 0.00 | 0.00 |
| 1596.84 | 4.03 | 0.02 | 446.46 | 4.19 | 0.04 | 788.82 | 4.20 | 0.00 | 0.00 | 4.42 | 0.00 | 0.00 |
| 1600.30 | 4.03 | 0.02 | 444.47 | 4.19 | 0.04 | 785.54 | 4.20 | 0.00 | 0.00 | 4.41 | 0.00 | 0.00 |
| 1603.77 | 4.03 | 0.02 | 442.42 | 4.19 | 0.04 | 782.28 | 4.20 | 0.00 | 0.00 | 4.41 | 0.00 | 0.00 |
| 1607.23 | 4.02 | 0.02 | 440.37 | 4.19 | 0.04 | 779.03 | 4.19 | 0.00 | 0.00 | 4.41 | 0.00 | 0.00 |
| 1610.70 | 4.02 | 0.02 | 438.33 | 4.19 | 0.04 | 775.79 | 4.19 | 0.00 | 0.00 | 4.41 | 0.00 | 0.00 |
| 1614.16 | 4.02 | 0.02 | 436.30 | 4.18 | 0.04 | 772.57 | 4.19 | 0.00 | 0.00 | 4.41 | 0.00 | 0.00 |
| 1617.63 | 4.02 | 0.02 | 434.35 | 4.18 | 0.04 | 769.36 | 4.19 | 0.00 | 0.00 | 4.40 | 0.00 | 0.00 |
| 1621.10 | 4.02 | 0.02 | 432.34 | 4.18 | 0.04 | 766.16 | 4.19 | 0.00 | 0.00 | 4.40 | 0.00 | 0.00 |
| 1624.56 | 4.02 | 0.02 | 430.41 | 4.18 | 0.04 | 763.06 | 4.19 | 0.00 | 0.00 | 4.40 | 0.00 | 0.00 |
| 1628.03 | 4.02 | 0.02 | 428.49 | 4.18 | 0.04 | 759.89 | 4.18 | 0.00 | 0.00 | 4.40 | 0.00 | 0.00 |
| 1631.50 | 4.02 | 0.02 | 426.50 | 4.18 | 0.04 | 756.81 | 4.18 | 0.00 | 0.00 | 4.40 | 0.00 | 0.00 |
| 1634.96 | 4.01 | 0.02 | 424.60 | 4.18 | 0.04 | 753.67 | 4.18 | 0.00 | 0.00 | 4.40 | 0.00 | 0.00 |
| 1638.43 | 4.01 | 0.02 | 422.70 | 4.17 | 0.04 | 750.62 | 4.18 | 0.00 | 0.00 | 4.39 | 0.00 | 0.00 |
| 1641.90 | 4.01 | 0.02 | 420.81 | 4.17 | 0.04 | 747.58 | 4.18 | 0.00 | 0.00 | 4.39 | 0.00 | 0.00 |
| 1645.37 | 4.01 | 0.02 | 418.93 | 4.17 | 0.04 | 744.55 | 4.18 | 0.00 | 0.00 | 4.39 | 0.00 | 0.00 |
| 1648.84 | 4.01 | 0.02 | 417.06 | 4.17 | 0.04 | 741.54 | 4.17 | 0.00 | 0.00 | 4.39 | 0.00 | 0.00 |
| 1652.31 | 4.01 | 0.02 | 415.20 | 4.17 | 0.04 | 738.54 | 4.17 | 0.00 | 0.00 | 4.39 | 0.00 | 0.00 |
| 1655.78 | 4.01 | 0.02 | 413.34 | 4.17 | 0.04 | 735.55 | 4.17 | 0.00 | 0.00 | 4.39 | 0.00 | 0.00 |
| 1659.25 | 4.01 | 0.02 | 411.49 | 4.17 | 0.04 | 732.57 | 4.17 | 0.00 | 0.00 | 4.38 | 0.00 | 0.00 |
| 1662.72 | 4.00 | 0.02 | 409.73 | 4.16 | 0.04 | 729.60 | 4.17 | 0.00 | 0.00 | 4.38 | 0.00 | 0.00 |
| 1666.19 | 4.00 | 0.02 | 407.89 | 4.16 | 0.04 | 726.65 | 4.17 | 0.00 | 0.00 | 4.38 | 0.00 | 0.00 |
| 1669.66 | 4.00 | 0.02 | 406.07 | 4.16 | 0.04 | 723.79 | 4.16 | 0.00 | 0.00 | 4.38 | 0.00 | 0.00 |
| 1673.13 | 4.00 | 0.02 | 404.32 | 4.16 | 0.04 | 720.86 | 4.16 | 0.00 | 0.00 | 4.38 | 0.00 | 0.00 |
| 1676.60 | 4.00 | 0.02 | 402.51 | 4.16 | 0.04 | 718.02 | 4.16 | 0.00 | 0.00 | 4.38 | 0.00 | 0.00 |
| 1680.08 | 4.00 | 0.02 | 400.78 | 4.16 | 0.04 | 715.19 | 4.16 | 0.00 | 0.00 | 4.37 | 0.00 | 0.00 |
| 1683.55 | 4.00 | 0.02 | 399.06 | 4.16 | 0.04 | 712.29 | 4.16 | 0.00 | 0.00 | 4.37 | 0.00 | 0.00 |
| 1687.02 | 4.00 | 0.02 | 397.34 | 4.15 | 0.04 | 709.49 | 4.16 | 0.00 | 0.00 | 4.37 | 0.00 | 0.00 |